\documentclass[draft,numberedheadings,nomathfonts]{aipproc}
\layoutstyle{6x9}
\usepackage{enumerate}
\usepackage{fancyhdr,afterpage}\fancypagestyle{frontfoot}{\fancyhf{}
\fancyfoot[L]{in M.~Novello and S.~E.~Perez Bergliaffa (eds), {\em
Cosmology and Gravitation: XVth Brazilian School of Cosmology and Gravitation}
(Cambridge Scientific Publishers, 2014) pp.~203-244.}}
\begin{document}\date{}
\title[]{\textbf{Cosmic structure, averaging and dark energy\footnote{Based on
5 lectures at {\em 15th Brazilian School on Cosmology and Gravitation},
Mangatariba, August 2012.}}}
\thispagestyle{frontfoot}
\author{David L. Wiltshire}{
address={Department of Physics \& Astronomy, University of Canterbury, Private
Bag 4800,\\ Christchurch 8140, New Zealand}
}
\font\bm=cmmib10
\def\B#1{\hbox{\bm#1}} \def\Bom{\B{\char33}}
\def\sbf{\footnotesize\bf}
\def\ns#1{_{\rm #1}}
\def\PRL#1{{\em Phys.\ Rev.\ Lett.}\ {\bf#1}}
\def\JCAP#1#2{{\em J.\ Cosmol.\ Astropart.\ Phys.}\ {#1} (#2) }
\def\ApJ#1{{\em Astrophys.\ J.}\ {\bf#1}}
\def\PL#1#2{{\em Phys.\ Lett.\ #1}~{\bf#2}}
\def\PR#1#2{{\em Phys.\ Rev.}\ #1~{\bf#2}}\def\MPL#1#2{{\em Mod}.\ \PL{#1}{#2}}
\def\AaA#1{{\em Astron.\ Astrophys.}\ {\bf#1}}
\def\MNRAS#1{{\em Mon.\ Not.\ R.\ Astr.\ Soc.}\ {\bf#1}}
\def\CQG#1{{\em Class.\ Quantum Grav.}\ {\bf#1}}
\def\GRG#1{{\em Gen.\ Relativ.\ Grav.}\ {\bf#1}}
\def\IJMP#1#2{{\em Int.\ J.\ Mod.\ Phys.}\ #1~{\bf#2}}
\def\JMP#1{{\em J.\ Math.\ Phys.}\ {\bf #1}}
\def\RMP#1{{\em Rev.\ Mod.\ Phys.}\ {\bf#1}}

\def\beq{\begin{equation}} \def\eeq{\end{equation}}
\def\bea{\begin{eqnarray}} \def\eea{\end{eqnarray}}
\def\Z#1{_{\lower2pt\hbox{$\scriptstyle#1$}}} \def\w#1{\,\hbox{#1}}
\def\X#1{_{\lower2pt\hbox{$\scriptscriptstyle#1$}}}\def\dOM{\dd{\Omega\Z2}^2}
\def\metric#1{g^{\raise1pt\hbox{$\scriptstyle\rm#1$}}_{\mu\nu}}
\def\Ns#1{_{\lower2pt\hbox{$\scriptstyle\rm#1$}}} \def\ave#1{\langle{#1}\rangle}
\def\lsim{\mathop{\hbox{${\lower3.8pt\hbox{$<$}}\atop{\raise0.2pt\hbox{$\sim$}}
$}}} \def\kms{\w{km}\w{s}^{-1}}\def\kmsMpc{\kms\w{Mpc}^{-1}}\def\bn{\bar n}
\def\dd{{\rm d}} \def\ds{\dd s} \def\etal{{\em et al}.}\def\deg{^\circ}
\def\al{\alpha}\def\be{\beta}\def\ga{\gamma}\def\de{\delta}\def\ep{\epsilon}
\def\et{\eta}\def\th{\theta}\def\ph{\phi}\def\rh{\rho}\def\si{\sigma}
\def\la{\lambda}\def\GA{\Gamma}\def\SI{\Sigma}\def\LA{\Lambda}
\def\LCDM{$\LA$CDM}\def\Az{{\cal A}}\def\Tz{{\mathbf T}}
\def\gsim{\mathop{\hbox{${\lower3.8pt\hbox{$>$}}\atop{\raise0.2pt\hbox{$
\sim$}}$}}} \def\ta{\tau} \def\ac{a} \def\tn{t\Z0}
\def\frn#1#2{{\textstyle{#1\over#2}}} \def\Deriv#1#2#3{{#1#3\over#1#2}}
\def\Der#1#2{{#1\hphantom{#2}\over#1#2}} \def\pt{\partial} \def\ab{{\bar a}}
\def\gb{\bar\ga} \def\BB{{\cal B}} \def\CC{{\cal C}} \def\Aa{{\cal A}}
\def\tv{\ta\ns{v}}\def\tw{\ta\ns{w}}\def\gw{\gb\ns w}\def\gv{\gb\ns v}
\def\av{{a\ns{v}\hskip-2pt}} \def\aw{{a\ns{w}\hskip-2.4pt}}\def\Vav{{\cal V}}
\def\DD{{\cal D}}\def\gd{{{}^3\!g}}\def\half{\frn12}\def\Rav{\ave{\cal R}}
\def\QQ{{\cal Q}}\def\rw{r\ns w}
\def\mean#1{{\vphantom{\tilde#1}\bar#1}} 
\def\bx{{\mathbf x}}\def\bH{\mean H}\def\Hb{\bH\Z{\!0}}\def\bq{\mean q}
\def\gb{\mean\ga}\def\OMMn{\OM\Z{M0}}
\def\rhb{\mean\rh}\def\OM{\mean\Omega}\def\etb{\mean\eta}
\def\fw{{f\ns w}}\def\fv{{f\ns v}} \def\goesas{\mathop{\sim}\limits}
\def\fvn{f\ns{v0}}\def\fvf{\left(1-\fv\right)}\def\Hh{H}
\def\OMM{\OM\Z M}\def\OmMn{\Omega\Z{M0}}\def\ts{t}\def\tb{\ts'}\def\Hm{H\Z0}
\def\Vv{{\cal V}}\def\Mm{{\cal M}}\def\K{{\cal K}}\def\NN{{\cal N}}
\def\Fi{\hbox{\footnotesize\it fi}}\def\etw{\eta\ns w} \def\etv{\eta\ns v}
\def\fvi{{f\ns{vi}}}\def\fwi{{f\ns{wi}}}\def\gbn{\gb_{\ns0}}
\def\Hx{H_{\lower1pt\hbox{$\scriptstyle0$}}}\def\OmBn{\Omega_{\ns{B0}}}
\def\LCDM{$\Lambda$CDM}\def\OmLn{\Omega\Z{\Lambda0}}\def\Omkn{\Omega\Z{k0}}
\def\OMBn{\OM_{\ns{B0}}}\def\etbH{\etb\X{\!\cal H}}\def\mr{{\bar r}}
\def\OMR{\OM\Z R}\def\OMk{\OM\Z k}\def\OMQ{\OM\Z{\QQ}}\def\OMRn{\OM_{\ns{R0}}}
\def\Dfb{D\Ns{TS}}\def\Dlcdm{D\Z{\Lambda{\rm CDM}}}
\def\dL{d\Z L}\def\dA{d\Z A} \def\TS{timescape}\def\OMkn{\OM_{\ns{k0}}}
\def\OmBn{\Omega\Z{B0}}\def\OMCn{\OM_{\ns{C0}}}\def\pR{P_{\ns R}}
\def\h{\,h^{-1}}\def\hm{\h\hbox{Mpc}}\def\Gcos{\GA_{cos}}
\def\rhM{\rh_{\ns M}}\def\rhR{\rh_{\ns R}}\def\kv{k\ns v}\def\abn{\ab_{\ns0}}
\def\rhbMn{\rhb_{\ns{M0}}}\def\rhbRn{\rhb_{\ns{R0}}}
\def\Hv{H\ns v}\def\Hw{H\ns w}\def\vp{\vphantom{\Bigl|}}
\def\RV{R\Z V}\def\Tb{\bar{T}}\def\nb{\bar{n}}\def\nbB{\nb_{\ns{B}}}
\def\etBg{\et\Z{B\ga}}\def\mpr{m_{\ns p}}\def\bD{\mean D}
\def\dAdec{d\Z{A\,{\rm dec}}}\def\zb{\bar{z}}
\def\OMPn{\OM\Z{\ga0}}\def\tad{\tau_{\ns d}}\def\tao{\tau_{\ns o}}
\def\Hx{H_{\lower1pt\hbox{$\scriptstyle0$}}}
\begin{abstract}
These lecture notes review the theoretical problems associated
with coarse-graining the observed inhomogeneous structure of the universe at
late epochs, of describing average cosmic evolution in the presence of
growing inhomogeneity, and of relating average quantities to physical
observables. In particular, a detailed discussion of the timescape scenario
is presented. In this scenario, dark energy is realized as a misidentification
of gravitational energy gradients which result from gradients in the kinetic
energy of expansion of space, in the presence of density and spatial
curvature gradients that grow large with the growth of structure. The
phenomenology and observational tests of the timescape model are discussed
in detail, with updated constraints from Planck satellite data. In addition,
recent results on the variation of the Hubble expansion on $\lsim100\hm$
scales are discussed. The spherically averaged Hubble law is significantly
more uniform in the rest frame of the Local Group of galaxies than in the
conventional rest frame assumed for the Cosmic Microwave Background. This
unexpected result supports a fundamental revision of the notion of the cosmic
rest frame, consistent with the expectations of the timescape scenario.
\end{abstract}\maketitle
\section{Introduction}

Present cosmological observations point to the need for a revolution in our
physical understanding. On one hand we have a very successful phenomenological
description of the universe based on the spatially homogeneous and
isotropic Friedmann--Lema\^{\i}tre--Robertson--Walker (FLRW) geometry.
However, this success comes at the price of the introduction of forms
of mass--energy that have never been directly observed, and which
constitute most of the stuff in the Universe: 27\% in the form of clumped
nonbaryonic dark matter, and 68\% in the form of a smooth dark energy
\cite{Pparm}.
Unknowns of this magnitude demand that we carefully re-examine the
assumptions of our physical models of the universe, and that we pay
careful attention to all observations.

The universe was certainly homogeneous to a high degree at the epoch of last
scattering, when the cosmic microwave background (CMB) radiation was laid down.
However, at the present epoch the matter distribution displays a complex
hierarchical structure with significant inhomogeneities up to scales of at
least $100\hm$, where $h$ is the dimensionless parameter related to the Hubble
constant by $\Hm=100h\kmsMpc$. The present universe is dominated in volume by
voids, with some 40\% of the volume in voids of a characteristic diameter
$\goesas30\hm$ \cite{hv02}--\cite{pan11} and populations of smaller minivoids
\cite{minivoids}. Galaxy clusters are grouped in sheets that surround the
voids, and filaments thread them, in a complex cosmic web \cite{web}.

A cosmological constant, $\LA$, as a source of dark energy might in itself not
pose a great theoretical puzzle, were it not for the {\em cosmic coincidence
problem}: why is the value of $\LA$ such that the
universe decelerates for much of its history and only begins to accelerate
relatively recently? In addition, there is another cosmic coincidence,
which some cosmologists view as a smoking gun: the onset of cosmic
acceleration also coincides with the epoch in which the large nonlinear
structures of the cosmic web begin to dominate, as the map of the time
history of universe \cite{map} clearly reveals.

The possibility that the phenomenon of dark energy is actually accounting
for the average effects of inhomogeneous structures on the expansion history
of the universe has led to an upsurge of interest in the averaging problem
in cosmology. This is a foundational question
since the physical ingredients of Einstein's theory have never been
precisely specified on all scales. There are many unsolved problems
relating to the coarse-graining, fitting, and averaging of geometry. In these
lectures I will discuss these issues, with an emphasis on the \TS\
cosmology, which does at least provide a phenomenologically viable
alternative to the standard model. In the \TS\ scenario, I have
attempted to address the key issue of gravitational energy which I
believe is intimately related to solving the riddle of ``dark energy''.
It is my hope that if we pay close attention to observations, and
think more deeply about fundamental concepts in light of new observations,
that we might develop better statistical notions of gravitational energy
and entropy, which may be important not only for cosmology at large
but also for the foundations of gravitational physics.

\section{The fitting problem: What is dust?}

\subsection{On what scale are Einstein's equations valid?}

In the standard FLRW cosmology, fundamental observers are defined to be
``comoving with the dust'' in geometries that are solutions to Einstein
equations with a dust or perfect fluid source. This poses two problems.
Firstly, it involves an extrapolation of Einstein's field equations
\beq
{G^\mu}_\nu={8\pi G\over c^4}{T^\mu}_\nu
\label{Efe}
\eeq
far beyond the scales on which they have been tested. General relativity
is only well tested for isolated systems -- such as the solar system or
binary pulsars -- for which ${T^\mu}_\nu=0$.

Secondly, the notion of what a ``dust particle'' is in cosmology is not
rigorously defined. The scale over which matter fields are coarse-grained
to produce the energy--momentum tensor on the r.h.s.\ of (\ref{Efe}) is not
prescribed, leaving an inherent ambiguity. Traditionally, galaxies have been
thought of as particles of dust. However, our observations show that galaxies
themselves are not homogeneously distributed. The largest typical nonlinear
structures are voids of diameter $30\hm$ \cite{hv02}--\cite{pan11}, so that
we must coarse grain on scales at least a few times larger to obtain a
notion of statistical homogeneity. This process of coarse-graining involves
unexplored statistical aspects of general relativity, which have barely been
studied.

There is no ambiguity in applying Einstein's equations (\ref{Efe})
to a fluid of particles with well-defined properties, such as ions, atoms
and molecules in the early phases of the universe's expansion. However, as
soon as gravitational collapse occurs then particle geodesics
cross. Phase transitions occur, so the definition of the particles in the
fluid approximation must change, giving rise to the following hierarchy of
coarse-grained `particles' in the epochs following last scattering:
\begin{enumerate}
\item Atomic, molecular, ionic or nuclear particles: applicable with
\begin{itemize}
\item dust equation of state within any expanding regions which have not yet
undergone gravitational collapse;
\item fluid equation of state within relevant collapsed objects (stars, white
dwarfs, neutron stars) for periods of time between phase transitions that
alter the non-gravitational particle interactions and the equation of state;
\end{itemize}
\item Collapsed objects such as stars and black holes coarse-grained as
isolated objects;
\item Stellar systems coarse-grained as dust particles within galaxies;
\item Galaxies coarse-grained as dust particles within clusters;
\item Clusters of galaxies coarse-grained as bound systems within expanding
filaments and walls;
\item Voids, walls and filaments combined as expanding regions of different
densities in a single smoothed out cosmological fluid.
\end{enumerate}

\subsection{Coarse--graining\label{grain}}
Any coarse-graining procedure
amounts to replacing the microphysics of a given spacetime region by
some collective degrees of freedom of those regions which are sufficient to
describe physics on scales larger than the coarse-graining scale. Einstein's
equations were originally formulated with the intent that the energy-momentum
tensor on the r.h.s.\ of (\ref{Efe}) should either describe fundamental fields,
such as the Maxwell field, or alternatively to the coarse-graining of the
purely nongravitational interactions described by such fields in the fluid
approximation.

Up to step 3 in the hierarchy, there are no real problems of principle
with coarse-graining since we are coarse-graining only over matter degrees
of freedom which appear exclusively in the energy-momentum tensor. In 1917 when
Einstein first applied general relativity to cosmology \cite{esu}, this was
sufficient since it had yet to be established that nebulae were distant
galaxies, and the prevailing view was that the density of the universe was the
density of the Milky Way. Decades later Einstein did consider the fitting
problem, when he built the Swiss cheese model in collaboration with Straus
\cite{ES}. However, this is a simple model which treats
inhomogeneities as Schwarzschild solutions placed in holes in a homogeneous
isotropic FLRW background. It deals with an idealized situation which is far
simpler than the actual cosmic web, which astronomers really only began to
uncover in the 1980s.

The fundamental problem then, is that since the universe is composed of a
hierarchy of long-lived structures much larger than those of stars, we must
also coarse-grain over the gravitational interactions within that hierarchy
to arrive at a fluid description for cosmology. With such a coarse-graining,
geometry no longer enters purely on the left hand side of Einstein's equations
but in a coarse-grained sense can be hidden inside effective fluid elements
of a smoothed out energy-momentum tensor on the right hand side of (\ref{Efe}).
We have a complex hierarchical fitting problem \cite{fit1,fit2} that must be
solved to relate the average geometry of the universe to the local geometry
to which our clocks and rulers are calibrated.

The fundamental quantities of interest as the sources of the right hand
side of Einstein's equations are those of mass--energy, momentum and angular
momentum. Effectively, if we demand that equations (\ref{Efe}) should
also apply in a coarse-grained version on cosmological scales, then it means
that we are seeking collective mass--energy parameters which average over
the rotational kinetic energies of galaxies, binding energies of galaxies
and clusters, kinetic energies of galaxies in virialized clusters, regional
spatial curvature etc. Furthermore we must approach the problem more than just
once, on a succession of scales. This necessarily involves the issue of
quasilocal gravitational energy, and more particularly statistical properties
of the gravitational interactions of bound systems.

Since we are no longer dealing with a fixed spatial metric this problem is far
more complicated than any equivalent problem in Newtonian theory, and indeed
it is largely unexplored territory. The physical degrees of freedom which we
must coarse grain are contained in the curvature tensor and the sources of the
field equations (\ref{Efe}).
In principle coarse-graining the curvature tensor might involve steps other
than simply coarse-graining of the metric. However, if a metric description of
gravity is assumed at each level, then schematically the hierarchy of
coarse-graining might be heuristically described as
\beq
\left. \begin{array}{r}
\metric{stellar}\to\metric{galaxy}\to\metric{cluster}\to
\metric{wall}\\ \vdots\quad \\ \metric{void} \end{array}
\right\}\to \metric{universe}
\label{coarse}\eeq
where the ellipsis denotes the fact that the metric of more than one type of
wall or void might possibly be relevant. In this scheme the lowest members
are assumed to be well modeled by exact solutions of Einstein's field
equations: $\metric{stellar}$ being a solution to the vacuum field equations
with a star or black hole source (given by the Schwarzschild or Kerr
solution), and $\metric{void}$ being that of a region filled with low density
ionic dust with whatever symmetries are relevant.

Within the hierarchy (\ref{coarse}) there are (at least) three steps that
involve the coarse-graining of gravitational degrees of freedom, which might
be summarized as
\begin{itemize}
\item {\em Galactic dynamics}: $\metric{stellar}\to\metric{galaxy}$;
\item {\em Cluster dynamics}: $\metric{galaxy}\to\metric{cluster}$;
\item {\em Cosmological dynamics}: $\metric{cluster}\to\left\{\metric{wall}
\oplus\cdots\oplus\metric{void}\right\}\to\metric{universe}$.
\end{itemize}
The gravitational degrees of freedom that are coarse-grained in galactic and
cluster dynamics involve the gravitational binding energies of bound systems
of different scales. By contrast cosmological dynamics deals with the
coarse-graining of expanding regions of different densities, i.e., with
the coarse-graining of the kinetic energy of expansion of space.

In the Newtonian cosmology \cite{Bondi} for particles of mass $m$ with
positions $r^i=a(t)x^i$, ($x^i=\,$const), relative to an arbitrary centre,
the Friedmann equation
\beq{\dot a^2\over a^2}+{kc^2\over a^2}={8\pi G\rho\over3}\label{F}\eeq
is obtained from the Newtonian energy equation, $T-U=-V$, where
$T=\half m\dot a^2 x^2$ is the kinetic energy per particle,
$V=-\frac43\pi G\rho a^2 x^2 m$ is the potential energy per particle
and $U=-\half k m c^2 x^2 $ is the total energy per particle, where $k$
is a constant. Eq.~(\ref{F}), the Hamiltonian constraint
of the full Einstein equations for the standard cosmology, is thus recognized
to contain terms related to gravitational potential and binding energy, $V$,
and the kinetic energy of expansion, $T$, in the Newtonian limit.

As long as the universe is perfectly homogeneous then these quantities
are the same for all observers. However, once there is inhomogeneity,
and in particular once there are gradients in spatial curvature, then
these concepts become entangled. On account of the strong
equivalence principle, spatial curvature cannot be defined at a point, and
any definition necessarily involves a regional {\em quasilocal} definition.
Gravitational binding energy and the kinetic energy of expansion are
thus quasilocal concepts tied to gradients in spatial curvature.

\subsection{Coarse--graining of bound systems}

A statistical description of cosmological relativity involves both
gravitational binding energy and the kinetic energy of expansion. It happens
that the coarse-graining of these respective gravitational degrees of freedom
relates to the scales at which the phenomena of dark matter and
dark energy are respectively observed. It is therefore possible that both
phenomena are related to different aspects of the same problem, namely
that the standard model incorporates a rigidity of spatial curvature
which is not demanded by full general relativity.

Since the coarse-graining of gravitational binding energy and the kinetic
energy of expansion involve different physical questions, it may be prudent
to investigate just one problem at a time. The \TS\ scenario
\cite{clocks}--\cite{obs} has been developed to deal with the problem
of the kinetic energy of expansion only: in the approach taken thus far all
coarse-grained regions are expanding ones. We will see in Sec.~\ref{TS} that
agreement with observation is obtained only by incorporating a fraction of
nonbaryonic dark matter. However, we must remain open-minded as to whether the
parameter found corresponds to actual new particles as in the standard $\LA$
Cold Dark Matter (\LCDM) model, or whether is a simply a phenomenological
parameter that accounts for the coarse-graining of binding energy that we
have not yet examined.

A piece of evidence in support of an alternative to conventional CDM is
the remarkable phenomenological success of Modified Newtonian Dynamics
(MOND) \cite{m83,sm02} at the level of galactic dynamics. This empirical
model works well at galactic levels, but fails at the cluster level.
While galactic and cluster dynamics both involve binding energy, the kinetic
energy degrees of freedom of the two situations are different. Galactic
dynamics typically involves stars in rotationally supported structures,
whereas cluster dynamics involves the less coherent motions of galaxies
which move in the combined potential but also interact with each other in
random pairwise encounters. Whereas the diameters of stars are very small
($<10^{-5}\,$\%) compared to their interparticle separations in galaxies, the
typical size of a galaxy is a more sizable fraction ($0.5\,$--$15\,$\%) of
typical intergalactic distances in a virialized cluster.

Newtonian dynamics is used almost exclusively in the treatment of both
galaxy and cluster dynamics. The rationale for this is that fields are
weak. However, even if spacetime is close to a Minkowski background, an
important question remains: {\em which} Minkowski background? There is no
global Minkowski background in the universe, and even if space is close to
Minkowski for small time intervals on a spatial 2-sphere encompassing a galaxy
or galaxy cluster, then the question remains: how do we calibrate the rulers
and clocks on that 2-sphere relative to another similar 2-sphere elsewhere?
Gravitational lensing calculations make use of a formula derived for the
ideal Schwarzschild geometry of an isolated point mass, which has an exact
timelike Killing vector. Are there pitfalls in applying such notions of mass
to circumstances in which there are no pure timelike Killing vectors, and no
truly isolated masses? To my knowledge, these questions have not been
rigorously posed in general relativity, let alone answered.

Some attempts have been made to understand galaxy rotation curves with
exact dust solutions \cite{CT1}--\cite{BG}. However, the applicability of
these solutions as alternatives to galactic dark matter has been debated
\cite{CT2,RS}. Furthermore, while new exact solutions to Einstein's
equations may offer new insights into the possibilities offered by general
relativity, they do not directly address the problems posed by
coarse-graining. They are also limited by the additional restrictions that must
be applied to reduce Einstein's equations to a soluble form. For example,
although galaxy clusters are often spherical in shape the spherically
symmetric dust Lema\^{\i}tre--Tolman--Bondi (LTB) solutions \cite{L}--\cite{B}
cannot be applied\footnote{By contrast LTB models are clearly applicable to
individual expanding spherical voids \cite{BKH} with ionic or molecular
sized dust.} to virialized clusters, since galaxies in
clusters do not collapse inwards in coherent spherical shells. Even if
the motion of individual galaxies is close to radial, the phases of the
galaxies relative to passage through the centre of the cluster are completely
uncorrelated. Individual galaxies will pass close to the core of the cluster
and emerge from the other side; but at any instant the
number of galaxies moving out from the centre might be comparable to the number
falling in.

In my view the problems of coarse-graining of galaxies and clusters are
difficult. However, in view of the phenomenological successes of MOND at
the galactic scale \cite{m83,sm02} we should be open to the possibility
that simplifying principles remain to be discovered.

\subsection{Coarse--graining the cosmological fluid}
The final step of coarse-graining
involves qualitatively new fundamental questions. If we require a
single model to describe the evolution of the universe from last scattering
to the present day, then we must coarse grain on scales over which the notion
of a dust `particle' has a meaning from last scattering up to the present.
The description of a galaxy composed of stars, or of a
virialized galaxy cluster composed of galaxies is only valid for those
epochs after which the relevant `particles' have formed and are themselves
relatively unchanging. Over cosmological timescales we do not have
well-defined invariant dust particles. The nature of galaxies and galaxy
clusters changes through growth by accretion of gas and by mergers.

To get around the problem of ill-defined particle-like building blocks, an
appropriate strategy is to coarse-grain the `dust' on scales large enough
that the {\em average} flow of mass from one cell to another is negligible up
to the present epoch. Although galaxy clusters vary greatly in size and
complexity, there are no common virialized structures larger than clusters.
Thus coarse-graining on scales larger than clusters necessarily means dealing
with fundamental objects that are themselves {\em expanding}, i.e., with
entities that resemble {\em fluid elements} in hydrodynamics rather than
point particles.

Another qualitative difference from the case of bound systems is that we have
to deal with expanding fluid elements that have vastly different densities at
the present epoch, and which evolve more or less independently.
Although we can receive signals from anywhere within our particle horizon,
the energy we receive in electromagnetic and gravitational waves, or
indeed in cosmic ray particles from distant galaxies, is
negligible in comparison with the rest-energy of the local density field.
The region which has contributed matter particles to define the local
geometry of our own galaxy is actually very small. This bounding sphere,
which Ellis and Stoeger \cite{ES09} call the {\em matter horizon}, is estimated
by them to be of order $2\,$Mpc for the Milky Way using assumptions about the
growth of perturbations from the standard cosmology. This scale also coincides
roughly with the scale at which the Hubble flow is believed to begin in the
immediate neighbourhood of the Local Group of galaxies. It is one way of
realizing the concept of {\em finite infinity}, introduced qualitatively by
Ellis in his first discussion of the fitting problem \cite{fit1}.

For galaxy clusters some sort of finite infinity notion -- which we will
better define in Sec.~\ref{TS} -- with a variable scale of order
$2$--$10\,$Mpc depending on the size of cluster might be useful for defining
the minimum smoothing scale containing bound structures. By combining such
regions we arrive at the walls and filaments that contain most of the mass of
the universe. However, to this we must also add the voids which dominate the
volume of the universe at the present epoch. These are the regions in which
structures have never formed, and which still contain the same ionic, atomic
and molecular dust content that has existed since very early epochs, only
greatly diluted by expansion.

If we set aside a few peculiar large wall structures \cite{chr}, then the
largest {\em typical} nonlinear structures are voids. Surveys indicate that
voids with characteristic mean effective radii\footnote{Voids display a degree
of ellipticity. The {\em mean effective radius} of a void is that of a sphere
with the same volume as occupied by the void \cite{hv02}--\cite{pan11}, which
is typically larger than the maximal sphere enclosed by the same void.} of
order $(15\pm3)\hm$ (or diameters of order $30\hm$), and a typical
density contrast of $\de\rh/\rh=-0.94\pm0.02$, make up 40\% of the volume of
the nearby universe \cite{hv02,hv04}. A recent study \cite{pan11} of the
Sloan Digital Sky Survey Data Release 7 finds a median effective void radius of
$17\hm$, with voids of effective radii in the range $10\hm$ to
$30\hm$ occupying 62\% of the survey volume. In addition to these there
are numerous smaller minivoids \cite{minivoids}, which combined with the
dominant voids ensure that voids dominate the present epoch universe by
volume.

\subsubsection{Coarse-graining at the statistical homogeneity scale}

Any minimal scale for the cosmological coarse-graining of the final smoothed
density distribution has to be substantially larger than the diameter of the
largest typical structures. Void statistics \cite{pan11} indicate an effective
cutoff of $60\hm$ for the largest mean effective diameters of voids,
i.e., twice the scale of the typical dominant void diameters. Thus
observationally, the relevant scale for coarse-graining appears to be of
order two to three times the dominant void diameter, e.g., of order $100\hm$.
Although the scale of transition to statistical homogeneity is
debated \cite{h05,sl09} recent results from the WiggleZ survey suggest that
the transition occurs in the range\footnote{This range represents the range of
bias-corrected values listed for various redshift ranges in Table 4 of
ref.~\cite{sdb12}. These estimates assume a \LCDM\ model and use data in the
redshift range $0.1<z<0.9$.} $70\h$--$100\hm$ \cite{sdb12}.

A statistical homogeneity scale of $\goesas100\hm$ also coincides roughly
with the Baryon Acoustic Oscillation (BAO) scale \cite{bao1,bao2}, $D\Ns{BAO}$.
Physically there is a good reason for this coincidence. The BAO scale is that
of the largest acoustic wave in the plasma at last scattering. For scales
larger than this the spectrum of initial density perturbations is roughly
scale invariant with a density contrast $\de\rh\Z B/\rh\Z B\goesas10^ {-5}$ in
baryons, and $\de\rh\Z C/\rh\Z C\goesas10^{-4}$ in cold dark matter.
Below the BAO scale initial density contrasts may be amplified by acoustic
waves in the plasma, so the amplitude of initial density contrasts is
somewhat larger, particularly at the scales associated with the higher order
odd acoustic peaks. The BAO scale therefore provides a demarcation
between the linear and nonlinear regimes of the subsequent growth of
structure.

At first, it might seem contradictory that the amplification of the
primordial density perturbations corresponding to the first acoustic peak
should give a small enhancement which in the standard \LCDM\ model can treated
in the linear regime of perturbation theory with good observational agreement,
whereas higher order peaks give enhancements which give rise to a nonlinear
regime. However, it must also be remembered that perturbations are nested,
so that in some cases we get amplifications on top of amplifications.
Characteristic features will arise from the fact that at last scattering the
odd acoustic peaks corresponds to compression in gravitational
potential wells and rarefaction in potential peaks,
whereas the even acoustic peaks correspond to rarefaction in potential
wells and compression in potential peaks. The odd peaks will thus produce
increasing amplifications of structure, while even peaks will somewhat
undo the amplifications (but not completely on account of baryon drag).

The fact that the diameter of the dominant voids at $D\Ns{void}\goesas30\hm$
\cite{hv02}--\cite{pan11} is close to one third of $D\Ns{BAO}$ is seen to be
a consequence of these structures growing from the additional amplification
provided by the third acoustic peak\footnote{This corrects statements made in
Sec.~V of ref.~\cite{equiv}, where on account of a confusion about the role of
the odd and even acoustic peaks it was incorrectly suggested the second peak
should provide a relevant scale to voids and the third peak to clusters of
galaxies. The scale of rich clusters of galaxies is more likely to result from
the nonlinear evolution of the fifth peak amplification.}. Since voids are
regions which appear to expand at faster rates\footnote{Here we refer to the
rate as measured by any one observer. Calibrating the expansion rate with
different canonical clocks is an essential ingredient of the timescape
cosmology which we will return to in Sec.~\ref{TS}} than walls, with density
contrasts growing to $\de\rh/\rh\to-1$ at late times, the exact ratios of
scales of the acoustic peaks at last scattering are not preserved in the
nonlinear regime today. In fact, a precise measurement of the difference
between the ratio $D\Ns{void}/D\Ns{BAO}$ and $1/3$ would provide useful
constraints on the variation of the Hubble parameter in the nonlinear regime.
Since voids are the dominant nonlinear structures in the cosmic web, the
beginning of an emergence of a notion of homogeneity at scales $\goesas70\hm$
\cite{sdb12} may be related cutoff in the statistics of void diameters at
$60\hm$ found by Pan \etal\ \cite{pan11}, rather than a scale related to the
second acoustic peak. Since the even acoustic peaks represent deamplified
initial density contrasts, they are unlikely to have very clear-cut signatures
in cosmic structure.

\subsubsection{Variations on scales larger than the statistical homogeneity
scale}

In the standard cosmology it is often assumed that as the domain
$\DD$ of a spatial average is made larger and larger at the present
epoch, the density contrast $\ave{\de\rh/\rh}\Z{\DD}$ will diminish to
small values which match those at the last scattering epoch. However,
this assumes constraints on the notion of statistical homogeneity over and
above\footnote{In particular, perturbations on different scales would need
to compensate each other in such a way as to maintain a homogeneous isotropic
universe which is at present within 1\% of being spatially flat overall.
In the standard model with FLRW evolution this is required to fit the overall
angular scale of the acoustic peaks in the CMB. We will see in
Sec.~\ref{cmb} that in the \TS\ model the acoustic scale can be
fit without such a restriction on spatial curvature.} those required for a
universe that has evolved from an initial density perturbation spectrum that
was close to scale-invariant, as is
consistent with the observed CMB anisotropies and primordial inflation.
Given initial nested density fluctuations on arbitrarily large spatial scales,
then if any arbitrarily large domain $\DD$ evolves independently by close to
FLRW evolution, its density at the present epoch will always have evolved
from a perturbation that was within the initial spectrum, but not necessarily
exactly the mean.

We can therefore crudely estimate the standard deviation of the density of
cells on scales larger than $100\hm$ by assuming that each cell evolves
as an independent Friedman universe from a smooth perturbation at the epoch of
last scattering. (This assumes that the backreaction contributions to be
discussed in Sec.~\ref{back} do not dominate the volume--average evolution.)
Using the Friedmann equation with pressureless dust only, for which
$$a\Z0^2\Hm^2(\OmMn-1)=a^2(t)H^2(t)[\Omega\Z M(t)-1],$$
we obtain a present epoch density contrast
\beq
\de\rh\Z0\simeq\left(H\over\Hm\right)^2{\de\rh_t\over(1+z)^2}\,.
\label{est}\eeq
Here the density contrast is relative to the critical density, so that
$\de\rh_t=\Omega\Z M(t)-1$ etc, where $\Omega\Z M$ is a density parameter
for the isolated region only. Thus if we take $\de\rh_t\simeq10^{-4}$ at last
scattering (for a CDM density contrast), when $z\simeq1090$ and when
evolution is roughly matter-dominated with $H\simeq 2/(3t)$ and
$t\simeq380,000\w{yr}$, we are led to $\de\rh\Z0\simeq0.025/h^2\simeq0.06$
if $h\simeq0.65$.

This crude estimate can be compared to the actual density variance determined
from large scale structure surveys \cite{h05,sl09}. Sylos Labini \etal\
\cite{sl09} determined the variance in the number density of luminous red
galaxies (LRGs) in the SDSS-DR7 by dividing the full sample of 53,066 galaxies
in the redshift range $10^{-4}<z <0.3$ into $N$ equal nonoverlapping volumes.
Over the range $4\le N\le15$, they found a standard deviation of order 8\%,
consistent with an earlier measurement of 7\% by Hogg \etal\ \cite{h05} in
a smaller LRG sample. These values are very close to our order of magnitude
estimate of 6\%, which has still not been corrected to include radiation
at last-scattering, or the effects of backreaction at late epochs.

Given a nearly scale--invariant spectrum of nested density perturbations, we
expect that the variance in density should not decrease appreciably if sample
volumes are increased at nearby redshifts. In principle, it should be possible
to calculate the variance as a function of scale, given the constraints from
the CMB anisotropy spectrum at long wavelengths. For spatial slices at higher
redshifts, looking further back in time, the variance would decrease in accord
with (\ref{est}) -- provided that a sample of objects such as LRGs can be
found which does not exhibit strong evolutionary effects over the range of
redshifts in question.

In summary, in order to coarse-grain fluid cells in such a way that the size of
a cell is larger than the largest typical nonlinear structures, with a mass
that does not change on average from last scattering until today,
observations show that we should coarse-grain fluid cells at a scale of order
$70\h$--$100\hm$. This scale will be called the statistical homogeneity scale
(SHS). Such a scale marks the transition from a nonlinear regime in which
there is a very large variance in $\de\rh/\rh$, to a regime in which
cosmological average evolution with a single Hubble parameter becomes well
defined. It does not mark a scale at which average evolution necessarily
becomes precisely FLRW, nor at which density contrasts become completely
negligible. Rather variations on spatial scales larger than the SHS at the
present epoch are bounded by a maximum $\de\rh/\rh\lsim0.1$, as is consistent
with observations.

In this subsection we have presented a summary of observational results to
be accounted for in cosmological coarse-graining, without assuming any details
about backreaction. There is one other observational puzzle which also requires
mention: the Sandage-de Vaucouleurs paradox\footnote{In the literature this has
been called the ``Hubble--de Vaucouleurs paradox'' \cite{paradox1,bary} and
alternatively the ``Hubble--Sandage paradox'' \cite{paradox2}. However, the
paradox never involved Hubble directly, but was originally raised by Sandage
and collaborators \cite{STH} in objection to de Vaucouleurs' hierarchical
cosmology \cite{deVau} before strong evidence for the cosmic web of voids,
sheets and filaments had amassed.}. This is the puzzle that in conventional
ways of thinking, we should expect large statistical scatter in the peculiar
velocities of galaxies below the SHS, if they are indeed ``particles of
dust''. In fact, on scales of order $20\,$Mpc the statistical scatter should
be so large that no linear Hubble law can be extracted. Yet $20\,$Mpc is the
very scale on which Hubble originally found his famous linear law. This
statistical quietness of the local Hubble flow is difficult to reconcile with
conventional understanding. In any FLRW universe which expands forever,
peculiar velocities do decay. However, the \LCDM\ parameters
required for the velocity dispersion predicted by structure formation
to match the observed velocity dispersion, do not coincide with the
concordance parameters \cite{ap02}.

\subsection{Approaches to coarse-graining}

The problem of coarse-graining in general relativity in a bottom-up fashion
is little studied. In principle, it is a very interesting question,
which should deal, e.g., at the galactic level with the problem of replacing
the Weyl curvature of individual Schwarzschild or Kerr solutions by a
coarse-grained Ricci curvature for a dust fluid. Higher levels of
coarse-graining in the hierarchy (\ref{coarse}) involve further physical
questions. Rather than dealing with multi-scale problems, the few existing
studies simplify the hierarchy (\ref{coarse}) to a single step.

\subsubsection{Covariant coarse-graining}
Korzy\'nski \cite{Ko09} has proposed a covariant coarse-graining
procedure, which could conceivably be applied to any step in the hierarchy
(\ref{coarse}) for which the starting point is the metric of a known dust
solution. Korzy\'nski's idea is to isometrically embed the boundary of a
comoving dust-filled domain -- required to have $S^2$ topology with positive
scalar curvature -- into a three-dimensional Euclidean space, and to construct
a ``fictitious'' three-dimensional fluid velocity which induces the same
infinitesimal metric deformation on the embedded surface as the ``true'' dust
flow does on the domain boundary in the original spacetime. This velocity field
is used to uniquely assign coarse-grained expressions for the volume expansion
and shear to the original domain. An additional construction using the
pushforward of the Arnowitt-Deser-Misner (ADM) shift vector \cite{ADM} is used
to similarly obtain a coarse-grained vorticity. The coarse-grained quantities
are quasilocal functionals which depend only on the geometry of the
boundary of the relevant domain. This formalism is at an early stage
of development, but could conceivably provide new methods for attacking
the fitting problem.

\subsubsection{Discretized universes}
The Lindquist--Wheeler model \cite{LW} is a lattice based approach, which has
received new interest recently \cite{CF}--\cite{CFO}. The coarse-graining
hierarchy (\ref{coarse}) is replaced by the simplified scheme
\beq
\metric{Schwarzschild}\to \metric{universe}
\eeq
with the proviso that $\metric{universe}$ does not represent a continuum metric
in the usual sense. Rather, by matching the spherical boundaries of radially
expanding geodesics in the Schwarzschild geometries of a regular lattice of
equal point masses, the Friedmann equations are obtained \cite{LW,CF}. The
matching is exact only at the points where the radial spheres intersect and
is approximate in the regions in which spheres overlap or are excluded. A
continuum cosmological geometry is thus realized only approximately.

This model is analogous to the Swiss cheese models \cite{ES} in the
sense that the point group symmetry of the lattice is a discretized version
of a system with overall global spatial homogeneity. Ray-tracing studies in the
spatially flat Lindquist--Wheeler model lead to results which are almost
identical to that of the exact Einstein-de Sitter solution\footnote{A
different result was first claimed in an earlier study \cite{CF}, but then
corrected \cite{CFO}. The review article \cite{dust} was written before
the corrected result \cite{CFO} was found, and it cites the earlier incorrect
result.}. While this demonstrates that the Lindquist--Wheeler does provide
a consistent lattice description of the FLRW models, it unfortunately does not
give any indication of how one should treat the problem of inhomogeneity,
without discrete symmetries.

\section{Averaging and backreaction}\label{back}

The terms ``coarse-graining'' and ``averaging'' are often used interchangeably
in a loose sense. However, whereas coarse-graining is generically a bottom-up
process, averaging is top-down: one is interested in the overall
average dynamics and evolution, usually without direct consideration of the
details of the course-graining procedure. Whereas coarse-graining is little
studied, considerably more attention has been paid to averaging. Several
approaches have been pursued and are discussed in many reviews including,
e.g., those of Buchert \cite{Brev,Brev2}, van den Hoogen \cite{vdH} and
Clarkson \etal\ \cite{CELU}.

Cosmological averaging typically starts from the assumption that a well-defined
average exists, with a number of assumed properties. If one assumes that
the Einstein field equations (\ref{Efe}) are valid for some general
inhomogeneous geometry, $\metric{}$, then given some as yet unspecified
averaging procedure denoted by angle brackets, the average of (\ref{Efe})
gives
\beq
\ave{{G^\mu}_\nu}=\ave{g^{\mu\la}R_{\la\nu}}-\frn12{\de^\mu}_\nu\ave{g^{\la\rh}
R_{\la\rh}}={8\pi G\over c^4}\ave{{T^\mu}_\nu}\,.
\label{aEfe}
\eeq
A number of choices are possible at this point since there is no {\em a priori}
reason to assume that $\ave{{G^\mu}_\nu}$ is the Einstein tensor
of an exact geometry.

In the {\em macroscopic gravity} approach, Zalaletdinov \cite{Z1}--\cite{Z3}
takes the average inverse metric $\ave{g^{\mu\nu}}$ and the average Ricci
tensor $\ave{R_{\mu\nu}}$ as basic variables, so that
\beq
\ave{g^{\mu\la}}\ave{R_{\la\nu}}-\frn12{\de^\mu}_\nu\ave{g^{\la\rh}}\ave{
R_{\la\rh}}+{C^\mu}_\nu={8\pi G\over c^4}\ave{{T^\mu}_\nu}\,,
\label{Zfe}
\eeq
where the correlation functions ${C^\mu}_\nu$ are defined by the difference
of the left hand sides of (\ref{Zfe}) and (\ref{aEfe}). Zalaletdinov
provides additional mathematical structure to prescribe a covariant averaging
scheme, thereby defining properties of the correlation functions.

Alternatively one can consider the difference
of the general inhomogeneous metric and the averaged metric
\beq
\metric{}=\bar\metric{}+\de\metric{},\qquad
\label{metdef}
\eeq
where $\bar\metric{}\equiv\ave{\metric{}}$, with inverse $\bar g^{\la\mu}\ne
\ave{g^{\la\mu}}$. One may now determine a connection $\bar\GA^\la_{\ \mu\nu}$,
curvature tensor ${\bar R^\mu}_{\ \;\nu\la\rh}$ and Einstein tensor $\bar G^
\mu_{\ \nu}$ based on the averaged metric, $\bar\metric{}$, alone.
The differences $\de\GA^\la_{\ \mu\nu}\equiv\ave{\GA^\la_{\ \mu\nu}}-\bar\GA^
\la_{\ \mu\nu}$, $\de R^\mu_{\ \nu\la\rh}\equiv\ave{R^\mu_{\ \nu\la\rh}}-\bar
R^\mu_{\ \nu\la\rh}$, $\de R_{\mu\nu}\equiv\ave{R_{\mu\nu}}-\bar R_{\mu\nu}$
etc, then represent the {\em backreaction} of the averaged inhomogeneities on
the average geometry determined from $\bar\metric{}$. Furthermore, the average
Einstein field equations (\ref{aEfe}) may be written
\beq
\bar G^\mu_{\ \nu}+\de G^\mu_{\ \nu}={8\pi G\over c^4}\ave{{T^\mu}_\nu}\,.
\label{bEfe}
\eeq
This expresses the fact that the Einstein tensor of the average metric is not
in general the average of the Einstein tensor of the original metric. The
processes of averaging and constructing the Einstein tensor do not commute.

Equation (\ref{aEfe}) and (\ref{bEfe}) are of course very similar, but may
differ in both the definition of the average represented by the angle brackets,
and also in the split of the background averaged Einstein tensor and the
correlation or backreaction terms. The manner in which averaging schemes
differ often relate to whether the effects of backreaction are assumed
to be weak or strong.

\subsection{Weak backreaction: the Friedmann--Lema\^{\i}tre universe as the
average\label{weak}}

The remarkable success of the standard cosmology, albeit with sources of
dark matter and dark energy which have not been directly observed, has
understandably led most researchers to assume that it must be correct, even
if only in an average sense. As a consequence, many researchers simply
begin from the starting point that the FLRW geometry must be the average,
or very close to the average evolution.

One can then either assume that\begin{itemize}
\item there is no backreaction on average evolution but inhomogeneities
are sufficiently large that they significantly affect the propagation of light,
as in the Swiss cheese \cite{FDU1,FDU2} and meatball \cite{MN11} models;
or
\item backreaction is sufficiently small that the changes to average
evolution can be treated perturbatively about a homogeneous isotropic
background, at least initially.
\end{itemize}

The second approach, {\em weak backreaction}, is of course intimately
related to standard cosmological perturbation theory. One assumes that the
average geometry $\bar\metric{}$ of (\ref{metdef}) is exactly FLRW,
and that the quantities $\de\metric{}$ can be treated as perturbative
corrections.

The issue of whether backreaction is significant or insignificant in the
perturbative FLRW context is a matter of much debate, with different authors
coming to different conclusions, which may be traced to various differences
in assumptions made. These issues are discussed in many reviews, such as those
of Clarkson \etal\ \cite{CELU} and Kolb \cite{K11}, and will not be discussed
in detail here. In my view this debate shows that there are potential problems,
which cannot be resolved by staying within perturbation theory.

In fact, all researchers are well aware that there is a nonlinear regime in
structure formation, which is explored by $N$-body simulations in the standard
cosmology. Since there is as yet no rigorous procedure for coarse-graining
the gravitational degrees of freedom which describe the small-scale structures,
the perturbative approach can only be valid given an implicit assumption that
there is no new physics to be found when coarse-graining the hierarchy
(\ref{coarse}).

Such an assumption underlies a typical argument against backreaction: if we
{\em assume} a FLRW geometry, and estimate the magnitude of the perturbations
using typical rotational and peculiar velocities of galaxies, then the
corrections are small \cite{Peebles}. However, at late epochs galaxies and
galaxy clusters are not homogeneously distributed, and cannot be considered
as randomly distributed gas particles on scales of tens of megaparsecs below
the SHS. The dominant structures on these
scales are voids of diameter $\goesas30\hm$ with density contrasts $\de\rh/
\rh\goesas-0.95$ \cite{hv02,hv04}. Using galaxy peculiar velocities as an
estimate of $\de\rh/\rh$ is therefore misplaced. There is no direct evidence
that a spatially homogeneous geometry is the correct one below the SHS.

It may thus simply be incorrect to assume that a FLRW model exactly
describes the average evolution of the universe at the largest scales for all
times. Approaches which do not make the restrictive assumption of average FLRW
evolution are those with {\em strong backreaction}.

\subsection{Strong backreaction: Spacetime and spatial averages}\label{strong}

If Einstein's equations for a single metric with a prescribed energy-momentum
tensor source are not the relevant equations for describing the average
evolution of the universe on cosmological scales, then new physical
ingredients are required, either explicitly in the averaging formalism
itself, or else implicitly in relating the results of a particular formalism
to observations. After all, our measurements involve physical rulers and
clocks adapted to a local geometry, and this local geometry must somehow
be matched to the statistical geometry that describes average cosmic
evolution.

Strong backreaction as a solution to the problem of dark energy elicits much
confusion in the community, as typified by the statement that dark energy
is {\em just} an issue of inhomogeneities, and that it is entirely solved
{\em within} general relativity. Even advocates of strong backreaction might
disagree with this statement, depending on what is meant by ``general
relativity''. There is a widely held view, particularly among those not
involved in general relativity research, that it consists solely of completed
old physics. However, those better acquainted with general relativity know that
even setting aside the regime of quantum gravity, general relativity is not a
final complete theory, but contains many open and unsolved questions -- in
particular in relation to gravitational energy and entropy and the averaging
problem.

Strong backreaction does involve new physics, but in my view the new physics
must involve a natural extension of the principles of relativity into
regimes which Einstein did not envisage when he wrote down his field equations
in 1915. Whether one wishes to call it ``general relativity'' or
``cosmological relativity'' or something else is therefore a matter of taste.
The essential point is that one is proposing new rules for the geometrical
structure of spacetime on cosmological scales.

The cosmological spacetime is to be a statistical average geometry. Any
process of taking an average will in general break the general covariance
of Einstein's equations. There are differing approaches to this, which
alternatively involve spacetime or spatial averages. Many mathematical
approaches exist, including Ricci flow \cite{CP}--\cite{CB}, group averaging
of the FLRW isometry group \cite{A09}, covariant frame-bundle averaging
\cite{BvdHC} and constant mean (extrinsic) curvature (CMC) flows
\cite{Re08,Re09}. Here I will just very briefly outline the two
approaches which have attracted the most attention, largely due to Zalaletdinov
and Buchert.

\subsubsection{Zalaletdinov's macroscopic gravity\label{secZ}}

Zalaletdinov has developed a theory called {\em macroscopic gravity} based on
spacetime averages \cite{Z1,Z2,Z3,MZ}. His aim is to consistently average the
Cartan equations from first principles, in analogy to the averaging of the
microscopic Maxwell--Lorentz equations in electromagnetism. However, whereas
electrodynamics is linear in the fields on the fixed background of Minkowski
spacetime, gravity demands an averaging of the nonlinear geometry of spacetime
itself and is considerably more complicated.

Additional mathematical structures are required to average tensors in a
covariant manner on a given manifold, $\Mm$. To this end Zalaletdinov
introduces bilocal averaging operators \cite{Z1}--\cite{Z3}, ${\Az^\mu}_{\al}
(x,x')$, with support at two points $x\in\Mm$ and $x'\in\Mm$, which allow one
to construct a bitensor extension, ${\Tz^\mu}_\nu(x,x')$, of a tensor ${T^\mu}_
\nu(x)$ according to
\beq
{\Tz^\mu}_\nu(x,x')={\Az^\mu}_{{\al'}}(x,x'){T^{\al'}}_{{\be'}}(x')
{\Az^{\be'}}_{\nu}(x',x)\,.
\eeq
The bitensor extension is then integrated over a 4-dimensional spacetime
region, $\SI\subset\Mm$, to obtain a regional average according to
\beq
\bar T^\mu_{\ \nu}(x)={1\over\Vv_\SI}\int_\SI\dd^4 x'\,\sqrt{-g(x')}\,
{\Tz^\mu}_\nu(x,x'),
\eeq
where $\Vv_\SI\equiv\int_\SI\dd^4 x\,\sqrt{-g(x)}$ is the spacetime volume
of the region $\SI$. The bitensor transforms as a tensor at every point but is
a scalar when integrated over a region for the purpose of averaging.

Macroscopic gravity is a general covariant averaging formalism, rather than an
approach which was specifically formulated with cosmology in mind. In order
to make contact with cosmology, additional assumptions have been made. For
example, Paranjape and Singh considered a spatial averaging limit \cite{PS07}.
Other studies have made the assumption, similar to the weak backreaction
approach, that the average geometry is FLRW \cite{CPZ}--\cite{vdH09}. In that
case it was found that the macroscopic gravity correlation terms take the form
of a spatial curvature, even though a spatially {\em flat} FLRW geometry was
assumed for the average geometry \cite{CPZ}.

In my view, although Zalaletdinov's formalism is mathematically elegant, it
has weaknesses as a physical theory. In particular, it has been designed to
closely resemble general relativity itself. Apart from the fact that it deals
with two geometric scales -- a microscopic one and a macroscopic one -- there
is no scale in the final theory. Cosmological observations suggest a
particular hierarchy of scales (\ref{coarse}), which may involve physical
issues more complex than simply taking one step from a microscopic theory
to a macroscopic theory of gravity. In particular, the coarse-graining of the
gravitational degrees of freedom involving binding energy at one level and
the kinetic energy of expansion at another, may give rise to qualitatively new
phenomena. Rather than seeking to mimic the steps involved in coarse-graining
matter degrees of freedom, we need to specify macroscopic scales and physical
principles relevant to coarse-graining in cosmology.

\subsubsection{Buchert's spatial averaging formalism\label{secB}}

In the late 1990s, building on earlier work \cite{CP,BE,EB}, Buchert
developed an approach \cite{buch1,buch2} for the spatial averaging of scalar
quantities associated with the Einstein field equations (\ref{Efe}), with
cosmological averages in a fully nonperturbative setting in mind at the outset.
He applied the $3+1$ ADM spacetime split \cite{ADM}, which is a natural
approach if the Einstein field equations (\ref{Efe}) are to be viewed as
evolution equations.

Rather than tackling the mathematically difficult problem of averaging tensors,
Buchert averaged scalar quantities in general inhomogeneous spacetimes with
perfect fluid energy--momentum sources. Such scalars include the density,
$\rh$, expansion, $\th$, and scalar shear, $\si^2=\frn12\si_{\al\be}\si^{\al
\be}$ etc. For an arbitrary manifold, one can always locally choose ADM
coordinates,
\beq\ds^2=-\Bom^0\otimes\Bom^0+g_{ij}(t,\bx)\,\Bom^i\otimes\Bom^j,
\label{split}\eeq
where $\Bom^0\equiv\NN(t,\bx)\,c\,\dd t$, and $\Bom^i\equiv\dd x^i+\NN^i(t,\bx)
\,c\,\dd t$ define the ADM lapse function, $\NN$, and shift vector, $\NN^i$.
Such coordinates can only be chosen globally if one restricts the evolution
problem to that of irrotational flow, as Buchert does. In that case
(\ref{split}) may be assumed to apply over global $t=\,$const spatial
hypersurfaces. For a dust source\footnote{Extensions to perfect fluid
\cite{buch2} and other matter sources \cite{BBR} have also been considered, as
well as to general hypersurfaces tilted with respect to the fluid flow
\cite{L09}--\cite{R10}.} we can then choose synchronous coordinates with $\NN=1
$ and $\NN^i=0$. With these choices, the Einstein equations may be averaged on
a domain, $\DD$, of the spatial hypersurfaces, $\SI$, to give
\bea
3{\dot\ab^2\over\ab^2}&=&8\pi G\ave\rh-\half c^2\Rav-\half\QQ,\label{b1}\\
3{\ddot\ab\over\ab}&=&-4\pi G\ave\rh+\QQ,\label{b2}\\
&&\hskip-45pt\pt_t\ave\rh+3{\dot\ab\over\ab}\ave\rh=0,
\label{b3}\eea
where an overdot denotes a $t$--derivative, and
\beq
\QQ\equiv\frac23\left\langle\Bigl(\th-\langle\th\rangle\Bigr)^2\right\rangle
-2\ave{\si^2}
=\frac23\left(\langle\th^2\rangle-\langle\th\rangle^2\right)-
2\ave{\si^2}\,,
\label{backr}\eeq
is the {\em kinematic backreaction}. In these equations angle brackets denote
the spatial volume average of a quantity, so that $\Rav\equiv\left(\int_\DD\dd
^3x\sqrt{\det\gd}\,{\cal R}(t,\bx)\right)/\Vav(t)$ is the average spatial
curvature, for example, with $\Vav(t)\equiv\int_\DD\dd^3x\sqrt{\det\gd}$ being
the volume of the domain $\DD\subset\SI$. Note that $\ab$ is {\em not}
the scale factor of any given geometry, but rather is defined in terms of the
average volume according to
\beq
\ab(t)\equiv\left[\Vav(t)/\Vav(t\Z0)\right]^{1/3}\,.
\eeq
It follows that the Hubble parameter appearing in
(\ref{b1})--(\ref{b3}) is related to the volume-average
expansion scalar, $\th$, by
\beq
{\dot\ab\over\ab}=\frn13\ave{\th}.\label{Hav}
\eeq
The condition
\beq \pt_t\left(\ab^6\QQ\right)+\ab^4c^2\pt_t\left(\ab^2\Rav\right)=0,
\label{intQ}\eeq
is required to ensure that (\ref{b1}) is the integral of (\ref{b2}).
In Buchert's scheme the non-commutativity of averaging and time evolution is
described by the exact relation \cite{BE,EB,buch1,RSKB}
\beq\pt_t{}\ave\Psi-\ave{\pt_t\Psi}=\ave{\Psi\th}
-\ave\th\ave\Psi\label{comm}\eeq
for any scalar, $\Psi$.

Eq.~(\ref{b2}) is suggestive since it implies that if the backreaction
term is large enough -- e.g., for a large variance in expansion
with small shear -- then the volume average acceleration, $\ab^{-1}\ddot\ab=
\frn13\Der\dd t\ave\th+\frn19\ave\th^2$, could be positive, even if the
expansion of all regions is locally decelerating. Although the fraction of the
volume occupied by the faster expanding regions is initially tiny, this
fraction may nonetheless become significant at late epochs, skewing the average
to give an illusion of acceleration during the transition epoch to void
domination. Whether this is observationally viable, however, depends crucially
on: (i) how large the variance in expansion rates can grow given realistic
initial constraints on density perturbations; and (ii) the operational
interpretation of the Buchert formalism. Since Buchert's formalism is a
statistical one, additional assumptions are required to relate solutions of the
Buchert equations to cosmological observations. The \TS\ cosmology, to
be discussed in Secs.~\ref{ideas}, \ref{TS}, provides such a scheme.

\subsection{Notions of average spatial homogeneity}

The relationship between average homogeneity and observations is crucial for
the interpretation any averaging scheme for inhomogeneous cosmology. The very
near isotropy of the CMB demonstrates that when photons travel to us from the
surface of last scattering, they traverse a geometry which to a very good
approximation must be isotropic in some average sense. If we assume a
statistical Copernican principle, then we must also expect some sort of
average notion of spatial homogeneity. The hard question is how to relate the
observed averaged isotropy of the geometry on our past light cone to an
appropriate notion of average spatial homogeneity.

Most cosmologists' physical intuition is guided largely by the FLRW models,
within which average homogeneity can be characterized in
(at least) three distinct ways:
\begin{enumerate}[(i)]
\item The notion of average spatial homogeneity is described by a class of
ideal comoving observers with synchronized clocks.
\item The notion of average spatial homogeneity is described by average
surfaces of constant spatial curvature (orthogonal to the geodesics
of the ideal comoving observers).
\item The expansion rate at which the ideal comoving observers separate within
the hypersurfaces of average spatial homogeneity is uniform.
\end{enumerate}
While these notions coincide for the FLRW geometries, it is not generally
the case once spatial homogeneity is only approximate rather than exact,
given that spatial curvature is characterized by more than a single scalar.

Already in perturbation theory about FLRW models, one can specialize to
spacetime foliations which preserve one of the notions (i)--(iii) of average
spatial homogeneity more fundamentally than the other two. Among the foliations
discussed in the classic work of Bardeen \cite{B80} we can recognize those of
each type above: the {\em comoving hypersurfaces} (and related synchronous
gauge) take property (i) as more fundamental; the {\em minimal shear
hypersurfaces}\footnote{For scalar perturbations this becomes a zero--shear
condition, i.e., $\K_{ij}-\frn13g_{ij}\K=0$, where $\K_{ij}$ is the extrinsic
curvature, $g_{ij}$ the intrinsic metric, and $\K\equiv{\K^\ell}_\ell$. For
general perturbations the hypersurfaces are defined by $\left(\K_{ij}-\frn13
g_{ij}\K\right)_{|ij}=0$, where the bar denotes a covariant derivative with
respect to the intrinsic 3--metric.} (and related Newtonian gauge) are one type
of foliation for which property (ii) is more fundamental; and finally the
{\em uniform Hubble flow hypersurfaces} take property (iii) as more
fundamental.

Bi\v{c}\'ak, Katz and Lynden-Bell \cite{BKL} have further analysed foliations
of perturbed FLRW models, with a view to enabling gauge choices in which the
rotations and accelerations of local inertial frames can be determined directly
from local energy--momentum perturbations $\de{T^\mu}_\nu$. They consider
uniform Hubble flow hypersurfaces; uniform intrinsic scalar curvature
hypersurfaces; and minimal shear hypersurfaces. The {\em uniform intrinsic
scalar curvature hypersurfaces} provide a foliation in addition to those
considered by Bardeen, which also take property (ii) as more fundamental.
Having chosen hypersurfaces Bi\v{c}\'ak, Katz and Lynden-Bell further fix the
gauge by adopting a condition similar to the minimal shift distortion
condition of Smarr and York \cite{SY}. For each choice of hypersurface it
then follows that the coordinates of local inertial frames are more or less
uniquely determined by the energy--momentum perturbations $\de{T^\mu}_\nu$.
In this sense these gauges might be seen as embodying Mach's principle. They
are substantially more restrictive than the commonly used synchronous gauge
or the generalized Lorenz--de Donder gauge \cite{BKL}.

In the nonlinear regime, below the SHS, not all of the conditions (i)--(iii)
can apply, even if they apply in some average sense on scales larger than the
SHS. The question is should any of these notions apply {\em below} the SHS?
The \TS\ scenario begins from the premise that a notion of uniform Hubble
flow can be applied below the SHS, in a way which takes Mach's
principle into the nonlinear regime, as we discuss in Sec.~\ref{ideas}.

This will involve a reinterpretation of the Buchert formalism
\cite{buch1,buch2}, which grew as a generalization of averaging in Newtonian
cosmology \cite{BE,EB}, and is based on an ADM approach on constant time
hypersurfaces of observers ``comoving with the dust''. Since the split of
space and time is unique in Newtonian theory, from the Newtonian viewpoint
this is the only natural choice one can make. However, this is not the
case in general relativity.

If particles of dust were invariant from the time of last scattering
until the present, then there would be no physical ambiguity about the notion
of ``comoving with the dust''. In such a case, a choice of constant time
hypersurfaces with a synchronous gauge is well motivated. However, as
discussed in Sec.~\ref{grain}, in order to consistently deal with both the
particles of ionic dust in voids, and also with `particles' of dust larger
than galaxies, we have to coarse-grain at the SHS over fluid elements which
are themselves expanding. This demands coarse-graining over the gravitational
degrees of freedom relating to spatial curvature, the kinetic energy of
expansion, and gravitational binding energy.

We will adopt the viewpoint that the Buchert time coordinate is a collective
degree of freedom of spacetime regions when coarse-grained at the SHS, and that
if we form thin sandwiches from such regions in the time direction then
they can be combined as effective hypersurfaces on which the Buchert formalism
can be applied. However, new physics applies within the coarse-grained
cells, as we will discuss next.

\section{Timescape scenario: Conceptual foundations}\label{ideas}

In considering the averaging problem, it is inevitable that at some
level one must deal with Mach's principle, which may be stated
\cite{Bondi,BKL}: {\em ``Local inertial frames (LIFs) are determined
through the distributions of energy and momentum in the universe by some
weighted average of the apparent motions''}. Mach's principle strongly
guided Einstein in developing general relativity as a theory in which
spacetime is a relational structure. As Einstein stated in his first
work on cosmology: ``In a consistent theory of relativity there can be
no inertia relatively to `space', but only an inertia of masses relatively
to one another'' \cite{esu}.

The refinement of the understanding of inertia that Einstein left us with in
relation to gravity, the Strong Equivalence Principle (SEP), only goes
part-way in addressing Mach's principle. The SEP tells us that we can always
remove the effects of gravity in a LIF in the neighbourhood of a point.
However, it says nothing about the average effect of gravity, and therefore
nothing about the ``weighted average of the apparent motions'' of
the matter in the universe.

The question of what gravitational mass--energy is in general relativity
is deeply subtle. On account of the SEP we can always get rid of gravity
in the neighbourhood of a point, so any reasonable definition is necessarily
quasilocal, involving integration over a bounding surface. The subject
of quasilocal gravitational energy has occupied many mathematical relativists
\cite{quasirev}, and there is no universally agreed definition. This may
reflect the fact that different notions of energy are applicable in different
circumstances, just as in thermodynamics we deal with internal energy, and
various free energies.

Two of the most familiar gravitational masses are the ADM mass \cite{ADM}
which is defined by an integral on a 2-sphere at spatial infinity for a
general asymptotically flat spacetime, and the Komar mass \cite{K59,H96}
\beq
M={-c^2\over8\pi G}\int_{S^2_\infty}{}^*\dd{\mathbf k}\label{kom}
\eeq
which is similarly defined for asymptotically flat spacetimes with an
asymptotically timelike Killing vector field, $\mathbf k$. The Komar mass
is identical to that appearing in the Newtonian gravitational potential
energy term, $\Phi=-GM/r$, in an asymptotic expansion, $g\Z{00}=-(1+2\Phi/c^2+
\dots)$, at spatial infinity \cite{H96}. Most of the
effort in the field of quasilocal gravitational energy has focused on
ways of defining general geometrical energy quantities which reduce to
familiar results in the case of isolated systems. Some interesting
examples\footnote{This subject has a long history going back to Einstein,
and a huge cast of mathematical relativists have made important contributions,
which I will not attempt to summarize here. See ref.\ \cite{quasirev} for
further details.} include the definitions of Brown and York \cite{BY}, and
Epp \cite{E00}. With a few exceptions of quasilocal energies calculated in
particular backgrounds, (e.g., \cite{CLN}--\cite{af09}), very little has
been done in a cosmological context, however.

\begin{figure}[htb]
\vbox{\centerline{\scalebox{0.5}{\includegraphics{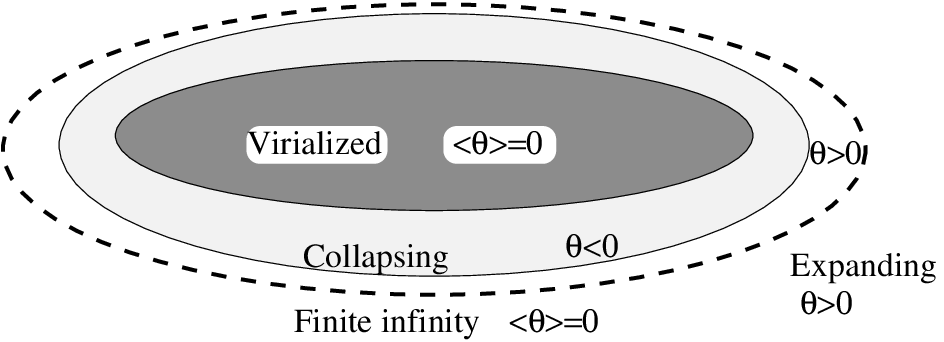}}}
\caption{\label{fig_fi}%
{A schematic illustration of the notion of finite infinity, $\hbox{\it fi}$
\cite{clocks}: the boundary (dashed line) to a region with average zero
expansion inside, and positive expansion outside.}}}
\end{figure}

For asymptotically flat geometries the average of the distribution of energy
and momentum in the external universe is zero. In the actual universe the
spacetime external to any concentrated mass also contains matter so that its
geometry does not have a time symmetry but is necessarily dynamically evolving.
In the \TS\ scenario it is proposed that in place of spatial infinity in
(\ref{kom}) the mass definition for the largest bound structure should be made
in reference to {\em finite infinity}, a timelike surface within which the
average volume expansion is zero.
In general there will be matter collapsing inwards around any virialized
regions, and thus the finite infinity surface will be expanding at the
boundary. (See Figure~\ref{fig_fi}.) The density of a shell at the finite
infinity surface defines the critical density. In a universe which is on
average underdense there must always be such a transition zone between the
overdense regions and the surrounding underdensity.

Accounting for the average effect of matter to address Mach's principle
effectively means specifying an appropriate definition of an asymptotic
region, such as finite infinity, containing any local geometry. Local
geometry should be determined by local matter, and if matter on a bounding
sphere obeys some symmetry principle then we should look no further in
determining the local standard of inertia. Since the very early universe
started out being very close to spatially homogeneous and isotropic and
cannot evolve arbitrarily far from from its initial conditions, it is
proposed that such symmetries can always be found at a {\em regional} scale.

\subsection{The cosmological equivalence principle}
In the \TS\ scenario we restrict the geometry of expanding regions (the
walls and voids) in the final stages of coarse-graining (\ref{coarse}) to an
average over domains which each obey the {\em cosmological equivalence
principle} (CEP) \cite{equiv}:

{\em In cosmological averages it is always possible to choose a suitably
defined spacetime region, the cosmological inertial region (CIR), on whose
boundary average motions (timelike and null) can be described by geodesics
in a geometry which is Minkowski up to some time-dependent conformal
transformation},
\beq \ds^2\Ns{CIR}=
a^2(\et)\left[-\dd\et^2+\dd r^2+r^2\dOM\right].
\label{cif}\eeq
A suitably defined region here refers to one which is smaller than the
scalar curvature scale within underdense voids, or alternatively is the
finite infinity scale for systems containing overdensities. Typically
this could be of order $2\h$--$10\hm$ for finite infinity regions
bounding small groups or rich clusters of galaxies.

Since the average geometry is a time--dependent conformal scaling of Minkowski
space, the CEP reduces to the standard SEP if $a(\et)$ is constant, or
alternatively over very short time intervals during which the time variation
of $a(\et)$ can be neglected. It is well--known that for the exchange of
photons between comoving observers in the background (\ref{cif}), to leading
order the observed redshift of one comoving observer relative to another yields
the same local Hubble law, whether the exact relation, $z+1=a\Z0/a$, is used
or alternatively the radial Doppler formula, $z+1=[(c+v)/(c-v)]^{1/2}$, of
special relativity is used, before making a local approximation. For a small
spacetime region in a spatially homogeneous isotropic background this is
a direct consequence of the SEP: it is impossible to distinguish whether
particles are moving radially in a flat space, or alternatively are at rest
in an expanding space.

The CEP makes the indistinguishability of radial motion from volume expansion
a feature of regional averages on scales up to $2\h$--$10\hm$, while allowing
for inhomogeneity between this scale and the SHS. However, it disallows global
coherent anisotropic expansion of the sort typified by Bianchi models. Bianchi
models single out preferred directions in the global background universe,
thereby imbuing spacetime with absolute qualities that go beyond an essentially
relational structure. To make general relativity truly Machian such backgrounds
need to be outlawed by principle, and the CEP achieves this while still
allowing inhomogeneity.

The CIR metric (\ref{cif}) is of course the spatially flat FLRW metric,
which in the standard cosmology is taken to be the geometry of the whole
universe. In our case the whole universe is inhomogeneous subject to the
restriction that it is possible to always choose (\ref{cif}) as a regional
average in expanding regions.

\subsection{Relative volume deceleration}\label{mach}

To understand the physical implications of taking an average geometry
(\ref{cif}) as the relevant average reference geometry for the relative
calibration of rulers and clocks in the absence of global Killing vectors,
let us construct a thought experiment analogy that I will call the {\em
semi-tethered lattice}. Take a lattice of observers in Minkowski space,
initially moving isotropically away from each nearest neighbour at uniform
initial velocities. The lattice of observers are chosen to be equidistant along
mutually oriented $\hat x$, $\hat y$ and $\hat z$ axes. Suppose that the
observers are each connected to six others by tethers of negligible mass and
identical tension along the mutually oriented spatial axes. The tethers are
not fixed but unwind freely from spools on which an arbitrarily long supply of
tether is wound. The tethers initially unreel at the same uniform rate,
representing a `recession velocity'. Each observer carries synchronized
clocks, and at a prearranged local proper time all observers apply brakes to
each spool, the braking mechanisms having been preprogrammed to deliver
the same impulse as a function of local time.

Applying brakes in the semi-tethered lattice experiment is directly analogous
to the decelerating volume expansion of (\ref{cif}) due to some average
homogeneous matter density, because it maintains the homogeneity and isotropy
of space over a region as large as the lattice. Work is done in applying the
brakes, and energy can be extracted from this -- just as kinetic energy of
expansion of the universe is converted to other forms by gravitational
collapse. Since brakes are applied in unison, however, there is {\em no net
force on any observer in the lattice}, justifying the {\em inertial frame}
interpretation, even though each observer has a nonzero 4-acceleration with
respect to the global Minkowski frame. The braking function may have an
arbitrary time profile; provided it is applied uniformly at every lattice site
the clocks will remain synchronous in the comoving sense, as all observers
have undergone the same relative deceleration.

Whereas the Strong Equivalence Principle allows us to define local inertial
frames, related to each other by local Lorentz transformations acting at a
point, the Cosmological Equivalence Principle refers to a {\em collective}
symmetry of the background. In defining the averaging region of the CIR we are
isolating just that part of the volume expansion which is regionally
homogeneous and isotropic.

\begin{figure}[htb]
\centerline{\ {\sbf(a)}\hskip-20pt
\includegraphics[width=2.8in]{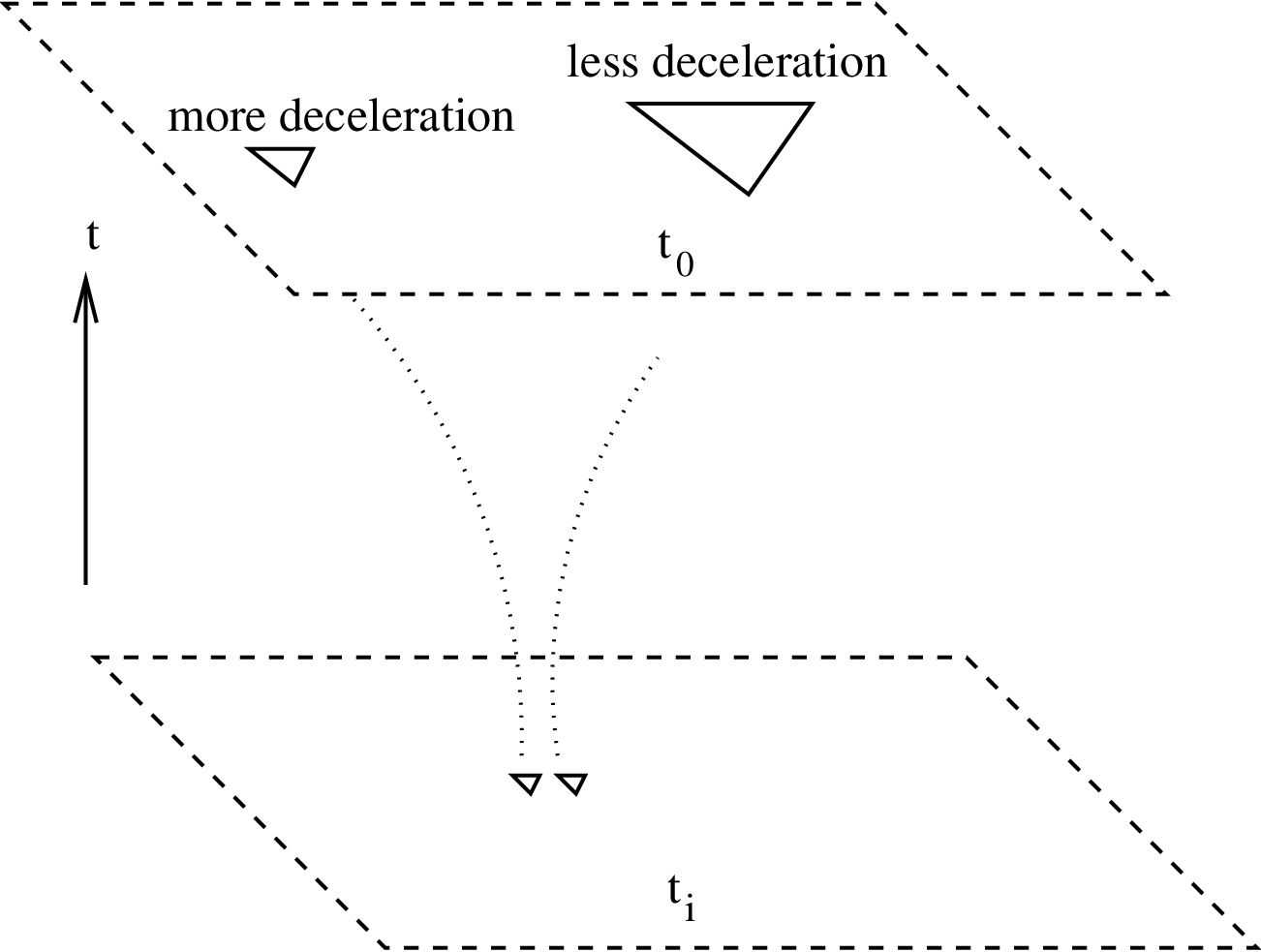}
\ {\sbf(b)}\hskip-20pt
\includegraphics[width=2.8in]{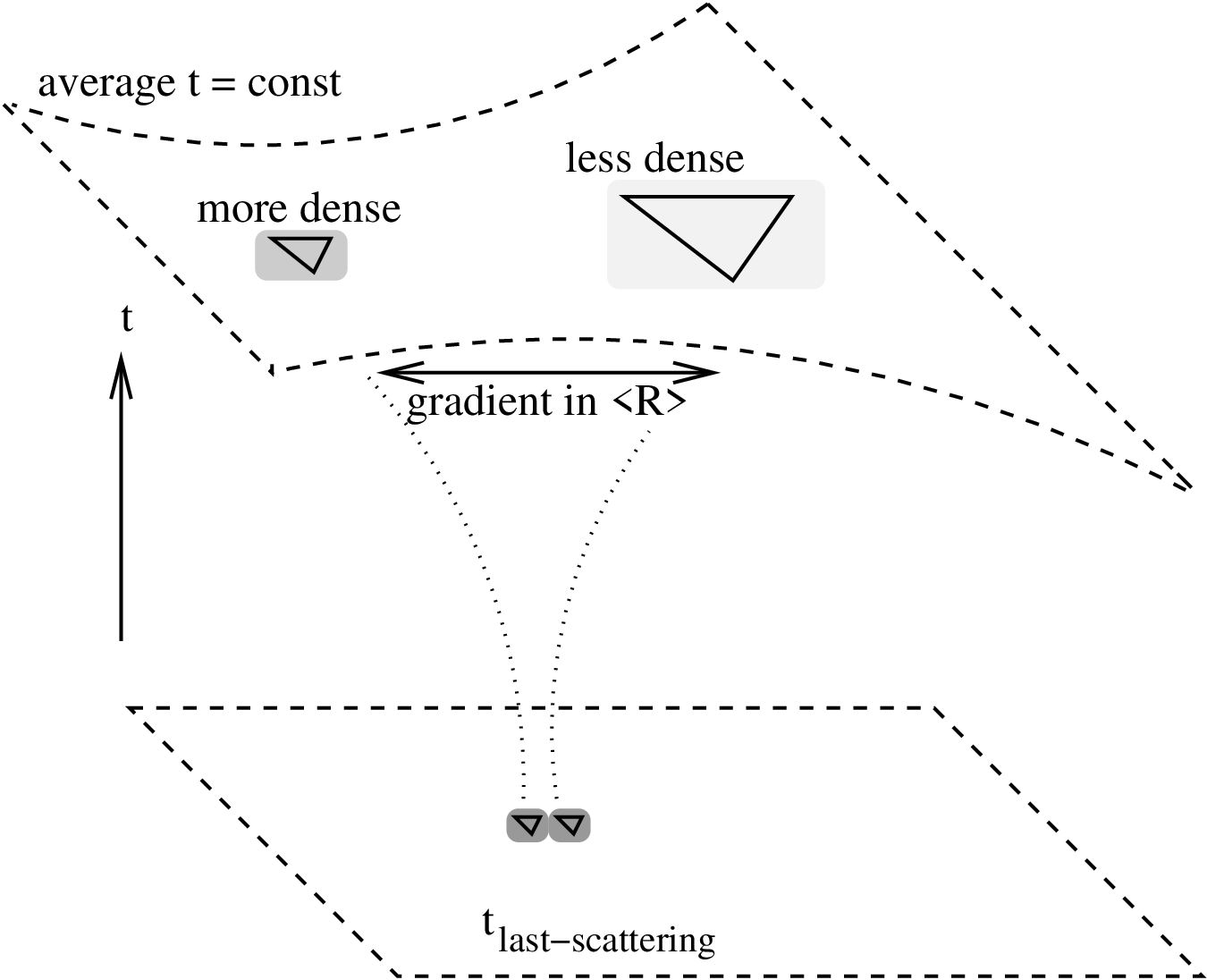}}
\caption{%
Two equivalent situations: {\bf(a)} in Minkowski space observers
in separate semi--tethered lattices, initially expanding at the same
rate, apply brakes homogeneously and isotropically within their respective
regions but at different rates;
{\bf(b)} in the universe which is close to homogeneous and isotropic at
last-scattering comoving observers in separated regions initially move away
from each other isotropically, but experience different locally homogeneous
isotropic decelerations as local density contrasts grow. In both cases there
is a relative deceleration of the observer congruences and those in the region
which has decelerated more will age less.}
\label{fig_equiv}
\end{figure}

Let us now consider two sets of disjoint semi-tethered lattices, with identical
initial local expansion velocities, in a background static Minkowski space.
(See Fig.~\ref{fig_equiv}(a).) Observers in the first congruence apply brakes
in unison to decelerate homogeneously and isotropically at one rate. Observers
in the second congruence do so similarly, but at a different rate. Suppose that
when transformed to a global Minkowski frame, with time $t$, that at each time
step the magnitudes of the 4--decelerations satisfy $\al\Z1(t)>\al\Z2(t)$ for
the respective congruences. By special relativity, since members of the first
congruence decelerate more than those of the second congruence, at any time
$t$ their proper times satisfy $\ta\Z1<\ta\Z2$. The members of the first
congruence age less quickly than members of the second congruence.

By the CEP, the case of volume expansion of two disjoint regions of different
average density in the actual universe is entirely analogous. The equivalence
of the circumstance rests on the fact that by the evidence of the CMB the
expansion of the universe was extremely uniform at the epoch of last
scattering. At that time all regions had almost the {\em same} density -- with
tiny fluctuations -- and the same uniform Hubble flow. At late epochs, suppose
that in the frame of any average cosmological observer there are expanding
regions of {\em different} density which have decelerated by different amounts
by a given time, $t$, according to that observer. Then by the CEP the local
proper time of the comoving observers in the denser region, which has
decelerated more, will be less than that of the equivalent observers in the
less dense region which has decelerated less. (See Fig.~\ref{fig_equiv}(b).)
Consequently the {\em proper time of the observers in the more dense CIR will
be less than that of those in the less dense CIR}, by equivalence of the two
situations.

The fact that a global Minkowski observer does not exist in the second case
does not invalidate the argument. The global Minkowski time is just a
coordinate label. In the cosmological case the only restriction is that the
expansion of both average congruences must remain homogeneous and isotropic in
local regions of different average density in the global average $t=$const
slice. Provided we can patch the regional frames together suitably, then if
regions in such a slice {\em are still expanding} and have a significant
density contrast we can expect a significant clock rate variance.

This equivalence directly establishes the idea of a {\em gravitational
energy cost for a spatial curvature gradient}, since the existence
of expanding regions of different density within an average $t=$const
slice implies a gradient in the average Ricci scalar curvature, $\Rav$,
on one hand, while the fact that the local proper time varies
on account of the relative deceleration implies a gradient in gravitational
energy on the other.

The variation of the normalization of asymptotic clocks due to a relative
volume deceleration is a new physical effect. We are familiar with boosts in
particular directions, which give significant effects only for large relative
velocities; e.g., as required to remain stationary in strong gravitational
fields. Since we only consider weak fields the relative deceleration of the
background is small. However, even if the relative deceleration is
typically of order $10^{-10}$m$\,$s$^{-2}$, cumulatively over the age of the
universe it leads to significant variation in the calibration of clocks,
as we will discuss at the end of Sec.~\ref{Bsol}.

\subsection{Statistical cosmological geometry}

\begin{figure}[htb]
\centerline{\quad{\sbf(a)}\hskip-20pt
\includegraphics[width=2.9in]{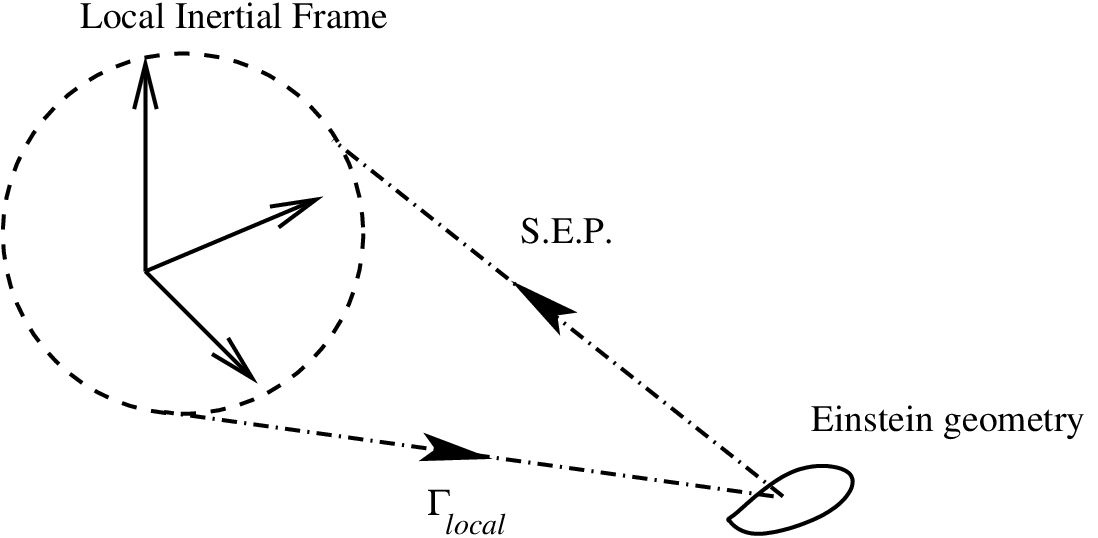}
{\sbf(b)}\hskip-15pt
\includegraphics[width=2.9in]{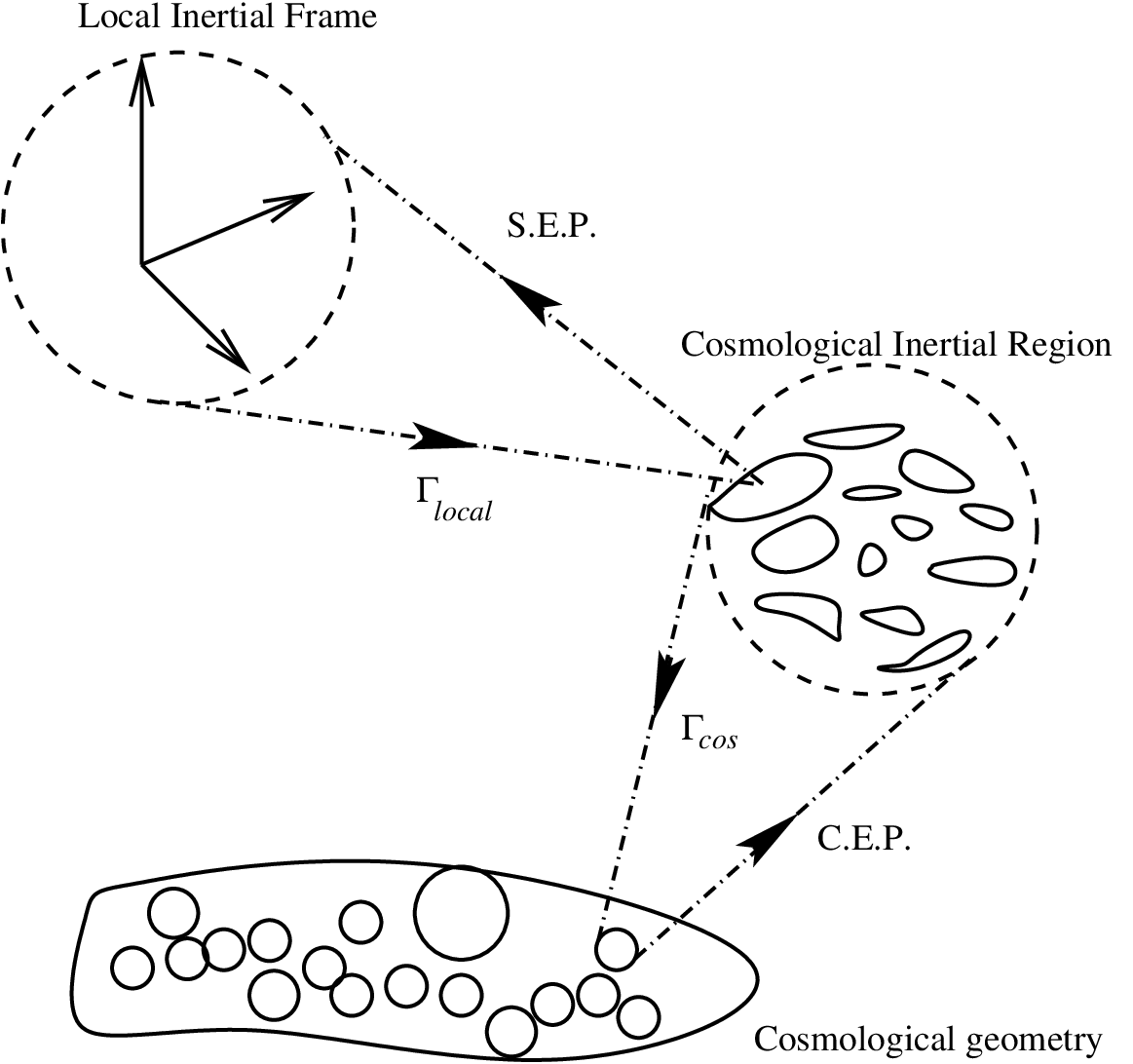}}
\caption{%
{\bf(a)} In general relativity the SEP allows one to relate a solution
of Einstein's equations with prescribed matter source to LIFs. The connection,
$\GA_{local}$, allows one to parallely transport tensorial
quantities from one LIF to another.
{\bf(b)} In the \TS\ scenario there is (at least) one additional
layer of geometrical structure. Local geometries with asymptotic regions
bounded by CIRs are combined in a cosmological average statistical geometry.
The CEP allows one to relate a solution for the statistical geometry to
the regional geometry of a CIR. The statistical cosmological geometry should
be equipped with an appropriate connection, $\Gcos$, to
allow parallel transport from one CIR to another.}
\label{fig_mod}
\end{figure}

The \TS\ scenario represents an extension of the concepts of general
relativity in the cosmological domain, as illustrated schematically in
Fig.~\ref{fig_mod}. In particular, it is recognized that once gravitational
degrees of freedom are coarse--grained then one is no longer dealing with
a simple solution of Einstein's equations with a prescribed matter source.
Rather than cutting and pasting exact solutions of Einstein's equations
as one does in the Swiss cheese \cite{ES} and meatball \cite{MN11} models,
we are dealing with a new statistical cosmological geometry in which
the relative volume deceleration provides a physical degree of freedom
to normalize canonical clocks. The relative phenomenological lapse function
provides a measure of the relative kinetic energy of expansion of CIRs.

The Buchert formalism clearly deals with statistical quantities. However,
although Buchert and Carfora \cite{BC2} realized early on that there will be
differences between the bare volume--average statistical parameters of the
Buchert formalism and dressed parameters as determined by any particular
observer, the relationship between observers and the statistical averages
requires additional inputs and assumptions. Likewise one must specify what
is understood by the phrase ``comoving with the dust'' once the dust
approximation has broken down, as is the case for observers in gravitationally
bound structures in overdense regions.

The \TS\ scenario seeks to address these questions by implementing
the CEP. To date a rigorous geometrical framework for the statistical
cosmological geometry has not been implemented. A phenomenological working
framework will be outlined in the next section. Ultimately the geometrical
framework should be one in which the Hubble parameter is to some extent
a gauge choice -- it corresponds to the first derivatives of the metric
of the statistical geometry encoded in the connection, $\Gcos$. This is of
course obvious in the early universe in which the FLRW and statistical
geometry are one and the same, on account of global homogeneity and isotropy.
How to develop an appropriate framework at late times is less obvious;
however, it seems likely that should involve the notion of {\em scale
invariance} of the statistical spatial 3-geometry.

\section{Timescape scenario: Phenomenological model}\label{TS}

In order to deal with the evolution from the epoch of last scattering up to
the present day, we assume that dust can be coarse-grained at the $\goesas100
\hm$ scale of statistical homogeneity over which mass flows can be neglected.
We apply the Buchert formalism, but interpret it in a different manner to
Buchert \cite{buch1,buch2}, who did not define the scale of coarse-graining of
the dust explicitly. We will assume that the Buchert average itself is
performed over our present horizon volume, to describe average cosmic
evolution on the largest scales accessible to our observations.

Prior to last scattering the universe is close to homogeneous, so that
\TS\ model is almost indistinguishable from the standard cosmology,
being very close to a standard matter plus radiation FLRW model with
negligible spatial curvature. At late epochs, the solutions to the Buchert
equations will differ substantially from a FLRW model. Assuming no dark
energy, then it is the matter density and its variance which drives the overall
evolution of the universe. While the radiation fluid certainly responds to
density gradients, this only affects questions such as gravitational lensing,
rather than the average cosmological evolution described by the Buchert
equations. We therefore treat the radiation fluid as a component with a
pressure $\pR=\frac13\rhR$ which commutes under the Buchert average,
\beq
\pt_t\ave{\pR}-\ave{\pt_t \pR}=\ave{\pR\th}-\ave{\pR}\ave{\th}=0,
\eeq
throughout the evolution of the universe, rather than using the more detailed
Buchert formalism that applies to fluids with pressure \cite{buch2}. The
relevant Buchert equations are then (\ref{b1}) with $\ave\rh\to\ave\rhM+
\ave\rhR$, (\ref{b2}) with $\ave\rh\to\ave\rhM+2\ave\rhR$, (\ref{b3})
with $\ave\rh\to\ave\rhM$, (\ref{intQ}), and
\beq
\pt_t\ave\rhR+4{\dot\ab\over\ab}\ave\rhR=0.
\label{b4}\eeq

To obtain a phenomenologically realistic solution consistent with observations
of voids in the cosmic web \cite{hv02}--\cite{pan11} we assume that the present
epoch horizon volume, $\Vav=\Vav\ns i\ab^3$, is a disjoint union of void and
wall regions characterized by scale factors $\av$ and $\aw$ related to the
volume-average scale factor by
\beq
\ab^3=\fvi\av^3+\fwi\aw^3\label{bav}
\eeq
where $\fvi$ and $\fwi = 1-\fvi$ represent the fraction of the initial volume,
$\Vav\ns i$, in void and wall regions respectively at an early unspecified
epoch. We may rewrite (\ref{bav}) as
\beq\fv(t)+\fw(t)=1,\eeq
where $\fw(t)=\fwi\aw^3/\ab^3$ is the {\em wall volume fraction} and $\fv(t)=
\fvi\av^3/\ab^3$ is the {\em void volume fraction}. Taking a derivative of
(\ref{bav}) with respect to the Buchert time parameter, $t$, we find that the
bare Hubble parameter is given by
\beq
\bH\equiv{\dot\ab\over\ab}=\fw\Hw+\fv\Hv\,,\label{bareH}
\eeq
where $\Hw\equiv\dot\aw/\aw$ and $\Hv\equiv\dot\av/\av$ are the Hubble
parameters of the walls and voids respectively as determined by the clocks
of volume--average observers.

The voids are assumed to have negative spatial curvature characterized
by $\Rav_{\ns v}\equiv6\kv/\av^2$ with $\kv<0$, while the wall regions
\cite{clocks} are on average spatially flat, $\Rav_{\ns w}=0$. It then
follows that
\beq
\Rav=\frac{6\kv\fvi^{2/3}\fv^{1/3}}{\ab^2}\label{Rav}
\eeq
We also assume that the kinematic backreaction vanishes separately within the
voids and walls\footnote{For spherical voids this is reasonable since the
average shear and vorticity are small. Shear and vorticity may be significant
within bound structures in the wall regions, but their contributions are of
the opposite sign in the Raychaudhuri equation and might be largely
self--canceling, giving
rise to second order effects. The Buchert formalism neglects vorticity, and
realistically this should be treated together with the effect of nonzero
shear. Since we smooth at the finite infinity scale we are neglecting the
gravitational physics associated with bound structures, where these effects
are likely to be important. In the model presented here is assumed that the
variation of the kinetic energy of expansion can be quantified independently
of the gravitational physics within nonexpanding regions.}
but not in the combined average.
One then finds that the kinematic backreaction (\ref{backr}) reduces to a
term depending on the relative expansion of voids and walls
\beq
\QQ=6\fv(1-\fv)\left(\Hv-\Hw\right)^2=\frac{2\dot\fv^2}{3\fv(1-\fv)}\,.
\label{Qav}\eeq

Since (\ref{b3}) and (\ref{b4}) are solved by $\ave{\rhM}=\rhbMn(\ab/\abn)^{-3}
$ and $\ave{\rhR}=\rhbRn(\ab/\abn)^{-4}$ respectively, where the subscript zero
refers to quantities evaluated at the present epoch, the remaining independent
Buchert equations may then be written as
\beq
\frac{\dot\ab^2}{\ab^2}+\frac{\dot\fv^2}{9\fv(1-\fv)}-\frac{\al^2\fv^{1/3}}
{\ab^2} = \frac{8\pi G}{3}\left( \rhbMn\frac{\abn^3}{\ab^3}+\rhbRn
\frac{\abn^4}{\ab^4}\right),\label{b5}
\eeq
\beq
\ddot\fv+\frac{\dot\fv^2(2\fv-1)}{2\fv(1-\fv)} + 3\frac{\dot\ab}{\ab}\dot\fv-
\frac{3\al^2 \fv^{1/3}(1-\fv)}{2\ab^2}=0,\label{b6}
\eeq
where $\al^2\equiv-\kv\fvi^{2/3}>0$.

Equation (\ref{b5}) may also be conveniently written
\beq
\OMM+\OMR+\OMk+\OMQ=1,\label{beq1}
\eeq
where
\beq\OMM\equiv{8\pi G\rhb\Z{M0}\abn^3\over 3\bH^2\ab^3}\,\quad
\OMR\equiv{8\pi G\rhb\Z{R0}\abn^4\over 3\bH^2\ab^4}\,,\quad
\OMk\equiv{\al^2\fv^{1/3}\over \ab^2\bH^2}\,,\quad
\OMQ\equiv{-\dot\fv^2\over 9\fv(1-\fv)\bH^2}\,,\label{om}
\eeq
are the volume--average or {\em bare} density parameters \cite{BC1,BC2} of
matter, radiation, average spatial curvature and kinematic backreaction
respectively. It is straightforward to add a cosmological constant term to
the r.h.s.\ of (\ref{b5}), giving rise to a further density parameter
$\OM\Z\LA= \LA/(3\bH^2)$, and in fact the equivalent solution with matter and a
cosmological constant (but no radiation) has been derived in \cite{V12,VM}.
Since we are interested in the possibility of a viable cosmology without dark
energy, we set $\OM\Z\LA=0$.

\subsection{Matching regional to statistical geometry}

Thus far we have simply set out the Buchert equations for a particular
ensemble of wall and void regions, leading to differential equations which
can be solved and possibly interpreted in many ways\footnote{Buchert and
Carfora \cite{BC3}, and Wiegand and Buchert \cite{WB}, have investigated a
very similar model, without radiation, which also allows the possibility of
internal kinematic backreaction within the walls and voids. They do not
directly consider the issue of gravitational energy.}.
Since the Buchert equations describe statistical averages, the
relationship of the statistical solutions to local geometry is crucial
to the physical interpretation of the Buchert formalism. Here I will outline
the phenomenological implementation of the principles of the \TS\ scenario
discussed in Sec.~\ref{ideas}.

The wall regions are a union of disjoint {\em finite infinity} regions
\cite{fit1,clocks} encompassing bound structures, with local average metric
(\ref{cif}), which can be rewritten as
\beq\ds^2\Z{\Fi}=-c^2\dd\tw^2+\aw^2(\tw)\left[\dd\etw^2+\etw^2\dOM\right]\,.
\label{wgeom}\eeq
in terms of the wall time, $\tw$, related to the wall conformal time by
$c\,\dd\tw=a\,\dd\etw$. Although each finite infinity region is distinct,
since they each represent a region within which the average density is
critical, evolved from the same initial conditions, the $\tw$ parameters
can be taken to be synchronous.

The voids are characterized by regional negatively curved metrics of the form
\beq\ds^2\Z{\DD\ns{v}}=-c^2\dd\tv^2+\av^2(\tv)\left[\dd\etv^2+\sinh^2(\etv)
\dOM\right]\,.\label{vgeom}\eeq
Generally the voids will have different individual metrics (\ref{vgeom}).
However, in the void centres the regional geometry will rapidly approach that
of an empty Milne universe for which the parameters $\tv$ can be assumed to
be synchronous. One could potentially use different curvature scales for
dominant voids and minivoids to characterize the average scalar curvature
$\ave{\cal R}$. However, in the two--scale approximation of \cite{clocks,sol}
a single negative curvature scale is assumed as a simplification.

Within the dust particles the metrics (\ref{wgeom}) and (\ref{vgeom}) are
assumed to be patched together with the condition of uniform quasilocal bare
Hubble flow \cite{clocks,equiv}
\beq
\bH={1\over\aw}\Deriv\dd\tw\aw={1\over\av}\Deriv\dd\tv\av,\label{uniH}
\eeq
discussed in Sec.~\ref{mach}. In particular, the regional Hubble parameters
are also equal to the bare Buchert Hubble parameter (\ref{bareH}). The Buchert
average parameters $\Hw$ and $\Hv$ refer to expansion rates with respect
to the volume--average time parameter $t$, so that (\ref{uniH}) may be
rewritten
\beq
\bH=\gw\Hw=\gv\Hv
\eeq
where
\beq\gw\equiv\Deriv\dd\tw{t\ },\qquad\gv\equiv\Deriv\dd\tv{t\ },
\label{clocks1}
\eeq
are phenomenological lapse functions of volume--average time, $t$, relative
to the time parameters of isotropic wall and void--centre observers
respectively. The ratio of the relative Hubble rates $h_r=\Hw/\Hv<1$ is related
to the wall lapse function by
\beq\gw=1+{(1-h_r)f_v\over h_r},\label{clocks2}\eeq
and $\gv=h_r\gw$.

As we ourselves live in a bound structure and can be considered to be wall
observers, there is no further need to refer to the void time parameter, $\tv$.
We will henceforth drop the subscript w from quantities defined in
(\ref{clocks1}) and replace $\tw\to\ta$, $\gw\to\gb$.

We may rewrite $\OMQ=-(1-\fv)(1-\gb)^2/(\fv\gb^2)$, and combine it with the
other density parameters (\ref{om}) to give
\beq
\gb={\sqrt{1-\fv}\,\Bigl[\sqrt{1-\fv}+\sqrt{\fv(\OM-1)}\;\Bigr]
\over1-\fv\OM}\,,
\label{gb1}\eeq
where
\beq \OM\equiv1-\OMQ=\OMM+\OMR+\OMk\,, \label{gb2}\eeq
which satisfies $\OM>1$ for the solutions of interest. As $t\to0$, $\fv\to0$,
$\OMQ\to0$, $\OM\to1$ and $\gb\to1$; i.e., initially the void fraction and
backreaction are negligible, and the wall time and volume-average time
parameters coincide.

Solutions of the Buchert equations are not directly related to any physical
metric. Since all cosmological information is obtained by a radial spherically
symmetric null cone average, given a solution of the Buchert equations we will
retrofit a spherically symmetric geometry relative to an isotropic observer
who measures volume-average time, according to
\beq
\dd\mean s^2=-c^2\dd t^2+\ab^2(t)\,\dd\etb^2+\Aa(\etb,t)\,\dOM.
\label{avgeom}
\eeq
Here the area quantity, $\Aa(\etb,t)$, satisfies
$\int^{\etbH}_0\dd\etb\, \Aa(\etb,t)=\ab^2(t)\Vav\ns{i}
(\etbH)/(4\pi)$, $\etbH$ being the conformal distance to
the particle horizon relative to an observer at $\etb=0$. The metric
(\ref{avgeom}) is spherically symmetric by construction, but is not a
LTB solution since it is not an exact solution of Einstein's equations, but
rather a phenomenological fit to the Buchert average of the Einstein equations.

In terms of the wall time, $\ta$, of finite infinity observers in walls
the metric (\ref{avgeom}) is
\beq\dd\mean s^2=-\gb^2(\ta)\,c^2\dd\ta^2+\ab^2(\ta)\,\dd\etb^2+\Aa(\etb,\ta)\,
\dOM\,.\label{avgeom2}\eeq
This geometry, which has negative spatial curvature is not the locally measured
geometry at finite infinity, which is given instead by (\ref{wgeom}). Since
(\ref{wgeom}) is not a statistical geometry, we match (\ref{wgeom}) to
(\ref{avgeom2}) to obtain a {\it dressed} statistical geometry. The matching is
achieved in two steps. Firstly we conformally match radial null geodesics of
(\ref{wgeom}) and (\ref{avgeom2}), noting that null geodesics are unaffected by
an overall conformal scaling. This leads to a relation
\beq
\dd\etw={\fwi^{1/3}\dd\etb\over\gb\fvf^{1/3}}\label{etarel}
\eeq
along the geodesics. Secondly, we account for volume and area factors by taking
$\etw$ in (\ref{wgeom}) to be given by the integral of (\ref{etarel}).

The wall geometry (\ref{wgeom}), which may also be written
\beq \ds^2\Z{\Fi}=-c^2\dd\ta^2+{\fvf^{2/3}\ab^2\over\fwi^{2/3}}
\left[\dd\etw^2+\etw^2\dOM\right]\,,
\eeq
on account of (\ref{bav}), is a local geometry only valid in spatially flat
wall regions. We now use (\ref{etarel}) and its integral to extend this
metric beyond the wall regions to obtain the dressed statistical metric
\bea
\ds^2&=&-c^2\dd\ta^2+{\ab^2\over\gb^2}\,\dd\etb^2+
{\ab^2\fvf^{2/3}\over\fwi^{2/3}}\,\etw^2(\etb,\ta)\,\dOM\nonumber\\
&=&-c^2\dd\ta^2+a^2(\ta)\left[\dd\etb^2+\rw^2(\etb,\ta)\,\dOM\right]
\label{dgeom}\eea
where $a\equiv\gb^{-1}\ab$, and
\beq\rw\equiv\gb\fvf^{1/3}\fwi^{-1/3}\etw(\etb,\ta).\eeq
While (\ref{wgeom}) represents a local geometry only valid in spatially flat
wall regions, the dressed geometry (\ref{dgeom}) represents an average
effective geometry extended to the cosmological scales,
parametrized by the volume--average conformal time which satisfies
$\dd\etb=c\,\dd t/\ab=c\,\dd\ta/ a$. Since the geometry on cosmological scales
does not have constant Gaussian curvature the average metric (\ref{dgeom}),
like (\ref{avgeom}), is spherically symmetric but not homogeneous.

Wall observers who try to fit a FLRW model with `cosmic time' synchronous to
wall time, $\ta$, are then effectively fitting the dressed geometry
(\ref{dgeom}), which is the closest thing there is to a FLRW geometry adapted
to the rulers and clocks of wall observers. The cosmological parameters we
infer from taking averages on scales much larger than the SHS will not then be
the bare parameters $\bH$, $\OMM$, $\OMk$, and $\OMQ$, but instead the {\em
dressed Hubble parameter}
\bea\Hh\equiv{1\over a}\Deriv\dd\ta a
={1\over\ab}\Deriv\dd\ta\ab-{1\over\gb}\Deriv\dd\ta\gb
=\gb\bH-\Deriv{\dd}t\gb\,,\label{42}
\eea
and the {\em dressed matter density parameter}
\beq \Omega\Z{M}=\gb^3\OMM\,.\eeq
There is similarly a dressed luminosity distance relation
\beq\dL=a\Z0(1+z)\rw,\label{eq:dL}\eeq where $a\Z0=\ab\Z0/\gbn$,
$1+z\equiv a\Z0/a=(\ab\Z0\gb)/(\ab\,\gbn)$, and
\beq\rw=\gbn\fvf^{1/3}
\int_t^{t\X0}{c\,\dd t'\over\gb(t')(1-\fv(t'))^{1/3}\ab(t')}\,,
\label{eq:rw}\eeq
We can also define an {\em effective angular diameter distance}, $\dA$, and an
{\em effective comoving distance}, $D$, to a redshift $z$ in the
standard fashion
\beq\dA={D\over1+z}={\dL\over(1+z)^2}\,.\label{dist}\eeq
\subsection{Cosmological solutions and their timescape interpretation}
\label{Bsol}

\begin{figure}[htb]
\centerline{\includegraphics[width=3.in]{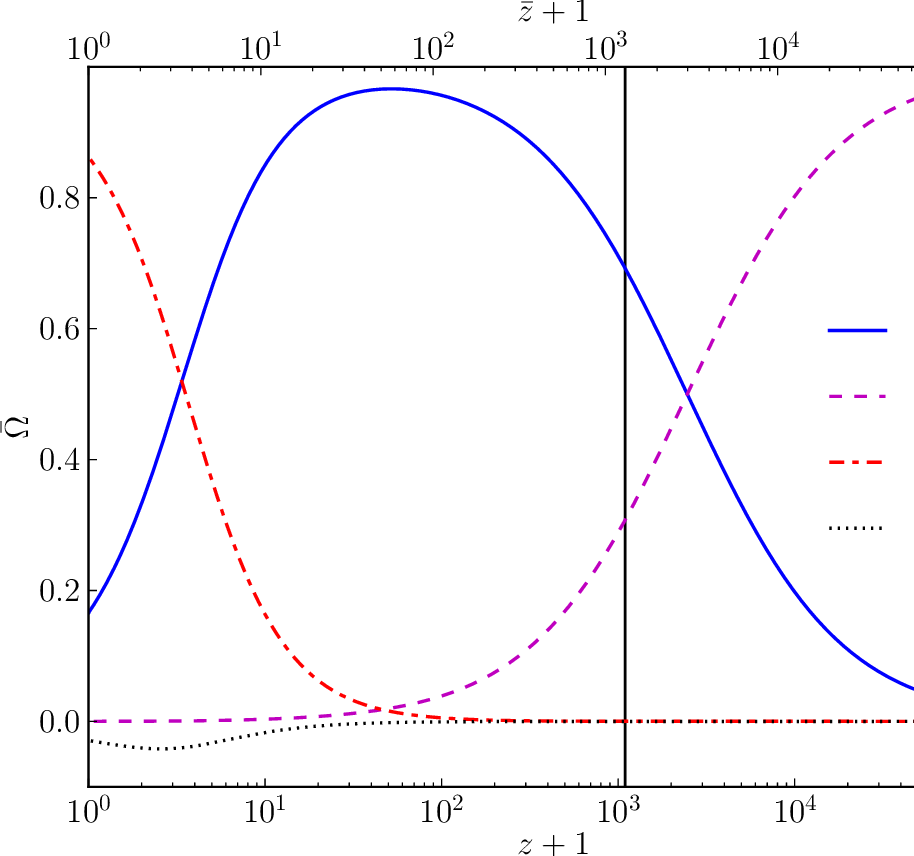}}
\caption{%
Bare density parameters (\ref{om}) for the full numerical
solution, as a function of dressed redshift $z+1=\bar\ga\,\abn/(\bar\ga\,\ab)$
(and bare redshift $\bar z+1=\abn/\ab$), for the dressed parameters $\Hm=61.7
\kmsMpc$, $\OmMn=0.410$. The vertical bar at $1094.88<z<1100.46$ corresponds to
the epoch of decoupling.}\label{fig_Om}
\end{figure}

We have recently obtained \cite{dnw} full numerical solutions of the Buchert
equations for a matter plus radiation fluid, evolved forward from an early
initial time when the solutions are well approximated by series solutions.
E.g., we begin integrations after the epoch of primordial nucleosynthesis,
at $\Hb t\simeq5\times10^{-11}$ when the universe is about a year old.
Here $\Hb= \bH(\tn)$ is the bare (volume-average) Hubble constant. Bare
density parameters (\ref{om}) for typical solutions are shown
in Fig.~\ref{fig_Om}.

While numerical solutions are needed to smoothly match solutions from the
radiation-dominated epoch to later epochs, the full numerical solution%
\footnote{The matter only solution, $\OMR=0$, is also analytically soluble
\cite{sol,obs}. However, the tracking limit is reached to within 1\% for
redshifts $z\lsim37$. For larger redshifts $z\gsim50$ one needs to include
radiation to obtain accurate solutions. Thus the full numerical solution is
actually required in that regime.} possesses a tracking limit with a simple
analytic form \cite{sol,obs} which is very accurate at epochs $z<10$. The
tracking corresponds to the walls expanding as an Einstein--de Sitter model,
$\aw=a\ns{w0}t$, and the voids as an empty Milne universe, $\av=a\ns{v0}t$,
in volume average time, so that $h_r=2/3$. The solution to the Buchert
equations is then given by
\bea
\ab&=&{\ab\Z0\bigl(3\Hb t\bigr)^{2/3}\over2+\fvn}\left[3\fvn\Hb t+
(1-\fvn)(2+\fvn)\right]^{1/3}\nonumber\\
\label{track1}\\
\fv&=&{3\fvn\Hb t\over3\fvn\Hb t+(1-\fvn)(2+\fvn)}\,.
\label{track2}
\eea

The density parameters (\ref{om}) and other quantities for the
tracking solution are all found to have simple analytic forms in terms of
the void fraction, $\fv$. For example, the bare Hubble parameter,
phenomenological lapse function, and dressed Hubble parameter satisfy $\bH=
(2+\fv)/(3t)$, $\gb=\half(2+\fv)$ and $H=(4\fv^2+\fv+4)/(6t)$ respectively.
(For further details, see ref.~\cite{obs}, Appendix B.) Parameters for the
full numerical solution with radiation differ from those of the tracker
solution by $0.3$\% or less at late times.

In the tracker limit the \TS\ wall time is related to volume average time
by
\beq
\ta=\frn23\ts+{4\OmMn\over27\fvn\Hb}\ln\left(1+{9\fvn\Hb t
\over4\OmMn}\right)\,, \label{tsol}
\eeq
where $\OmMn=\frn12(1-\fvn)(2+\fvn)$ is the present epoch dressed matter
density. In general the two parameters will differ substantially at late
epochs -- in fact by some billions of years -- meaning that the age of
the universe is observer--dependent. Nonetheless, we and all the objects we
observe are necessarily in regions of greater than critical density, where
the asymptotic time parameter is wall time, $\ta$. Consequently this radical
departure from conventional assumptions does not lead to any immediate
conflict with observation, on account of our mass--biased view of the
universe.

A present epoch large variation of clock rates, of order $35$\%, is the
cumulative effect of an instantaneous relative volume deceleration between
walls and voids which can be defined as \cite{equiv}
\beq
{\al\over c}={1\over\left[\gb^2-1\right]^{1/2}}\Deriv\dd\ta\gb=\Der\dd t
\left[\gb^2-1\right]^{1/2}\,.
\label{a2}\eeq
This is the deceleration that would arise from treating $\gb$ as the
$\ga$--factor of a purely transverse Lorentz boost. The phenomenological lapse
function relates to an isotropic regional volume deceleration, and is not
associated with any particular spatial direction, which is why the transverse
Lorentz boost formula is applied. For the late time tracker solution
\beq
{\al\over c}={3(1-\fvn)(2+\fvn)\fv(t)\bH(t)\over2\sqrt{3\fvn\Hb t
\left[15\fvn\Hb t+4(1-\fvn)(2+\fvn)\right]}}\,.\label{a3}
\eeq

\begin{figure}[htb]
\centerline{\ {\sbf(a)}\hskip-10pt
\includegraphics[width=2.4in]{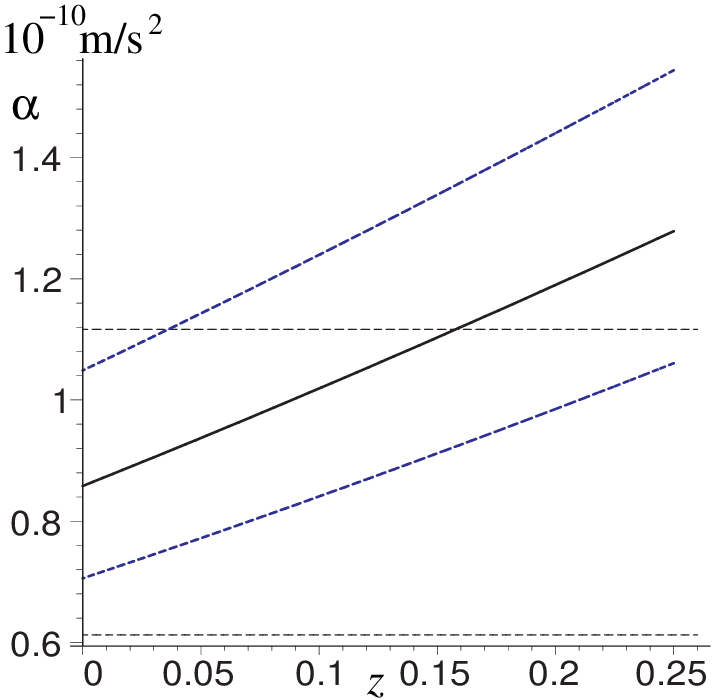}
\qquad {\sbf(b)}\hskip-20pt
\includegraphics[width=2.4in]{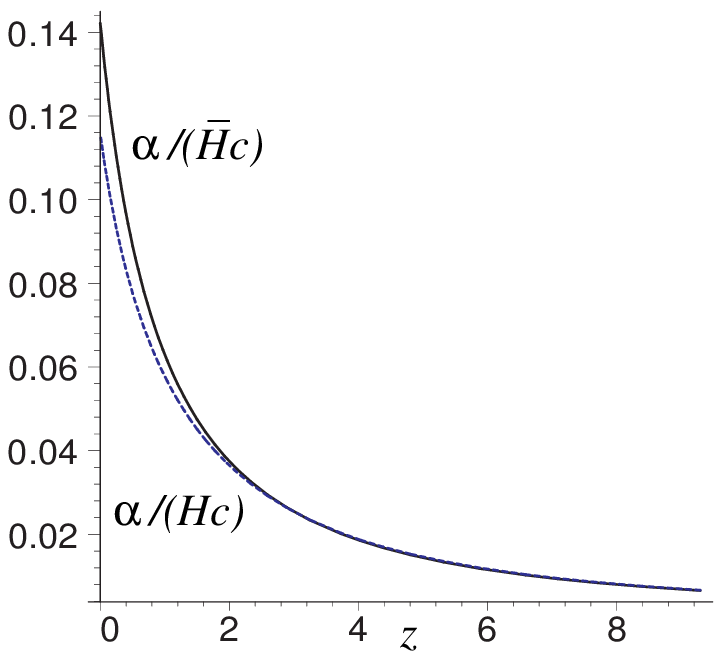}}
\caption{%
The magnitude of the relative deceleration scale \cite{equiv}, $\al$: {\bf(a)}
in terms of its absolute value for redshifts $z<0.25$; {\bf(b)} in terms of the
dimensionless ratios $\al/(c\bar H)$ (solid curve) and $\al/(cH)$ (dashed
curve) for redshifts $z<10$. In panel (b) just the best fit value $\fvn=0.695$
is shown, whereas in panel (a) the solid and dashed represent the best fit
value and 1$\si$ uncertainties from Table~\ref{parms}. The narrower range of
uncertainties obtained from the Planck data gives a smaller range of
uncertainty in $\al$ as compared to earlier work \cite{equiv}. In panel
{\bf(a)} the horizontal dotted lines indicate the upper and lower bounds of the
empirical acceleration scale of MOND when normalized to $H\Z0=61.7\pm3.0\kmsMpc
$.}
\label{fig_acc}
\end{figure}
The relative deceleration parameter is plotted in Fig.~\ref{fig_acc}, in
absolute terms at small redshifts, $z<0.25$, and as a fraction of $c\bH$ and
$cH$ over a larger range of redshifts. Although $\al$ is larger in absolute
terms at earlier times, the Hubble expansion is much larger at early times, so
that the ratio $\al/(c\bH)$ or $\al/(cH)$ is in fact small at large redshifts.
Using the parameter values from Table~\ref{parms} in Sec.~\ref{cmb}, we find
$\al\Z0=8.6^{+1.9}_{-1.5}\times10^{-11}$m$\,$s$^{-2}$ at $z=0$, which is well
within the weak field regime. Intriguingly, this coincides with the empirical
acceleration scale of MOND, $\al\ns{mond}=1.2_{-0.2}^{+0.3}\times10^{-10}h_{75}
^2$ms$^{-2}$ \cite{McGaugh}, where $h_{75}=H\Z0/(75\kmsMpc)$. For the values of
$\Hm$ given in Table~\ref{parms}, $\al\ns{mond}=8.1^{+3.0}_{-2.0}\times10^{-11}
$ms$^{-2}$. It has been often observed that the value $\al\ns{mond}$ is close
to $c\Hm$ \cite{sm02}. However, $c\Hm$ is actually one order of magnitude
larger than $\al\ns{mond}$, whereas here $\al\Z0$ and $\al\ns{mond}$ agree
precisely within the uncertainty. Furthermore, $\al$ is a relative deceleration
scale obtained from {\em derivatives} of quantities related to the Hubble
parameter, meaning that one should not simply expect a numerical coincidence
based on the value of $\Hm$. Since the physics of bound systems has not been
investigated in the \TS\ scenario, it remains to be seen whether the $\al
\Z0\simeq\al\ns{mond}$ coincidence has any deeper significance.

\subsection{Apparent acceleration and Hubble expansion variance}

The gradient in gravitational energy and cumulative differences of clock
rates between wall observers and volume average observers has an important
consequence for apparent cosmic acceleration. A volume average isotropic
observer, namely one whose local geometry has the same spatial curvature
as the volume average, would infer an effective {\em bare deceleration
parameter}
$\bq\equiv-\ddot\ab/(\bH^2\ab)$. Using the tracker solution approximation
$\bq=2\fvf^2/(2+\fv)^2$, which is always positive, meaning that there is no
actual acceleration. However, a wall observer infers a {\em dressed
deceleration parameter}
\beq
q={-1\over H^2 a}{\dd^2 a\over\dd\ta^2}=
{-\fvf(8\fv^3+39\fv^2-12\fv-8)\over\left(4+\fv+4\fv^2\right)^2}\,,
\label{qtrack}\eeq
where again we have used the tracker solution in the last step.
At early times, when $\fv\to0$, both the bare and dressed deceleration
parameters take the Einstein--de Sitter value $q\simeq\bq\simeq\half$. However,
unlike the bare parameter which monotonically decreases to zero, the dressed
parameter becomes negative when $\fv\simeq0.59$ and $\bq\to0^-$ at late times.

The origin of apparent cosmic acceleration in the \TS\ scenario
differs from that envisaged in some other interpretations of the Buchert
formalism, since $|\OMQ|\lsim0.042$ at all times which means that the
backreaction is never large enough to make $\bq$ negative.
Cosmic acceleration is recognized as an apparent effect which arises
due to the cumulative clock rate variance of wall observers relative to
volume--average observers. It becomes significant only when the voids
begin to dominate the universe by volume, which occurs at low redshifts.
Since the epoch of onset of apparent acceleration is directly related to the
void fraction, $\fv$, this solves the cosmic coincidence problem.

In addition to apparent cosmic acceleration, another important apparent effect
will arise if one considers scales below the SHS. By any one set of clocks it
will appear that voids expand faster than wall regions. Thus a wall observer
will see galaxies on the far side of a dominant void of diameter $\goesas30\hm$
to have a greater local Hubble parameter than the dressed global average $\Hm$,
while galaxies within an ideal wall have a local Hubble parameter lower than
$\Hm$. The local maximum Hubble parameter across a void seen by a wall
observer is $H\ns{vw0}={1\over\av}\Deriv\dd\ta\av=h_r^{-1}\bH\simeq\frn32\bH$.
Furthermore, since the bare Hubble parameter $\bH$ provides a measure of the
uniform quasilocal flow, it must also be the minimum `local' value within an
ideal wall at any epoch. With a dressed Hubble constant $\Hm=61.7\pm3.0\kmsMpc$
(see Table~\ref{parms}), we can expect a local Hubble expansion that varies
between a minimum $50.1\pm1.7\kmsMpc$ within our local filament (towards the
Virgo cluster), and a maximum $75.2^{+2.0}_{-2.6}\kmsMpc$ across local voids.
Averaging over many structures in spherical shells will reduce the variation,
as will be discussed in Sec.~\ref{hvar}.

\section{Timescape scenario: Observational tests}

There are three types of potential cosmological tests of the timescape
scenario:
\begin{enumerate}[(i)]
\item tests of the average expansion history on scales larger
than the SHS, involving quantities derived from luminosity and angular
diameter distance measures;
\item tests of cosmological averages on scales larger than the SHS that include
contributions from the growth of structures (late epoch integrated
Sachs--Wolfe effect, cosmic shear, weak lensing, redshift space distortions
etc);
\item tests of the local expansion history below the SHS.
\end{enumerate}
Class (iii) deals with scales which are in the nonlinear regime of perturbation
theory in the standard model, and it is quite possible that this regime needs
to be understood before one can make progress with class (ii).
Tests in class (i) will include equivalents to every cosmological test of the
standard FLRW model. We will consider class (i) tests in
Secs.~\ref{lumd}--\ref{EoS}, \ref{timed}; tests which require the treatment
of redshift space distortions and therefore fall into class (ii) in
Secs.~\ref{bao}, \ref{ccl}; and finally a class (iii) test in Sec.~\ref{hvar}.
\subsection{Luminosity distances: supernovae, gamma ray bursts}
\label{lumd}
The luminosity distance relations (\ref{eq:dL}), (\ref{eq:rw}) have been
tested extensively with type Ia supernovae (SneIa) data \cite{LNW,SW} and
with gamma--ray bursters \cite{grb}. In the case of the supernovae, it turns
out that the luminosity distance is so close to that of the standard model
that the question of whether a better fit is provided by the \TS\ model
or by the spatially flat \LCDM\ model depends on the manner in which the
data is reduced \cite{SW}. In other words, the differences between the two
models are at the level of current systematic uncertainties in SneIa data
reduction -- supernovae being standardizable candles, rather than perfect
standard candles.

Two empirical methods commonly used to reduce SNeIa data are the Multicolor
Light Curve Shape fitter MLCS2k2 \cite{JRK}, and the Spectral Adaptive Light
curve Template SALT/SALT-II methods \cite{guy05,guy07}. MLCS2k2 calibration
uses a nearby training set of SNeIa assuming a close to linear Hubble law,
whereas SALT/SALT-II uses the whole dataset to calibrate empirical light
curve parameters. Since SneIa from beyond the range in which the Hubble law is
linear are used, a cosmological model must be assumed\footnote{In refs.\
\cite{2mod,mg12} it is incorrectly stated that in the SALT/SALT-II methods
data is reduced ``assuming the Friedmann equation''. In fact, any cosmological
model can be used in applying the SALT/SALT-II method, and in ref.\ \cite{SW}
we have applied it to the \TS\ model. However, it is true that {\em very
often} data is reduced using the standard cosmology with the Friedmann
equation to produce tables of apparent magnitudes and redshifts. Data reduced
in this fashion cannot be used to test non-standard cosmologies; one must
perform a separate SALT/SALT-II data reduction for each nonstandard model that
one investigates.}. We find that the \TS\ model provides a better fit
to SneIa data than the standard spatially flat \LCDM\ model if the MLCS2k2
method is used, while conversely the standard model provides a better fit
if the SALT-II method is used \cite{SW}. However, the Bayesian evidence for
these conclusions is not very strong.

\begin{figure}[htb]
\vbox{\centerline{\scalebox{0.6}{\includegraphics{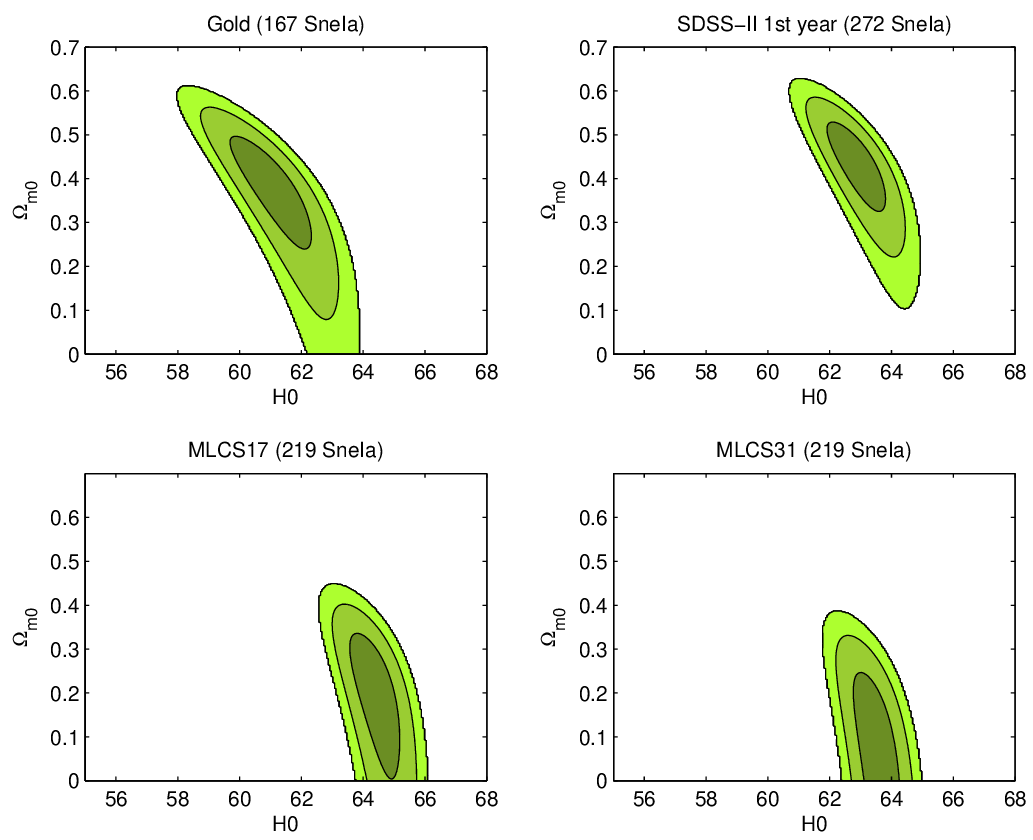}}}
\caption{\label{SW:sn}%
Confidence limits \cite{SW} for \TS\ model fits
to $z\ge0.033$ cut samples of Gold07 ($\RV=3.1$) \cite{R07}, SDSS-II
($\RV=2.18$) \cite{Kessler09}, MLCS17 ($\RV=1.7$) and MLCS31 ($\RV=3.1$)
\cite{H09}. In each case an overall normalization of the Hubble constant from
the published dataset is assumed.}}
\end{figure}
One important issue that arises in the \TS\ model is that the
luminosity distance relation (\ref{eq:dL}), (\ref{eq:rw}) only applies on
scales larger than the SHS. In some SneIa compilations data below this
scale is included. Such data needs to be removed when testing the timescape
model. It was found that even when such systematics are accounted for, there
are still marked differences in the cosmological parameters deduced\footnote{It
should be noted that in the MLCS method the value of $\Hm$ depends on an
overall calibration of the distance scale; e.g., from Cepheid distances.
There is therefore a freedom to shift the contours along the $\Hm$ axis in
Fig.~\ref{SW:sn} depending on that normalization. The relative value of
$\Hm$ for different fits is more important than the absolute values.} depending
on additional assumptions made in data reduction, as is seen in
Fig.~\ref{SW:sn} in which 4 different implementations of MLCS2k2 are
considered. There is a known degeneracy between intrinsic colour variations
in SneIa and reddening by dust in the host galaxy, determined by the parameter
$\RV$. However, the differences seen between the different panels in
Fig.~\ref{SW:sn} involve more than simply the value of this parameter.
Much remains to be done to resolve these systematic issues.

In recent years correlations of empirical properties of gamma-ray bursters have
been used to determine Hubble diagrams at larger redshifts than those probed
by SneIa \cite{GRB1}--\cite{GRB4}. A recent analysis of 69 GRBs \cite{grb}
found that the \TS\ model gave a better fit than the spatially flat \LCDM\
model, 
but not by a margin that is statistically significant. Further improvement in
understanding of the systematic issues is required before GRB can provide
tight constraints.
\subsection{Cosmic microwave background anisotropies}
\label{cmb}
A complete analysis of the CMB anisotropy spectrum in the \TS\ cosmology
is highly nontrivial, since the standard model analysis includes the late time
integrated Sachs-Wolfe effect, which requires a from first principles
reinvestigation in the \TS\ model. While such an analysis has not yet
been completed, we are nonetheless able to compute the angular diameter
distance of the sound horizon at any epoch, and to independently compute
the epochs of matter--radiation decoupling, photon--electron decoupling and
the baryon drag epoch, and substantial constraints on model parameters
\cite{dnw} can already be made using the Planck data \cite{Pparm}.

Since the early universe is extremely close to being spatially homogeneous and
isotropic, in the \TS\ model there is no change to
physical processes at those epochs, but rather in the calibration of
parameters. In our case, there are two sets of observers -- wall
observers such as ourselves, and the volume average observers to whom the
average cosmological parameters (\ref{om}) are most directly
related. Computations are most readily performed from the point of view
of the volume-average observers, if we account for the fact that they
determine a cooler CMB temperature than us at the present epoch. There is
a focusing and defocusing of light between walls and voids, and the
number density of CMB photons in the negatively curved voids is less than
in the walls.

The volume-average CMB temperature, $\Tb$, is related to wall temperature, $T$,
by
\beq
\Tb=\gb^{-1}T \,,
\eeq
at any epoch. The difference is negligible at early times when $\gb\simeq1$;
however, at the present epoch $\Tb\ns0=\gbn^{-1}2.275\,$K is typically 35\%
lower than the temperature we measure. The bare baryon number density is then
given by
\beq
\nbB={3\Hb^2\OMBn\over8\pi G\,m_{\ns p}}\left(\Tb\over\Tb\ns0\right)^3\
\,,
\label{nv_b}\eeq
where $\OMBn$ is the present epoch bare baryon matter density parameter
and $\mpr$ is the proton mass.

The standard analysis of early universe physics applies when
calibrated in terms of volume-average parameters.
One very important consequence of this is that the baryon--to--photon ratio,
$\etBg$, is recalibrated as compared to the standard cosmology, and we can
potentially obtain a fit with no primordial lithium abundance anomaly
\cite{pill}. In particular, \TS\ fits have been performed
\cite{clocks,dnw,LNW} for the range $\etBg=(5.1 \pm0.5)\times10^{-10}$ favoured
by constraints from light element abundances alone\footnote{A higher value is
assumed in \LCDM\ fits of CMB data, giving rise to the lithium abundance
anomaly. While there is an intrinsic tension in the light element data between
abundances of deuterium and lithium-7 \cite{bbn2}, for the range of $\etBg$ we
adopt here all abundances fall within 2$\si$.} \cite{bbn1,bbn2}.

\begin{figure}[htb]
\vbox{\centerline{\scalebox{0.48}{\includegraphics{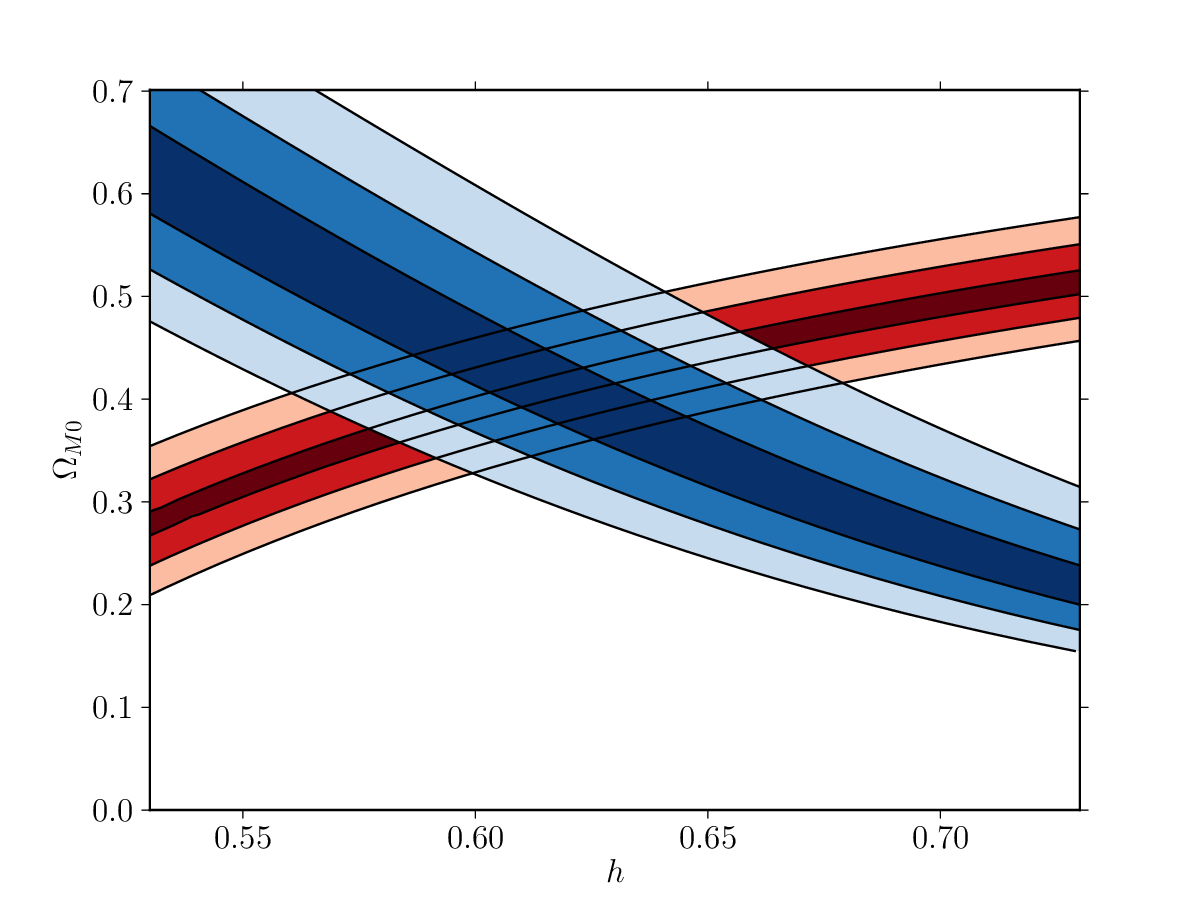}}}
\caption{\label{cmbbao}%
Contours of ($h$, $\OmMn$) parameter values for which the angular diameter
of the sound horizon at decoupling matches the angular scale $\th_*=0.0104139$
\cite{Pparm} to within $\pm2$\%, $\pm4$\% and $\pm6$\% are shown in blue
(upper left to lower right). Contours of parameter values for which the
present-day effective comoving scale of the sound horizon at the baryon drag
epoch matches the value $98.88\hm$ \cite{Pparm} are shown in red (lower
left to upper right). In each case the baryon--to--photon ratio is assumed to
be in the range $4.6<10^{10}\etBg<5.6$, for which there is no primordial
lithium abundance anomaly \cite{pill}.}}
\end{figure}
The volume--average sound horizon scale at any epoch is given by
\beq
\bD_s={\ab(t)\over\ab\Z0}{c\over\sqrt{3}}\int_0^{x\ns{dec}}{\dd x\over x^2
\bH\sqrt{1+0.75\,x\,\OMBn/\OMPn}}\,,\label{Ddec}
\eeq
where $\OM\Z{\ga0}=2g_*^{-1}\OMRn$ is the present epoch volume-average photon
density parameter, $g_*=3.36$ is the relative degeneracy factor of relativistic
species, $x\ns{dec}=\zb_{\ns{dec}}+1\equiv\gbn(1+z_{\ns{dec}})/\gb_{\ns{dec}}$
is the value of $\ab/\abn$ at photon--electron decoupling, and $\OMBn=\etBg\mpr
\bn\Z{\ga0}$ is fixed in terms of $\mpr$, $\etBg$ and the present epoch
volume--average photon density, $\bn\Z{\ga0}$.

We compute the comoving scale of the sound horizon at photon-- electron
decoupling, $\bD_s(t\ns{dec})$, from (\ref{Ddec}), and its angular diameter
distance, $\dAdec$, from (\ref{eq:dL})--(\ref{dist}) using the numerical
solutions to (\ref{b5}), (\ref{b6}) at the same time
as solving the Peebles equation to determine the ionization fraction
\cite{dnw}. The angular scale $\th_*=\bD_s(t\ns{dec})/\dAdec$ can then be
constrained to match the measured value \cite{Pparm} to any desired
accuracy.

For BAO measurements, the relevant comoving size of the sound horizon is that
at the baryon drag epoch, which occurs at $t=t\ns{drag}$ when $c\,\tad\simeq1$,
where
\beq
\tad(t)\equiv\int_{t}^{t_0}{\dot\tao\dd t\over\ab R}=
\int_{t}^{t_0}{\sigma_{\ns{T}}\nb_{\ns{e}}\dd t\over\ab R}\label{tad}
\eeq
is the drag depth, $\tao$ is the optical depth, $\sigma_{\ns{T}}$ is the
Thomson scattering cross-section, $\nb_{\ns e}=\nb_{\ns p}$ is the bare free
electron density, and $R\equiv0.75\rh\Z B/\rh_\ga=0.75\,(\OMBn\ab)/
(\OMPn\abn)$. Since we are not yet able to constrain the BAO scale directly
from galaxy clustering statistics, we determine $\bD_s(t\ns{drag})$ at the
same time as other numerical integrations, and constrain it using Planck
satellite estimates \cite{Pparm}.

In Fig.~\ref{cmbbao} we display two sets of contours in the ($\Hm$, $\OmMn$)
parameter space obtained in ref.~\cite{dnw}: firstly, parameters which match
the acoustic scale of the sound horizon $\th_*=0.0104139$ \cite{Pparm} to
within $\pm2$\%, $\pm4$\% or $\pm6$\%; and secondly parameters which similarly
match the present effective comoving scale of the sound horizon at the baryon
drag epoch as determined by the standard \LCDM\ model analysis of the Planck
data, namely\footnote{Since the Hubble constant $\Hm=67.11\kmsMpc$ determined
from the Planck satellite is a fit to the \LCDM\ model, any effective present
comoving scale must be given in units $\hm$, as the \TS\ model will
generally yield a different value for $\Hm$.} $98.88\hm$ \cite{Pparm}.
\begin{table}
\begin{tabular}{lll}
\hline
Parameter &&Range\\
\hline
Present void fraction&$\fvn$&$0.695^{+0.041}_{-0.051}$\\
Bare Hubble constant&$\Hb$&$50.1\pm1.7\kmsMpc$\\
Dressed Hubble constant&$\Hm$&$61.7\pm3.0\kmsMpc$\\
Local maximum Hubble constant&$H\ns{vw0}$&$75.2^{+2.0}_{-2.6}\kmsMpc$\\
Present phenomenological lapse function&$\gbn$&$\vp1.348^{+0.021}_{-0.025}$\\
Dressed matter density parameter&$\OmMn$&$0.41^{+0.06}_{-0.05}$\\
Dressed baryon density parameter&$\OmBn$&$\vp0.074^{+0.013}_{-0.011}$\\
Bare matter density parameter&$\OMMn$&$0.167^{+0.036}_{-0.037}$\\
Bare baryon density parameter&$\OMBn$&$\vp0.030^{+0.007}_{-0.005}$\\
Bare radiation density parameter&$\OMRn$&$\vp\left(5.00^{+0.56}_{-0.48}\right)
\times10^{-5}$\\
Bare curvature parameter&$\OMkn$&$\vp0.862^{+0.024}_{-0.032}$\\
Bare backreaction parameter&$\OM\Z{\QQ0}$&$-0.0293^{+0.0033}_{-0.0036}$\\
Nonbaryonic/baryonic matter densities ratio&$\OMCn/\OMBn$&$\vp4.6^{+2.5}_{-2.1}
$\\ Age of universe (galaxy/wall observer)&$\ta\ns{0}$&$14.2\pm0.5\,$Gyr\\
Age of universe (volume-average observer)&$\tn$&$17.5\pm0.6\,$Gyr\\
Apparent acceleration onset redshift&$z\ns{acc}$&$0.46^{+0.26}_{-0.25}$\\
\hline
\end{tabular}
\caption{\label{parms}Estimates of the cosmological parameters of the
\TS\ model \cite{dnw} obtained from a $\pm2$\% match to the angular scale,
$\th_*$, of the sound horizon at decoupling; and to a $\pm6$\% match to
the effective comoving scale, $r\ns{drag}$, of the sound horizon at the
baryon drag epoch, using recent values from the Planck satellite analysis
\cite{Pparm}. A tighter constraint is applied to $\th_*$ as it is
purely geometrical, whereas the calibration of $r\ns{drag}$ involves additional
uncertainty since the ratio of nonbaryonic to baryonic matter densities may
differ between the \TS\ and \LCDM\ models.}
\end{table}

The full numerical solutions \cite{dnw} provide tighter constraints than
earlier analyses \cite{LNW}, leading to the parameters
listed\footnote{A recent phenomenologically motivated analysis \cite{rob13}
using a completely different approach produces a void fraction which agrees
with that found here.}
in Table~\ref{parms}. Particular parameters can be ruled out on the basis
that matter--radiation equality must occur before last scattering, so
that $\OMM/\OMR>1$ at $z\ns{dec}$. In particular, we can rule out a dressed
matter density parameter $\OmMn<0.2$ if $\Hm<65\kmsMpc$. If we compare
Fig.~\ref{SW:sn} we see that the SneIa data reduction methods used in
the Gold07 \cite{R07} and SDSS-II \cite{Kessler09} samples remain consistent
with the new constraints, whereas those of ref.~\cite{H09} do not.

A detailed treatment of the acoustic peaks in the CMB data may of course
still challenge the \TS\ cosmology, as it will certainly further
tighten the constraints. Work on this problem, which
requires a revisiting of CMB data analysis from first principles, is in
progress.
\subsection{The effective `equation of state'}
\label{EoS}
\begin{figure}
\vbox{\centerline{\quad{\sbf(a)}\hskip-15pt
\includegraphics[width=3.03in]{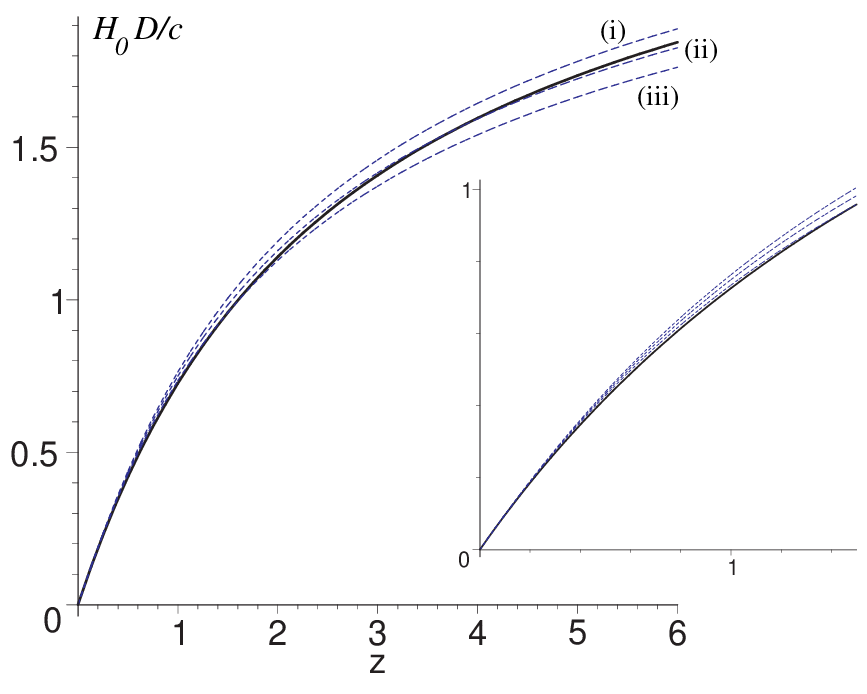}
\quad{\sbf(b)}\hskip-15pt
\includegraphics[width=2.55in]{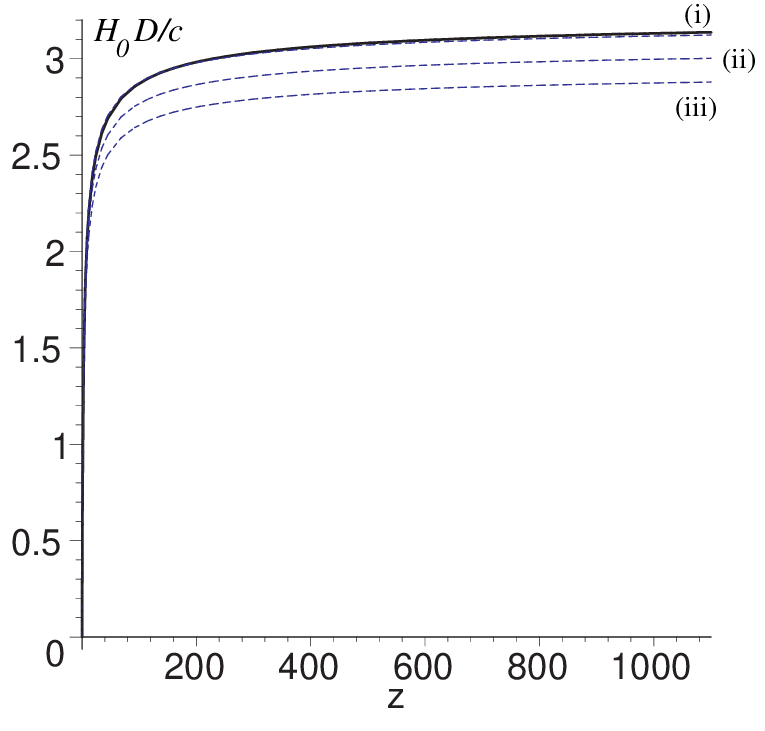}
}
\caption{\label{fig_coD} The effective comoving distance $c^{-1}\Hx D(z)$ is
plotted for the \TS\ model which best fits Planck \cite{Pparm}, with $\fvn=
0.695$ (solid line) \cite{dnw}; and for various spatially flat \LCDM\ models
(dashed lines).
The parameters for the dashed lines are (i) $\OmMn=0.3175$ (best-fit \LCDM\
model to Planck \cite{Pparm}); (ii) $\OmMn=0.35$; (iii) $\OmMn=0.388$. Panel
{\bf(a)} shows the redshift range $z<6$, with an inset for $z<1.5$, which is
the range tested by SneIa data. Panel {\bf(b)} shows the range $z<1100$ to the
surface of last scattering, tested by Planck.}}
\end{figure}

A direct method for comparing the expansion history with those of homogeneous
models with dark energy, is to observe that for a standard spatially flat
cosmology with dark energy obeying an equation of state $P\Z D=w(z)\rh\Z D$,
the quantity
\beq
{\Hm D\over c}=\int_0^z{\dd z'\over\left[\OmMn(1+z')^3+\Omega\Z{D0}
\exp\left(3\int_0^{z'}{(1+w(z''))\dd z''\over 1+z''}\right)\right]^{1/2}}\,,
\label{rFLRW}\eeq
does not depend on the value of the Hubble constant, $\Hm$, but only directly
on $\OmMn=1-\Omega\Z{D0}$. Since the best-fit values of $\Hm$ are potentially
different for different models, a comparison of $\Hm D/c$ curves as a function
of redshift for the \TS\ model versus the \LCDM\ model gives a good
indication of where the largest differences can be expected, independently of
the value of $\Hm$. Such a comparison is made in Fig.~\ref{fig_coD}.

As the redshift range changes the \TS\ model interpolates between \LCDM\
models with different values of ($\OmMn$,$\OmLn$). If we consider the
\TS\ model that is a best fit to the Planck data, then for the largest
redshifts $50\lsim z\lsim1100$, $\Dfb$ is essentially indistinguishable
from the $\Dlcdm$ for model (i) with parameter values $(\OmMn,\OmLn)=(0.3175,
0.6825)$ which best-fit the Planck data \cite{Pparm}. By contrast over the
range $2\lsim z\lsim6$ a close fit is provided by model (ii) with
$(\OmMn,\OmLn)=(0.35,0.65)$. For the closest redshifts, $z<1.5$, $\Dfb$
becomes indistinguishable from $\Dlcdm$ for model (iii) with $(\OmMn,\OmLn)=
(0.338,0.721)$. It is this feature which makes it difficult to distinguish
the \TS\ model from the \LCDM\ model on the basis of SneIa data alone.
However, with complementary tests over the full range of redshifts
the expansion histories should be distinguishable.

Fig.~\ref{fig_coD} shows just one value of $\fvn$. If we compare Fig.~2
of Ref.\ \cite{obs}, we see that with $\fvn=0.76$, $\Dfb$ similarly
interpolates between \LCDM\ models with $(\OmMn,\OmLn)=(0.34,0.64)$ at
low redshift and $(\OmMn,\OmLn)=(0.25,0.75)$ at high redshift. I.e., as
the present epoch void fraction is increased the width of the range of
equivalent \LCDM\ $\OmMn$ values increases, as well as the overall values
being less.

The shapes of the $\Hm D/c$ curves depicted in Fig.~\ref{fig_coD} represent
the actual observable quantity one is measuring in tests that some researchers
loosely refer to as `measuring the equation of state'. For spatially flat
dark energy models, with $\Hm D/c$ given by (\ref{rFLRW}), one finds that the
function $w(z)$ appearing in the fluid equation of state $P\Z D=w(z)\rh\Z D$
is related to the first and second derivatives of (\ref{rFLRW}) by
\beq
w(z)={\frn23(1+z)D'^{-1}D''+1\over\OmMn(1+z)^3\Hm^2D'^2c^{-2}-1}
\label{eos}\eeq
where prime denotes a derivative with respect to $z$. Such a relation
can be applied to observed distance measurements, regardless of whether
the underlying cosmology has dark energy or not. Since it involves
first and second derivatives of the observed quantities, it is actually
much more difficult to determine observationally than directly fitting
$c^{-1}\Hm D(z)$.

\begin{figure}
\centerline{{\sbf(a)}\hskip-1pt
\scalebox{0.45}{\includegraphics{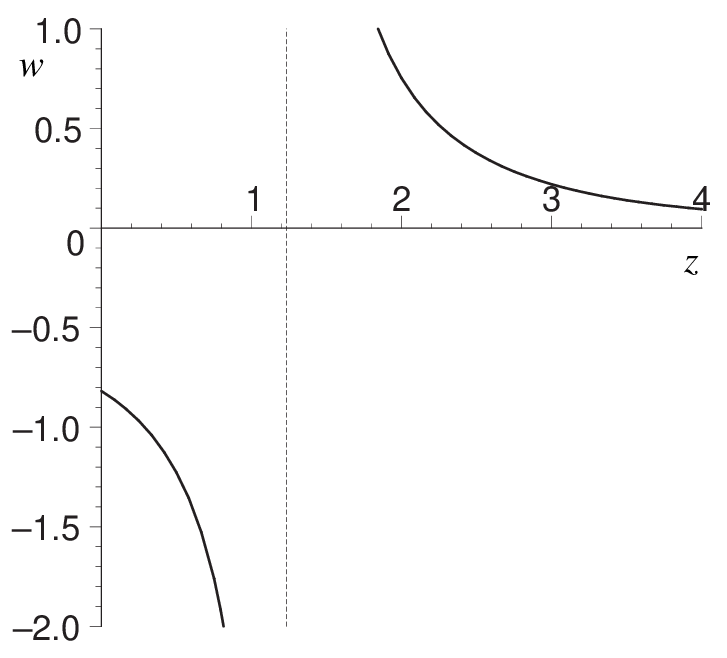}}\
{\sbf(b)}\hskip-1pt
\scalebox{0.428}{\includegraphics{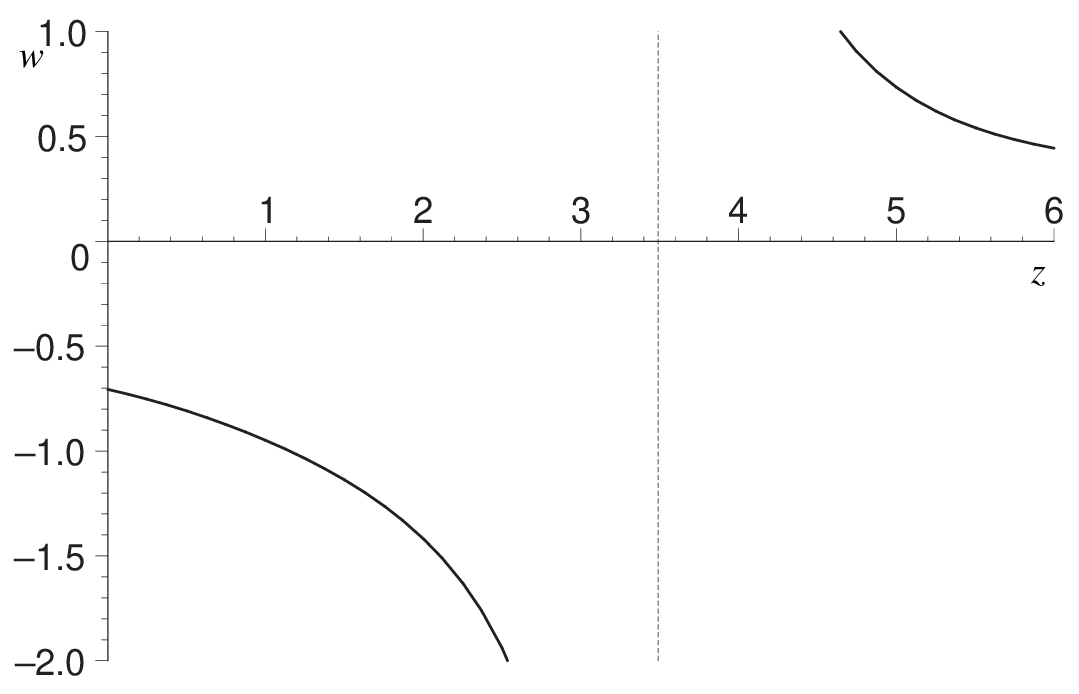}}
}
\caption{The artificial equivalent of an equation of state
constructed using the effective comoving distance
(\ref{eos}), plotted for the \TS\ tracker solution with best-fit value
$\fvn=0.695$, and two different values of $\OmMn$: {\bf(a)} the canonical
dressed value $\OmMn=\frn12(1-\fvn)(2+\fvn)=0.41$; {\bf(b)} $\OmMn=0.3175$.}
\label{fig_wz}
\end{figure}

The equivalent of the equation of state, $w(z)$, for the \TS\ model
is plotted in Fig.~\ref{fig_wz}. The fact that $w(z)$ is undefined at
a particular redshift and changes sign through $\pm\infty$ simply reflects
the fact that in (\ref{eos}) we are dividing by a quantity which goes
to zero for the \TS\ model, even though the underlying curve of
Fig.~\ref{fig_coD} is smooth. As we are not dealing with a dark energy
fluid in the \TS\ model, $w(z)$ simply has no physical meaning.

Nonetheless, phenomenologically the results do agree with the usual inferences
about $w(z)$ for fits of standard dark energy cosmologies to SneIa data. In
particular, for low redshifts the average value of $w(z)$ is close to -1, but
it eventually it crosses `the phantom divide' to $w(z)<-1$. For fundamental
homogeneous dark energy fluids, $w<-1$ signals a violation of the dominant
energy condition and with that a breakdown of standard laws of physics. Here it
is simply a consequence of an inappropriate parametrization of the expansion
history of a universe which does not evolve according to the Friedmann
equation.

The redshift at which `$w=-1$ is crossed' in the \TS\ model depends on the
value of $\OmMn$ that is assumed in the FLRW style analysis. For the canonical
model of Fig.~\ref{fig_wz}(a), with $\OmMn=0.41$ one finds that `$w=-1$ is
crossed' at $z=0.29$, with $\OmMn=0.388$ `$w=-1$ is crossed' at $z=0.40$,
and with $\OmMn=0.3175$ (the \LCDM\ value from Planck \cite{Pparm} in
Fig.~\ref{fig_wz}(b)), `$w=-1$ is
crossed' at $z=1.15$. For the same value of $\fvn$, taking a lower value of
$\OmMn$ in a FLRW--style analysis leads to $w(z)$ being closer to $w=-1$ for a
larger range of redshifts. Thus if a \TS\ model luminosity distance
relation is correct then one can easily be led to different conclusions about
`dynamical dark energy' \cite{ZZ,SCHMPS} over the range of
redshifts, $z<1.5$, probed by SneIa, depending on prior assumptions about the
value of $\OmMn$ from other datasets.

What appears as an $\OmMn$ dependent varying $w(z)$ from the FLRW perspective
actually reflects the fact that the effective energy density assumed in the
standard analysis is not scaling as $\OMM\propto(1+z)^3$, as would be the case
for any homogeneous model. Consequently the \TS\ model simply lies outside
the class of models typically contemplated for dark energy diagnostics
\cite{GCC}--\cite{SSS}. For example, the $Om(z)$ diagnostic of Sahni, Shafieloo
and Starobinsky \cite{SSS,SSS2} is designed to be a constant, $\OmMn$, at all
redshifts for a spatially flat FLRW model, but to differ for other $w(z)$
functions. One can compute a formula for the $Om(z)$
diagnostic \cite{obs}, although this is not particularly useful since the
\TS\ model has a singular $w(z)$ and lies outside the class of empirical
functions usually used to analyse the diagnostic. Existing analyses can only
be applied in asymptotic limits such as $z\to0$, when \cite{obs}
\beq
Om(0)=\frn23\left.H'\right|_0={2(8\fvn^3-3\fvn^2+4)(2+\fvn)\over(4\fvn^2+\fvn
+4)^2}
\label{intc}\eeq
For $\fvn=0.695^{+0.041}_{-0.051}$, $Om(0)=0.643^{+0.008}_{-0.004}$. In fact,
this coincides with the intercept of Fig.~3 in ref.\ \cite{SSS2},
determined from SneIa, BAO and CMB data.
\subsection{The Alcock--Paczy\'nski test and baryon acoustic oscillations}
\label{bao}
The BAO scale provides a convenient standard ruler which can be detected
both in the radial ($z$) and transverse directions ($\th$) leading to a
determination of the quantity
\beq F(z)\equiv\left|\Deriv\de\th z\right|={(1+z)H(z)\dA(z)\over c}
={H(z)D(z)\over c}\label{FAP}
\eeq
related to the Alcock--Paczy\'nski test\footnote{Alcock and Paczy\'nski
\cite{AP} originally defined their test statistic as $f\Ns{AP}=z^{-1}F(z)$.
Since $D(z)\to0$ as $z\to0$, the original Alcock--Paczy\'nski test function is
actually the derivative $F'(z)$ in the limit $z\to0$, rather than $F(z)$. As
seen in Fig.~8 of ref.\ \cite{obs} this statistic has a greater power to
discriminate between the \TS\ and \LCDM\ models. However, taking a derivative
with respect to $z$ requires better quality data, and for the time being one is
limited to testing quantities such as (\ref{FAP}) or $D\Ns V =(zD^2H^{-1})^{1/
3}$.} \cite{AP}. The BAO scale has now been detected at several redshifts
in galaxy clustering statistics \cite{wz,gal} and the Lyman--$\al$ forest
\cite{boss}, and provide a promising geometric test of the expansion history.

\begin{figure}[htb]
\centerline{\ {\sbf(a)}\hskip-20pt
\includegraphics[width=2.9in]{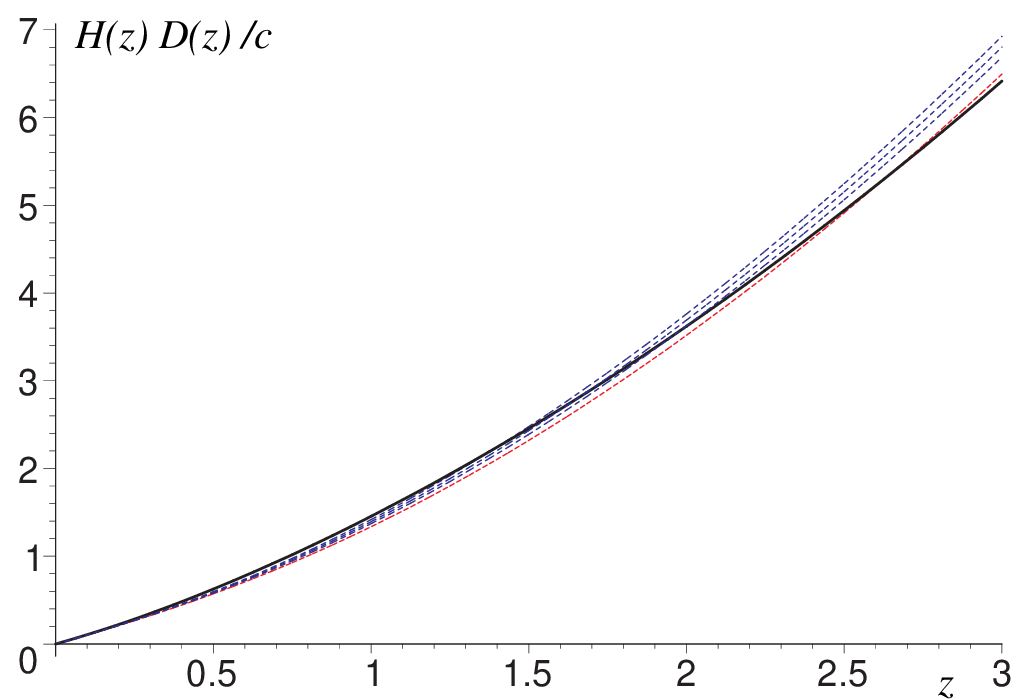}
\quad {\sbf(b)}\hskip-30pt
\includegraphics[width=2.9in]{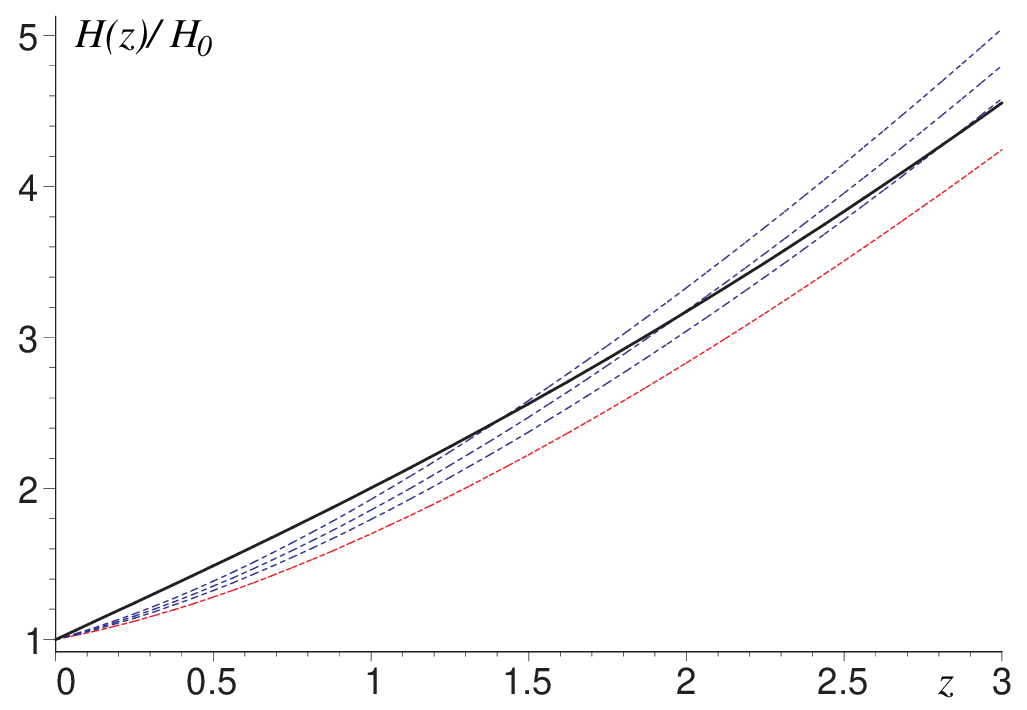}}
\caption{%
{\bf(a)} $F(z)=c^{-1}H(z)D(z)$; {\bf(b)} $H(z)/\Hm$. In each we display curves
for the \TS\ model with $\fvn=0.695$ (solid line), and comparison spatially
flat \LCDM\ models (dashed lines): for the 3 values of $\OmMn$ shown in
Fig.~\ref{fig_coD}, and also the value $\OmMn=0.27$ used in the fits of
\cite{wz}-\cite{boss}. In both panels the \LCDM\ curves are arranged from
bottom to top by the values of $\OmMn=0.27$, $0.3175$, $0.35$, $0.388$.}
\label{APH}
\end{figure}
In Fig.~\ref{APH} we show the test function $F=HD/c$ and also the function
$H(z)/\Hm$ (with dressed Hubble parameter) for \TS\ and \LCDM\ examples,
over the range of redshifts tested to date \cite{wz}--\cite{boss}. In fact,
at the effective redshifts tested in the WiggleZ survey, for $\fvn=0.695$ the
\TS\ values $F(0.21)=0.246$, $F(0.41)=0.496$, $F(0.60)=0.776$, $F(0.78)=
1.067$ all agree with the Alcock--Paczy\'nski fits of this quantity in
Table~1 of ref.~\cite{wz}, within uncertainties. While this is encouraging,
the methods of analysis used for the BAO scale assume the standard model,
both in applying Fourier space techniques, and in treating redshift space
distortions. These aspects of the data analysis need to be revisited from
first principles in the \TS\ model before we can be completely confident
in using constraints from these tests.

From Fig.~\ref{APH}(a) we see that the expectations for $H(z)D(z)/c$ for the
\TS\ and \LCDM\ models are very close for most of the redshift range
currently considered. A more discriminating test can in principle be
obtained by dividing the curve of Fig.~\ref{APH}(a) by that of
Fig.~\ref{fig_coD} to produce the quantity $H(z)/\Hm$ shown in
Fig.~\ref{APH}(b). The most notable feature is that the slope of $H(z)/\Hm$
is less than in the \LCDM\ cases, as is to be expected for a model whose
(dressed) deceleration parameter varies more slowly than for \LCDM.
Two different measurements are required to produce this information, however,
both the BAO measurement to determine $H(z)D(z)/c$, and luminosity distance
measurements to determine $\Hm D(z)/c$. In addition to examining the
model--dependent issues in BAO measurements, it also necessitates sorting
out the systematics of SNeIa that currently limit model comparison, as
discussed in Sec.~\ref{lumd}.
\subsection{Test of (in)homogeneity}
\label{ccl}
Clarkson, Bassett and Lu \cite{CBL} have constructed a test statistic based on
the observation that for homogeneous, isotropic models which obey the Friedmann
equation, the present epoch curvature parameter, a constant, may be written as
\beq
\Omkn={[c^{-1}H(z)D'(z)]^2-1\over[c^{-1}\Hm D(z)]^2}\label{ctest1}
\eeq
for all $z$, irrespective of the dark energy model or any other model
parameters. Consequently, taking a further derivative, the quantity
\beq
\CC(z)\equiv1+c^{-2}H^2(DD''-D'^2)+c^{-2}HH'DD'\label{ctest2}
\eeq
must be zero for all redshifts for any FLRW geometry. A deviation of $\CC(z)$
from zero, or of (\ref{ctest1}) from a constant value, would therefore mean
that the assumption of FLRW evolution is violated.

\begin{figure}
\centerline{{\sbf(a)}
\includegraphics[width=2.4in]{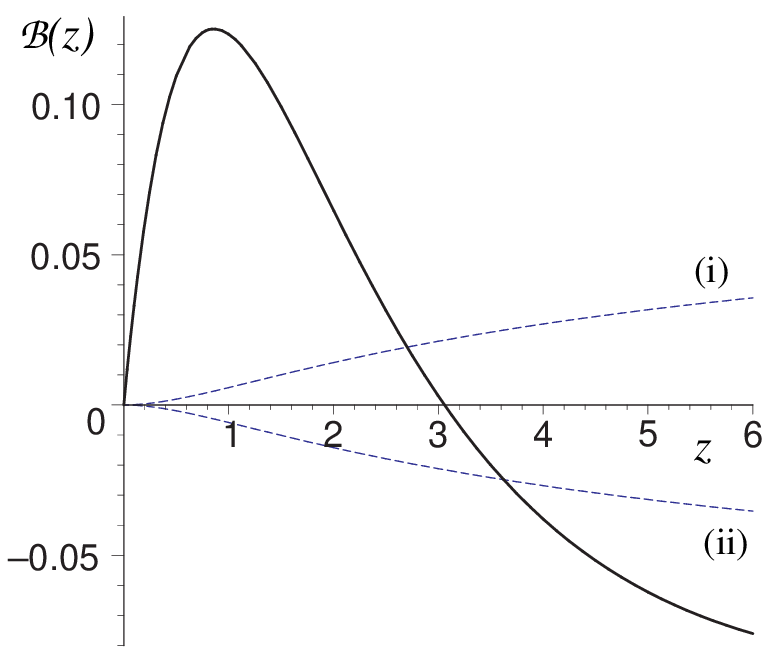}
\qquad
{\sbf(b)}
\includegraphics[width=2.4in]{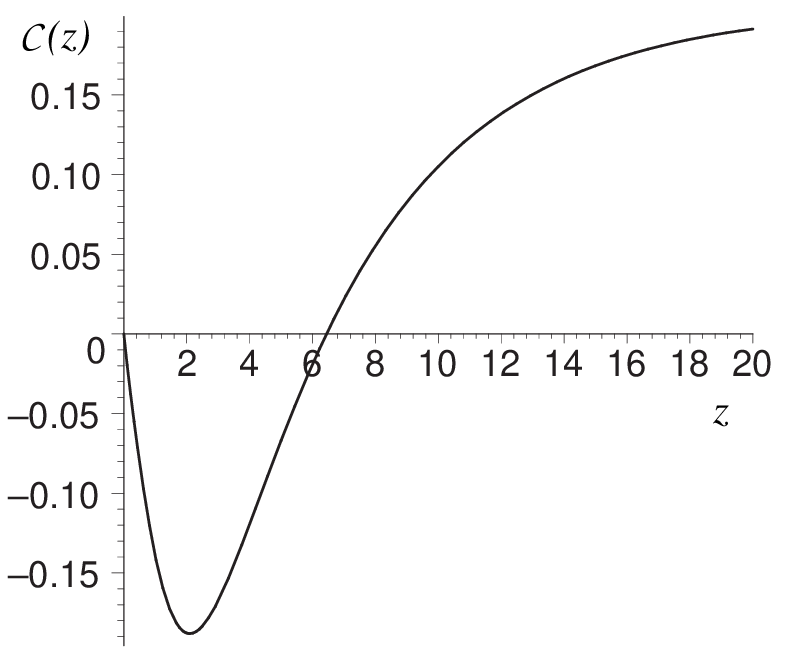}}
\caption{{\bf (a)}
The (in)homogeneity test function $\BB(z)=[c^{-1}HD']^2-1$ is plotted for
the \TS\ tracker solution with$\fvn=0.695$ (solid line), and
compared to equivalent curves $\BB=\Omkn(c^{-1}\Hm D)^2$ for two
\LCDM\ models: {\bf(i)} $\OmMn=0.32$, $\OmLn=0.67$,
$\Omkn=0.01$; {\bf(ii)} $\OmMn=0.32$, $\OmLn=0.69$, $\Omkn=-0.01$.
{\bf(b)} The (in)homogeneity test function $\CC(z)$ is plotted for
the $\fvn=0.695$ tracker solution.}
\label{fig_Bex}
\end{figure}

The functions (\ref{ctest1}) and (\ref{ctest2}) are computed in ref.\
\cite{obs}. It is more feasible to fit (\ref{ctest1}) than which involves one
derivative less of redshift. In Fig.\ \ref{fig_Bex} we show both $\CC(z)$, and
also the function $\BB(z)=[c^{-1}HD']^2-1$ from the numerator of (\ref{ctest1})
for the \TS\ model, as compared to two \LCDM\ models with a small amount of
spatial curvature. A spatially flat FLRW model would have $\BB(z)\equiv0$.
The \TS\ $\BB(z)$ function is easily distinguishable from the FLRW cases.
However, this requires better quality data than is currently available.
As noted in Sec.~\ref{bao}, present BAO data is able to constraint $H(z)D(z)$
but not yet $H(z)D'(z)$. Therefore, while the Clarkson, Bassett and Lu test
\cite{CBL} is a powerful one, it may be some time before it can be implemented.
\subsection{Time drift of cosmological redshifts}
\label{timed}
As noted in Sec.~\ref{bao}, the combined measurements of $\Hm D(z)/c$ and $H(z)
D(z)/c$ provide a means to determine $H(z)$ which at present is subject to
model dependencies and many systematic uncertainties. A model
independent determination of $H(z)$, which is also needed to determine the
quantity $\BB(z)$ in the (in)homogeneity test of Sec.~\ref{ccl}, is
provided by a measurement of the real time variation of the redshifts of
distant sources over a long time period \cite{S62}--\cite{L98}.
Although extremely challenging, such a measurement may be
possible over a 20 year period by precision measurements of the Lyman-$\al$
forest in the redshift range $2<z<5$ with the next generation of
Extremely Large Telescopes \cite{ELT1,ELT2}.

\begin{figure}[htb]
\centerline{\includegraphics[width=2.2in]{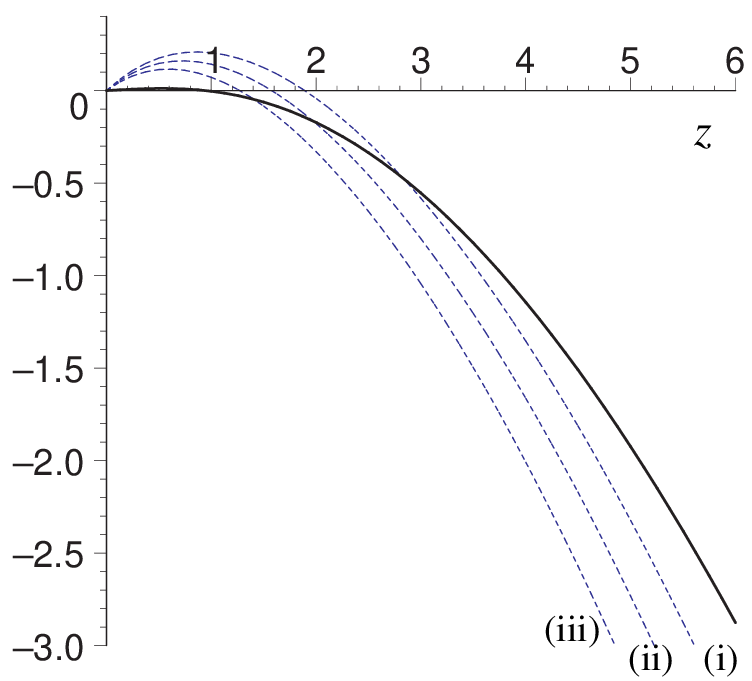}}
\caption{The function $\Hm^{-1}\Deriv\dd\ta z$ for the \TS\ model with
$\fvn=0.695$ (solid line) is compared to $\Hm^{-1}\Deriv\dd\ta z$ for the three
spatially flat \LCDM\ models shown in
Fig.~\ref{fig_coD} (dashed lines).}
\label{fig_zdot}
\end{figure}

In ref.\ \cite{obs} an analytic expression for $\Hm^{-1}\Deriv\dd\ta z$
is determined, the derivative being with respect to wall time for observers
in galaxies. The resulting function is displayed in Fig.~\ref{fig_zdot} for
the \TS\ model with $\fvn=0.695$, and is compared to those of
three spatially flat \LCDM\ models.
The \TS\ model curve is considerably flatter
than those of the \LCDM\ models. This is a consequence of
the magnitude of the apparent acceleration being considerably
smaller in the \TS\ model, as compared to the magnitude of the acceleration
in \LCDM\ models. For cosmologies with no apparent acceleration,
$\Hm^{-1}\Deriv\dd\ta z$ is always negative.
If there is cosmic acceleration at late epochs, real or apparent, then
$\Hm^{-1}\Deriv\dd\ta z$ will become positive at low redshifts, though
at a somewhat larger redshift than of the onset of (apparent) acceleration.
For $\fvn=0.695$, $\Hm^{-1}\Deriv\dd\ta z>0$ for $0<z<0.946$,
but with a tiny amplitude compared to the \LCDM\ models.

The very clear differences in redshift time drift for low redshifts $z\lsim2$
could lead to a decisive test of the \TS\ model versus \LCDM\ models.
Observationally, however, it is expected that measurements will be best
determined for sources in the Lyman $\al$ forest in the range, $2\lsim z\lsim5
$. At such redshifts the magnitude of the drift is somewhat
more pronounced in the case of the \LCDM\ models. For a source at $z=4$,
over a period of $\de\ta=10$ years we would have $\de z=-7.2\times10^{-10}$
for the \TS\ model with $\fvn=0.695$ and $\Hm=61.7\kmsMpc$. By comparison,
for a spatially flat \LCDM\ model with $\Hm=67.1\kmsMpc$ and $\OmMn$ we
have $\de z=-9.3\times10^{-10}$ for the same source over 10 years. Different
values of ($\Hm$, $\OmMn$) can produce degeneracies at particular redshifts.
However, a large sample of sources over the whole range $2\lsim z\lsim5$
should be able to constrain the shape of the $\Hm^{-1}\Deriv\dd\ta z$ curve
sufficiently to determine $H(z)$ in that range, and to distinguish the \TS\
and \LCDM\ cosmologies.
\section{Variation of the Hubble expansion}
\label{hvar}
Potentially the most interesting tests of the \TS\ model are those
below the SHS, since here we should find variation of the Hubble expansion
but with a scale--dependent amplitude constrained by the uniform quasilocal
Hubble flow condition.

\subsection{Problems and puzzles of bulk flows}\label{flows}
Traditionally astronomers have almost always analysed the variation of the
Hubble expansion in terms of {\em peculiar velocities}, namely as deviations
from a linear Hubble law
\beq
v\ns{pec}=cz-\Hm r\,\label{vpec}
\eeq
where $r$ is an appropriate distance measure. Such a definition implicitly
makes a strong assumption about spacetime geometry: on the scales of interest
spatial curvature can be neglected and the redshift associated with the
Hubble expansion can be treated in the manner of a recession velocity as in
special relativity. A linear Hubble law is observed to hold out to redshifts
$z\goesas0.1$, though on very small scales $z\lsim0.02$ below the SHS the
Hubble flow enters into a `nonlinear regime'.

For some decades astronomers have sought the scale on which peculiar velocities
converge to the flow indicated by the CMB temperature dipole. The dipole
is usually assumed to arise solely from a special relativistic boost, and
in addition to the known motion of our Sun with respect to the barycentre of
the Local Group (LG) of galaxies, this suggests that the LG itself is moving at
$635\pm38\kms$ in a direction $(\ell,b)=(276.4\deg,29.3\deg)\pm3.2\deg$ in
galactic coordinates. This direction defines a {\em clustering dipole}, namely
a direction in which it is expected we should find an overdensity which
gravitationally attracts the LG, galaxies between the LG and the overdensity
and galaxies on the other side of the overdensity.

There is no {\em a priori} reason why such Newtonian concepts of gravitational
attraction should persist on very large scales on which space is expanding.
Nonetheless, even though the very local Hubble flow on scales of tens of
megaparsecs is nonlinear, a linear Newtonian approximation is assumed to apply
at larger scales, and the amplitude of the peculiar velocities of galaxies is
estimated in a linearly perturbed FLRW model according to \cite{p93,l05}
\beq {\mathbf v}({\mathbf r})={\Hm\OmMn^{0.55}\over4\pi}\int\dd^3{\mathbf r'}\,
\delta_m({\mathbf r'})\,{({\mathbf r'}-{\mathbf r})\over|{\mathbf r'}-
{\mathbf r}|^3}\label{vlin}\eeq
where $\delta_m({\mathbf r})=(\rho-\bar\rho)/\bar\rho$ is the density contrast.

The search for convergence of bulk flows within this framework has a three
decade history summarized in refs.\ \cite{ltmc,bcmj}. Contrary to earlier investigations
\cite{el06}, Lavaux \etal\ \cite{ltmc} failed to find convergence in the 2MASS
survey on scales up to 120$\hm$: less than half the amplitude was generated on
scales $40\hm$, and whereas most of the amplitude was generated within 120$\hm$
the direction did not agree. Bilicki \etal\ \cite{bcmj} analysed a larger
sample in the 2MASS survey using a different methodology and failed to find
convergence within 150$\hm$. Some studies have found persistent bulk flows
extending to very large scales \cite{wfh09}--\cite{kash10}, and their
consistency with the \LCDM\ model is much debated \cite{dn11}--\cite{thf12}.

Recent attention has focused on the influence of the Shapley Concentration on
our local motion, as this is a particularly dense concentration of galaxies in
the clustering dipole direction. However, Shapley is at a distance of $138\hm$,
well beyond the SHS, and an influence at our location would represent an
unusually large scale correlation. A very recent study \cite{sfact} using SneIa
fails to find a significant turnover in peculiar velocities on the other side
of the Shapley Concentration, casting further doubt on the attractor model.

\subsection{Model independent analysis of Hubble expansion variation}
In general relativity it is well-known that every exact dust solution of the
Einstein equations which is not spatially homogeneous and isotropic exhibits
differential expansion of space. Furthermore, by the SEP the concept of a
special relativistic boost applies only in a LIF in the neighbourhood of a
point, and a general expansion of space cannot always be reduced to simple
boosts. Consequently the conceptual framework we have just described in
Sec.~\ref{flows} represents an extrapolation of Newtonian concepts into a
regime in which they cannot obviously be expected to be valid.

\begin{figure}[htb]
\vbox{\vskip\baselineskip\centerline{\scalebox{0.32}{\includegraphics
{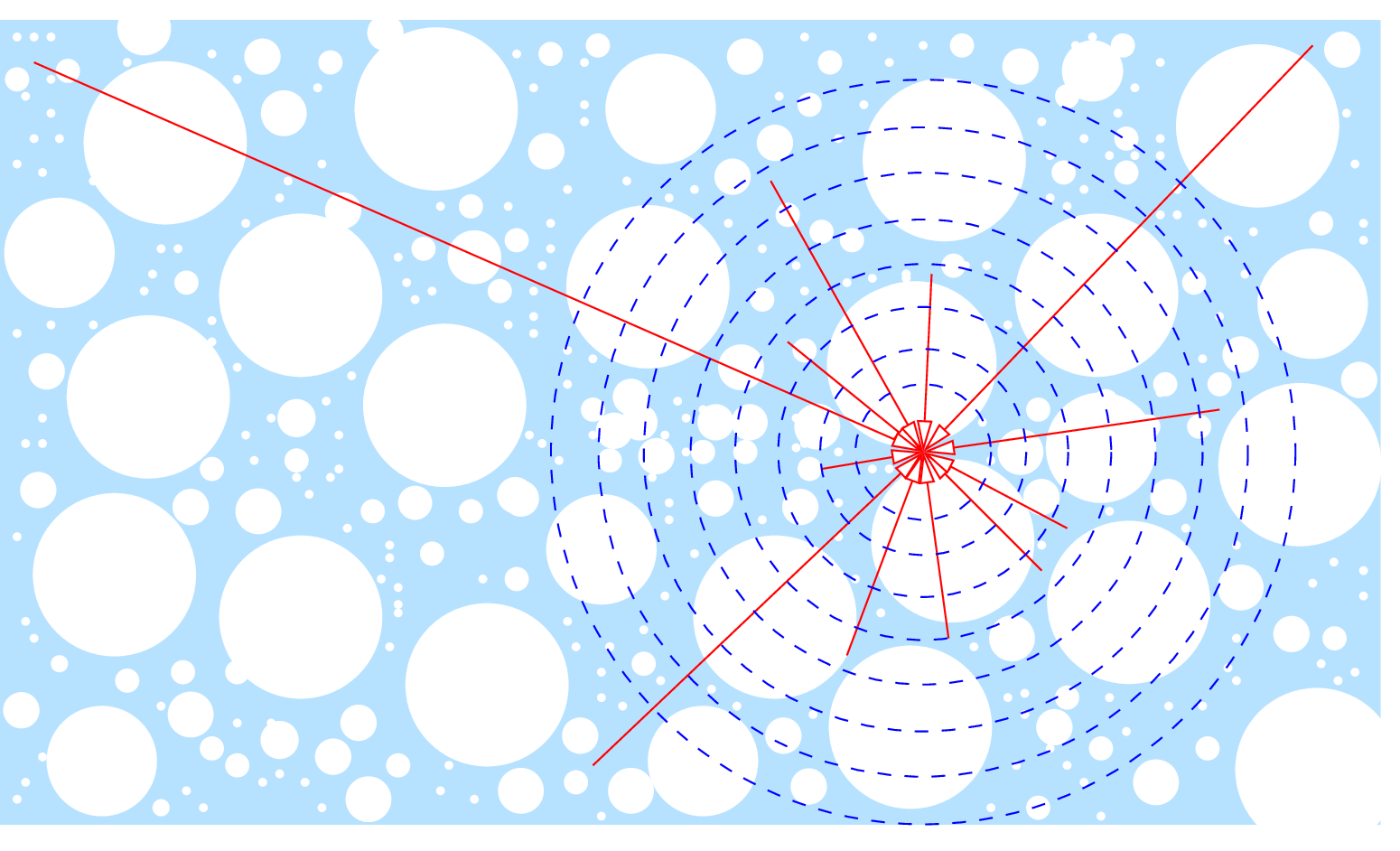}}}\caption{\label{Hshells}%
Schematic diagram of spherical averaging. The universe is described as ensemble
of filaments, walls and voids: expanding regions of different density which
have decelerated by different amounts and therefore experience different local
expansion rates at the present epoch. If one averages $cz/r$ in spherical
shells (dotted lines) about a point then once the shells are a few times
larger than the typical nonlinear structures
\cite{hv02}--\cite{pan11}, an average Hubble law with small statistical scatter
is obtained, whereas there are considerable deviations for shells on scales
comparable to the typical nonlinear structures.}}
\end{figure}

In the \TS\ scenario the greatest variations in spatial curvature occur
below the SHS, and a spatially flat geometry cannot be assumed to apply at
every scale. In recent work \cite{wsmw} we have analysed the variation of the
Hubble expansion in a model independent manner, with no geometrical
assumptions. We simply assumed that a linear average Hubble law exists in the
leading approximation, and then determined the best-fit Hubble law in
independent spherical shells, even in the regime of the nonlinear Hubble flow.
The conceptual picture behind such averages is illustrated in
Fig.~\ref{Hshells}.

Null geodesics (indicated by arrowed lines converging on a centre in
Fig.~\ref{Hshells}) which traverse scales larger than the SHS experience an
average expansion with $cz/r$ defining a Hubble constant equal to that
determined in spherical shells whose inner boundary is at least a few times
larger than the largest typical nonlinear structures. Below the SHS null
geodesics which traverse a single void will experience a higher expansion rate
than those that only traverse wall regions. We thus expect considerable
variation in the average values of $cz/r$ for sources in shells whose diameters
are comparable to the largest typical nonlinear structures. Since the largest
typical nonlinear structures are $\goesas30\hm$ diameter voids
\cite{hv02}--\cite{hv04}, and since these occupy a greater volume of space than
walls and filaments, we expect that a spherical average of $cz/r$ should in
general produce larger than average values of the Hubble `constant' on
scales below the SHS. Furthermore, if the results of \cite{sdb12} are
correct then an asymptotic average value of $\Hm$ should emerge on
$70\h$--$100\hm$ scales.

Finally, there is the question of the choice of cosmic rest frame. Since space
is differentially expanding below the SHS (as measured by one set of clocks),
the expansion law can be expected to differ from that of a spatially flat
geometry with rigid expansion plus local boosts. In the \TS\ scenario the
{\em finite infinity} scale defines the appropriate notion of a rest frame
(the CIR), and for bound systems this should be a scale on which space is
marginally expanding bounding a critical density volume. In addition to
determining averages in the conventional CMB rest frame, we have also
performed averages in the rest frames of the Local Group and the Local
Sheet\footnote{Since our galaxy is in a thin filamentary sheet in a local
environment dominated by voids \cite{t08}, the finite infinity scale should be
relatively near. For rich clusters of galaxies the scale is larger.} (LS)
\cite{t08}.

\begin{figure}[htb]
\centerline{\qquad\vbox{\halign{#\hfil\cr
\scalebox{0.36}{\includegraphics{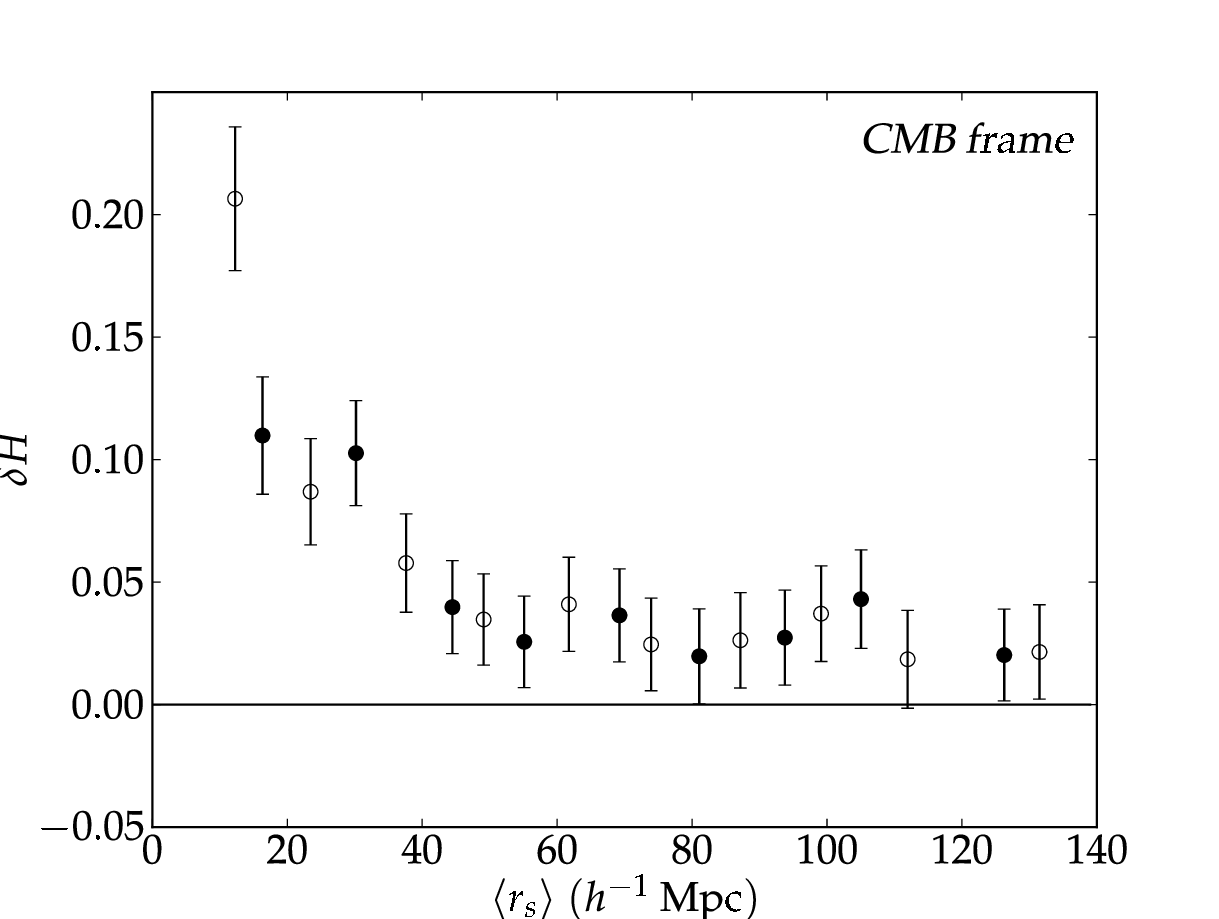}}\cr
\noalign{\vskip-5pt}\quad{\sbf(a)}\cr}}
\vbox{\halign{#\hfil\cr
\scalebox{0.36}{\includegraphics{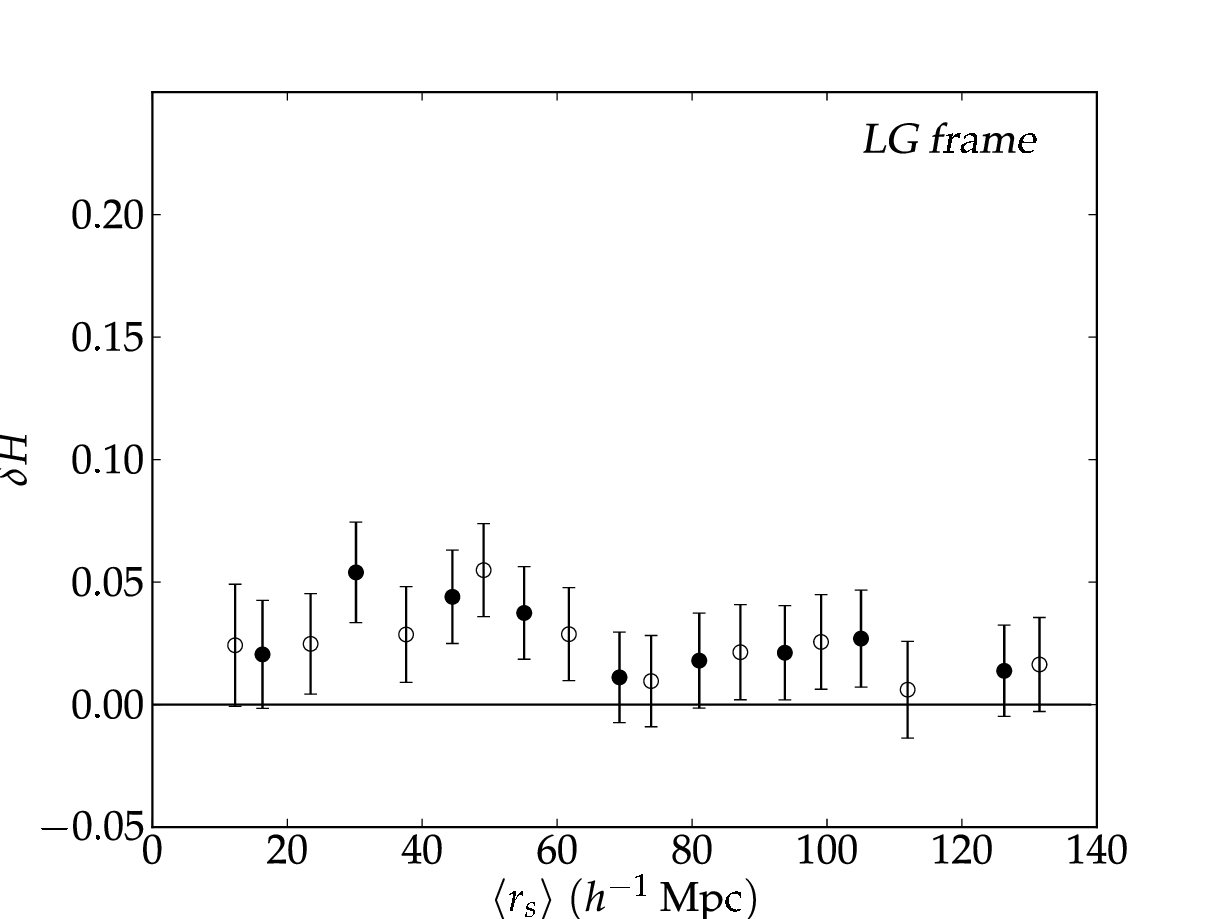}}\cr
\noalign{\vskip-5pt}\quad{\sbf(b)}\cr}}
}\caption{\label{dHs}%
Fractional variation in the Hubble flow $\de H_s=\left(H_s-{\bar H}\Z0\right)/
{\bar H}\Z0$ in spherical shells as a function of weighted mean shell distance:
{\bf(a)} CMB frame; {\bf(b)} LG frame. In each case the filled data points
represent one choice of shells boundaries, and the open data points a second
choice of shell boundaries. Each filled (open) data point is thus correlated to
the two nearest neighbour open (filled) data points.}
\end{figure}

In ref.\ \cite{wsmw} we analysed variation of the Hubble flow in the
COMPOSITE sample of 4,534 galaxies compiled by Watkins, Feldman and Hudson
\cite{wfh09,fwh10}. Spherical averages were computed in independent
shells\footnote{In earlier work, Li and Schwarz \cite{ls08} performed a
similar analysis of a subset of 54 distances from the Hubble Space Telescope
(HST) Key project data, in the CMB rest frame only. With a very small sample
they divided it into an inner and outer shell, with a moving boundary,
producing a correlated average.} with a minimum width of $12.5\hm$. We
minimized the sum $\chi^2=\sum_i\left[\si_i^{-1}(r_i-cz_i/H)\right]^2$ with
respect to $H$, where $z_i$, $r_i$ and $\si_i$ denote individual redshifts,
distances and distance uncertainties (in units $\hm$) respectively. This leads
to a value of the Hubble constant in the $s$th shell,
\beq
H_s=\left(\sum_{i=1}^{N_s}{(cz_i)^2\over\si_i^2}\right)\left(\sum_{i=1}^{N_s}
{cz_i r_i\over\si_i^2}\right)^{-1}\,.\label{Hs}
\eeq
Results for the fractional variation, $\de H_s=\left(H_s-\bH\Z0\right)/\bH\Z0$,
are plotted in Fig.~\ref{dHs} in the CMB and LG frames. Here $\bH\Z0$ is the
asymptotic value of the Hubble constant, determined from all the data in
the sample beyond $r>156.25\hm$. Results in the LS frame values are very
similar to the LG frame.

We see that the spherically averaged Hubble law is significantly more
uniform in the LG frame than in the CMB frame. In the inner shells the Bayesian
evidence in favour of the LG frame expansion being more uniform is very very
strong with $\ln B>10$ \cite{wsmw}. If the cosmic rest frame is defined as the
one in which the Hubble expansion is most uniform, with minimal statistical
variations, then from our vantage point the LG frame is much closer to having
this character. Such a result is completely unexpected and surprising from
the viewpoint of the standard cosmology, but does accord with the expectation
of the \TS\ scenario that the local finite infinity scale should define the
standard of rest for observers within a bound system. The frame of minimum
Hubble expansion variance still remains to be determined, and this may still
differ somewhat from the LG frame.

As discussed in ref.\ \cite{wsmw}, if one performs a random boost on each data
point, it involves replacing $cz_i\to cz_i'=cz_i+v\cos\ph_i$, where $\ph_i$
is the angle on the sky between the data point and the boost direction. In a
dataset with uniform sky coverage, terms linear in the boost velocity will be
roughly self-canceling inside the sums in (\ref{Hs}), leaving a leading order
average difference
\beq
H'_s-H_s\goesas{v^2\over2\Hm\ave{r_i^2}}\,.
\eeq
The differences between panels (a) and (b) in Fig.~\ref{dHs} do indeed
appear to have this character. This suggests that the persistent large scale
bulk flows seen in the standard peculiar velocity framework may arise largely
as a systematic error from choosing a cosmic rest frame which has a
significant boost with respect to the frame in which statistical variations of
the Hubble expansion are minimal.

\begin{figure}[htb]
\vbox{\halign{\hfil#\hfil\cr \noalign{\vskip10pt}
\scalebox{0.36}{\includegraphics{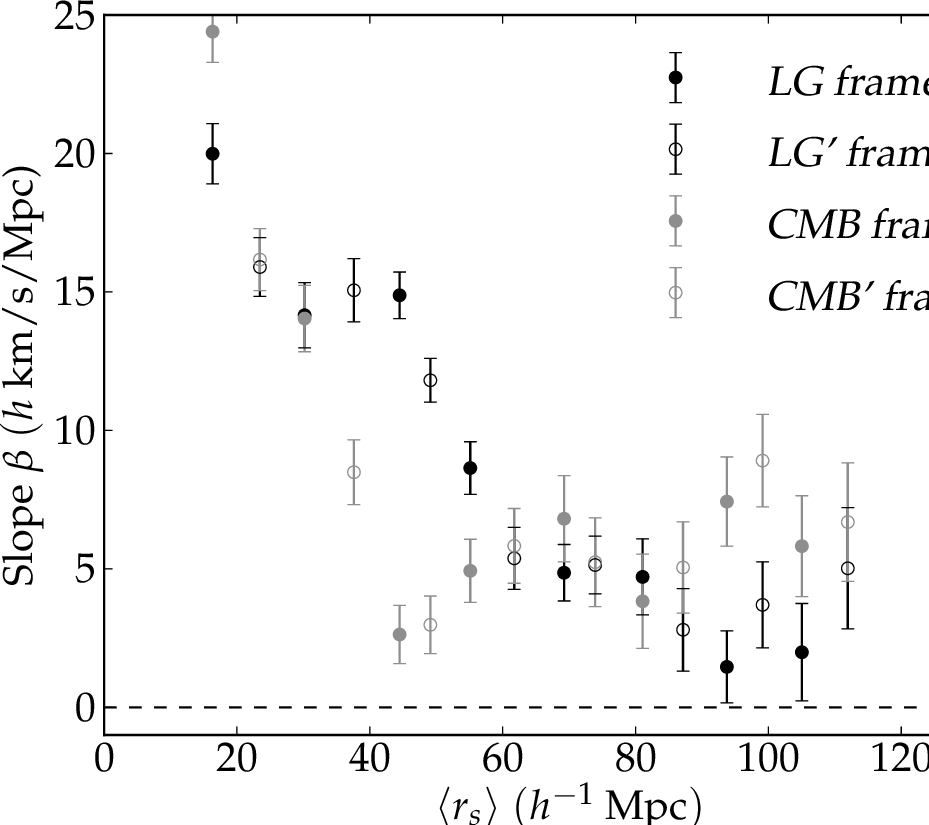}}\cr}}
\caption{\label{sdipfit}%
The slope $\be$ of the linear dipole relation $cz/r=H_d+\be\cos\ph$, plotted in
the same shells as Fig.~\ref{dHs} \cite{wsmw} in the CMB and LG rest frames.}
\end{figure}

An exception to the rule that $|\de H_s|$ is smaller in the LG frame than in
the CMB frame does occur for shells roughly in the range $40\h\lsim r\lsim
60\hm$. It turns out that there is also a LG frame dipole associated with
structures in this range.
In ref.\ \cite{wsmw} in addition to studying radial spherical variations we
also investigated angular variations by adapting a Gaussian window averaging
method of McClure and Dyer \cite{md07}. This established that a dipole is the
strongest angular multipole feature in both frames, but particularly in the LG
frame, until one reaches distances $r\gsim90\hm$. We then fitted a simple
dipole Hubble law
\beq\frac{cz}r=H_d+\be\cos\ph\,,\label{di}\eeq
in the same independent spherical shells used for the spherical averages.
This gave a dipole amplitude, $\be$, shown in Fig.~\ref{sdipfit}.

The magnitudes of the dipoles in the two frames coincide in the shell with mean
radius $\mr=30.2\hm$, and also in the shell with $\mr=61.7\hm$, but the dipoles
exhibit very different behaviour for the shells in between. In particular, the
CMB frame dipole magnitude reaches a minimum of $\be=(2.6\pm0.6)h\kmsMpc$
(close to zero) at $\mr=44.5\hm$, whereas for the LG frame $\be=(14.9\pm0.8)h
\kmsMpc$ in the same shell. The CMB frame dipole then increases while the LG
frame dipole decreases. The dipole directions for independent shells within
each frame are strongly consistent in the range $37.5\h\le r\le62.5\hm$. Beyond
$\mr=61.7\hm$, the CMB frame dipole maintains statistically significant
residual levels, while the LG frame dipole drops to a level statistically
consistent with zero around $90\h$--$100\hm$\footnote{Beyond this scale the
quadrupole becomes as strong as the dipole, and one needs to fit higher
multipoles. There is a hint of a small feature in the LG frame at scale
corresponding to one BAO distance beyond the nearest wall. However,
significantly more data is required in the outer shells to verify this.}.

It therefore appears that the boost to the CMB frame is largely compensating
for the effect of structures in the range $37.5\h\le r\le62.5\hm$, which
are also responsible for monopole variations of the Hubble `constant' in the
LG frame shown in Fig.~\ref{dHs}. The results
are consistent with a foreground density gradient producing an anisotropy in
the distance--redshift relation which is almost, but not exactly, of the
same nature as the Doppler shift produced by a Lorentz boost. Rather than
thinking purely about overdensities as in the attractor model, what is
important below the SHS are the peculiar foregrounds created by voids as well
as by superclusters. Those directions in which void--filled foregrounds are
opposed to wall regions on the opposite side of the sky will lead to the
strongest density gradients. Relevant structures are identified in
Sec.~III$\,$C of ref.~\cite{wsmw}.

\subsection{Origin of the CMB dipole}
It was further established in ref.\ \cite{wsmw} that the Gaussian window
averaged sky map of angular Hubble flow variation in the LG frame has a
correlation coefficient of $-0.92$ with the map that would be produced by the
residual CMB temperature dipole in the LG rest frame. The correlation
coefficient is insensitive to the choice of the Gaussian window smoothing
angle in the range $15\deg<\si_\th<40\deg$.

The strong correlation of the two sky maps is consistent with
the hypothesis that the CMB temperature dipole is only partly due to a
Lorentz boost. The portion usually attributed to the motion of the LG might
be largely due to the differential expansion of space produced by
peculiar foregrounds below the SHS. In the LG frame the residual temperature
dipole is $\de T=\pm(5.77\pm0.36)\,$mK. For the \LCDM\ model with Planck
best-fit parameters, this would be produced by an anisotropy $\de D=\mp(0.30
\pm0.02)\hm$ in the distance to the surface of last scattering. The value for
the \TS\ model is essentially the same. If produced by the differential
expansion of foreground structures within a mean distance of $60\hm$, this
amounts to a 0.5\% anisotropy on these scales, which is entirely
plausible\footnote{As is discussed in Sec.~VI$\,$A of ref.\ \cite{wsmw} a
multipole decomposition of the Hubble expansion should be developed, in
spherical shells, to determine whether the dipole converges to the required
amplitude.}.

There is differential expansion of space below the SHS at all locations, of
a magnitude bounded by the growth of structure from the initial perturbations
at the surface of last scattering. CMB photons which traverse large distances
see an average of all of these variations, producing an average distance to
the last scattering surface. However, the last stretch of the journey produces
slight differences that depend on peculiar density foregrounds. The same
small residual anisotropies will apply to all cosmic distance measurements on
scales much greater than the SHS, but at a level smaller than the uncertainties
in many large scale measurements.

Although it might generally be expected that a differential expansion of space
would produce higher order multipole CMB anisotropies of a magnitude
comparable to the CMB dipole, these are in fact highly suppressed for
off--centre observers in voids \cite{aa06}, which represent a good first
approximation for our actual local environment \cite{t08}. Ray tracing in a
simple LTB void with parameters matched to the LG frame Hubble dipole of
Fig.~\ref{sdipfit} gives a CMB quadrupole/dipole ratio of less than 0.1\%
\cite{wsmw}. Using the Szekeres models \cite{Sz,bw09} one can perform ray
tracing through exact solution geometries which more closely mimic our peculiar
foregrounds; while a higher quadrupole/dipole CMB ratio of order 1\% is found,
this is still observationally viable \cite{bnw}.

The suggestion that a large fraction of the CMB dipole is not purely due to
a boost is, of course, a radical departure for observational cosmology.
However, a number of potential anomalies have been observed in the large
angles multipoles of the CMB anisotropy spectrum for a decade now, and
their significance has increased with the release of the Planck data
\cite{Piso}. A study by Freeman \etal\ \cite{fgm06} found that of several
possible systematic errors, a 1--2\% error in the CMB dipole subtraction stood
out as the one possible effect which could potentially resolve the power
asymmetry anomaly.

As is discussed in ref.\ \cite{wsmw} a nonkinematic contribution to
the foreground Hubble expansion may also explain why attempts to measure the
effects of aberration and frequency modulation in the Doppler boosting of the
CMB spectrum yield a boost direction which moves across the sky when only large
angle multipoles are considered \cite{Pboost}. Aberration and frequency
modulation can also be readily tested in the radio spectrum. Rubart and
Schwarz \cite{rs13} have recently found that the assumption of a kinematic
origin for the cosmic radio galaxy dipole is inconsistent at the 99.5\%
confidence level, using the NRAO VLA Sky Survey data. The direction of the
radio dipole is consistent with that of the Hubble variance dipole we find
in the LG frame.

\section{Conclusion}

Observations over the last few decades have revealed a universe much more
complex, varied and interesting than had been previously imagined. The
observations are at present well in advance of our theoretical understanding.
The phenomenon of apparent cosmic acceleration demands that we
think more deeply about one of the central unsolved problems of general
relativity: the nature of gravitational mass--energy, which cannot be localized
on account of the equivalence principle. The standard \LCDM\ model adds
cold dark matter to make gravity stronger at some scales and then adds dark
energy to make gravity weaker at larger scales, while keeping space rigidly
expanding. Both phenomena may be indicative of a renormalization of the notion
of gravitational mass in a hierarchy of nonrigid spacetime structures.

The \TS\ scenario is a from--first--principles attempt to come to grips
with the essential physics of the fitting problem \cite{fit1,fit2}, and to
specify a physically viable interpretation of the Buchert averaging scheme
\cite{Brev,Brev2,buch1} without a smooth ``dark energy''. A
phenomenological model has been developed \cite{clocks,obs,dnw}, which had
remained observationally viable over the six years since its conception
\cite{clocks}. Much work remains to be done. In particular, while the
tests of Secs.~\ref{lumd}--\ref{timed} give a means of comparing
average cosmological quantities with those of the \LCDM\ model, the
most exciting developments are to be made by considering physics
below the SSH. This may also inform the development of tests related to
the average growth of structure.

It is below the SSH, where the structures are most inhomogeneous, that the
most interesting differences between the \TS\ scenario and the standard
cosmology are to be found. In the \LCDM\ model spacetime is spatially flat
on these scales, while in the \TS\ scenario its spatial curvature varies
greatly. A rigorous mathematical description of the statistical geometry
on these scales remains to be determined. However, to develop such a
description I believe we should not be guided simply by mathematical
elegance but by physical principles and observations.

The simple idea that the finite infinity scale should define the cosmic
rest frame for bound system observers led to the idea of testing the Hubble
expansion variation in the LG and LS frames, as well as in the CMB frame,
with a result that was much more definitive than we ourselves anticipated
\cite{wsmw}. Since the analysis of Sec.~\ref{hvar} is model--independent it
is not a direct verification of the \TS\ scenario; but it is consistent
with the \TS\ model and it is extremely hard to reconcile with the
standard cosmology. As discussed in ref.\ \cite{wsmw}, a change to our
understanding of the local Hubble expansion variation and its effect on the CMB
dipole may have some impact on many different aspects of observational
cosmology, including not only CMB anomalies but also SneIa systematics and
the calibration of the distance scale.

\section*{Acknowledgments} This work was supported by the Marsden fund of the
Royal Society of New Zealand. I thank Krzysztof Bolejko, Teppo Mattsson, Ahsan
Nazer, Peter Smale, Nezihe Uzun and Rick Watkins for helpful discussions and
correspondence. I also wish to thank M\'ario Novello and Santiago
Perez Bergliaffa for their support and for
organizing a very memorable cosmology school.
\bibliographystyle{aipproc}

\begin{thebibliography}{99}
\bibitem{Pparm}
P.~A.~R.~Ade, N.~Aghanim, C.~Armitage-Caplan, \etal, 
\AaA{571}, A16 (2014).

\bibitem{hv02}
F.~Hoyle and M.~S.~Vogeley,
\ApJ{566}, 641 (2002).

\bibitem{hv04}
F.~Hoyle and M.~S.~Vogeley,
\ApJ{607}, 751 (2004).

\bibitem{pan11}
D.~C.~Pan, M.~S.~Vogeley, F.~Hoyle, Y.~Y.~Choi, and C.~Park,
\MNRAS{421}, 926 (2012).

\bibitem{minivoids}
A.~V.~Tikhonov and I.~D.~Karachentsev,
\ApJ{653} (2006) 969. 

\bibitem{web}
J.~E.~Forero-Romero, Y.~Hoffman, S.~Gottloeber, A.~Klypin and G.~Yepes,
\MNRAS{396}, 1815 (2009).

\bibitem{map}
J.~R.~Gott, M.~Juric, D.~Schlegel, F.~Hoyle, M.~Vogeley, M.~Tegmark,
N.~Bahcall and J.~Brinkmann,
\ApJ{624}, 463 (2005).

\bibitem{esu}
A.~Einstein,
Sitzungsber.\ Preuss.\ Akad.\ Wiss., 142 (1917).
[English translation in {\em``The collected papers of Albert Einstein.\ Vol.\
{\bf6}''}, (Princeton University Press, 1997) pp.\ 421--432]

\bibitem{ES}
A.~Einstein and E.~G.~Straus,
\RMP{17}, 120 (1945); Err.\ {\bf18}, 148 (1946).

\bibitem{fit1}
G.~F.~R.~Ellis,
in B.~Bertotti, F.~de Felice and A.~Pascolini (eds), {\em General
Relativity and Gravitation}, (Reidel, Dordrecht, 1984) pp.~215--288.

\bibitem{fit2}
G.~F.~R.~Ellis and W.~Stoeger,
\CQG{4}, 1697 (1987).

\bibitem{Bondi}
H.~Bondi,
{\em Cosmology},
(Cambridge University Press, 1961)

\bibitem{clocks}
D.~L.~Wiltshire,
{\it New J.\ Phys.}\ {\bf9}, 377 (2007).

\bibitem{sol}
D.~L.~Wiltshire,
\PRL{99}, 251101 (2007).

\bibitem{equiv}
D.~L.~Wiltshire,
\PR D{78}, 084032 (2008).

\bibitem{obs}
D.~L.~Wiltshire,
\PR D{80}, 123512 (2009).

\bibitem{m83}
M.~Milgrom,
\ApJ{270}, 365 (1983).

\bibitem{sm02}
R.~H.~Sanders and S.~S.~McGaugh,
Ann.\ Rev.\ \AaA{40}, 263 (2002).

\bibitem{CT1}
F.~I.~Cooperstock and S.~Tieu,
\MPL A{21}, 2133 (2006).

\bibitem{CT2}
F.~I.~Cooperstock and S.~Tieu,
\IJMP A{22}, 2293 (2007).

\bibitem{BG}
H.~Balasin and D.~Grumiller,
\IJMP D{17}, 475 (2008).

\bibitem{RS}
A.~Raki\'c and D.~J.~Schwarz,
{\it PoS} {\bf IDM2008}, 096 (2008) [arXiv:0811.1478].

\bibitem{L}
G.~Lema\^{\i}tre,
{\it Ann.\ Soc.\ Sci.\ Bruxelles} A {\bf53}, 51 (1933)
[\GRG{29}, 641 (1997)].
\bibitem{T}
R.~C.~Tolman,
Proc.\ Nat.\ Acad.\ Sci.\ {\bf 20}, 169 (1934);
\bibitem{B}
H.~Bondi,
\MNRAS{107}, 410 (1947).

\bibitem{BKH}
K.~Bolejko, A.~Krasi\'nski and C.~Hellaby,
\MNRAS{362}, 213 (2005).

\bibitem{ES09}
G.~F.~R.~Ellis and W.~Stoeger,
\MNRAS{398}, 1527 (2009).

\bibitem{chr}
R.~G.~Clowes, K.~A.~Harris, S.~Raghunathan, \etal,
\MNRAS{429}, 2910 (2013).

\bibitem{h05}
D.~W.~Hogg, D.~J.~Eisenstein, M.~R.~Blanton,
N.~A.~Bahcall, J.~Brinkmann, J.~E.~Gunn, and D.~P.~Schneider,
\ApJ{624}, 54 (2005).

\bibitem{sl09}
F.~Sylos Labini, N.~L.~Vasilyev, L.~Pietronero, and Y.~V.~Baryshev,
Europhys.\ Lett.\ {\bf 86}, 49001 (2009).

\bibitem{sdb12}
M.~Scrimgeour, T.~Davis, C.~Blake, \etal,
\MNRAS{425}, 116 (2012).

\bibitem{bao1}
D.~J.~Eisenstein, I.~Zehavi, D.~W.~Hogg, \etal,
\ApJ{633} 560, (2005).

\bibitem{bao2}
S.~Cole, W.~J.~Percival, J.~A.~Peacock, \etal, 
\MNRAS{362}, 505 (2005).

\bibitem{paradox1}
F.~Sylos Labini, M.~Montuori and L.~Pietronero,
Phys.\ Rep.\ {\bf293}, 61 (1998).

\bibitem{bary}
Y.~Baryshev,
AIP Conf.\ Proc.\ {\bf822}, 23 (2006).

\bibitem{paradox2}
A.~D.~Chernin, I.~D.~Karachentsev, M.~J.~Valtonen, V.~P.~Dolgachev,
L.~M.~Domozhilova and D.I.~Makarov,
\AaA{415}, 19 (2004).

\bibitem{STH}
A.~Sandage, G.A.~Tammann and E.~Hardy,
\ApJ{172}, 253 (1972).

\bibitem{deVau}
G.~de~Vaucouleurs,
Science {\bf167}, 1203 (1970).

\bibitem{ap02}
M.~Axenides and L.~Perivolaropoulos,
\PR D{65}, 127301 (2002).

\bibitem{Ko09}
M.~Korzy\'nski,
\CQG{27}, 105015 (2010).

\bibitem{ADM}
R.~Arnowitt, S.~Deser and C.~W.~Misner,
in L.~Witten (ed), {\it Gravitation: An introduction to current research},
(Wiley, New York, 1962) pp.~227--265.

\bibitem{LW}
R.~W.~Lindquist and J.~A.~Wheeler,
\RMP{29}, 432 (1957); Err.\ {\bf31}, 839 (1959).

\bibitem{CF}
T.~Clifton and P.~G.~Ferreira,
\PR D{80}, 103503 (2009).

\bibitem{Cl10}
T.~Clifton,
\CQG{28}, 164011 (2011)

\bibitem{UEL}
J.~P.~Uzan, G.~F.~R.~Ellis and J.~Larena,
\GRG{43}, 191 (2011).

\bibitem{CFO}
T.~Clifton, P.~G.~Ferreira and K.~O'Donnell,
\PR D{85}, 023502 (2012).

\bibitem{dust}
D.~L.~Wiltshire,
\CQG{28}, 164006 (2011).

\bibitem{Brev}
T.~Buchert,
\GRG{40}, 467 (2008).

\bibitem{Brev2}
T.~Buchert,
\CQG{28}, 164007 (2011).

\bibitem{vdH}
R.~J.~van den Hoogen,
in T.~Damour, R.~T.~Jantzen and R.~Ruffini (eds), {\it Proceedings of the 12th
Marcel Grossmann Meeting on General Relativity}, (World Scientific,
Singapore, 2012) pp.~578-589.

\bibitem{CELU}
C.~Clarkson, G.~F.~R.~Ellis, J.~Larena and O.~Umeh,
{\em Rept.\ Prog.\ Phys.}\ {\bf74}, 112901 (2011).

\bibitem{Z1}
R.~M.~Zalaletdinov,
\GRG{24}, 1015 (1992).

\bibitem{Z2}
R.~M.~Zalaletdinov,
\GRG{25}, 673 (1993).

\bibitem{Z3}
R.~M.~Zalaletdinov,
{\it Bull.\ Astron.\ Soc.\ India} {\bf 25}, 401 (1997).

\bibitem{FDU1}
P.~Fleury, H.~Dupuy and J.~P.~Uzan,
\PR D{87}, 123526 (2013).

\bibitem{FDU2}
P.~Fleury, H.~Dupuy and J.~P.~Uzan,
\PRL{111}, 091302 (2013).

\bibitem{MN11}
V.~Marra and A.~Notari,
\CQG{28}, 164004 (2011).

\bibitem{K11}
E.~W.~Kolb,
\CQG{28}, 164009 (2011).

\bibitem{Peebles}
P.~J.~E.~Peebles,
{\em AIP Conf.\ Proc.}\ {\bf1241}, 175 (2010). 

\bibitem{CP}
M.~Carfora and K.~Piotrkowska,
\PR D{52}, 4393 (1995). 

\bibitem{CM}
M.~Carfora and A.~Marzuoli,
\PRL{53}, 2445 (1984).

\bibitem{BC1}
T.~Buchert and M.~Carfora,
\CQG{19}, 6109 (2002).

\bibitem{CB}
M.~Carfora and T.~Buchert,
in {\em``Waves and Stability in Continuous Media''}
eds. M.~Nanganaro, M.~Ronaco and R.~Sionero, (World Scientific, Singapore,
2008) pp.~118-127 [arXiv:0801.0553].

\bibitem{A09}
C.~Anastopoulos,
\PR D{79}, 084029 (2009).

\bibitem{BvdHC}
J.~Brannlund, R.~J.~van den Hoogen and A.~A.~Coley,
\IJMP D{19}, 1915 (2010).

\bibitem{Re08}
M.~Reiris,
\CQG{25}, 085001 (2008).
\bibitem{Re09}
M.~Reiris,
\GRG{41}, 1083 (2009).

\bibitem{MZ}
M.~Mars and R.~M.~Zalaletdinov,
\JMP{38}, 4741 (1997).

\bibitem{PS07}
A.~Paranjape and T.~P.~Singh,
\PR D{76}, 044006 (2007).

\bibitem{CPZ}
A.~A.~Coley, N.~Pelavas and R.~M.~Zalaletdinov,
\PRL{95}, 151102 (2005).
\bibitem{P08}
A.~Paranjape,
\PR D{78}, 063522 (2008).
\bibitem{PS08}
A.~Paranjape and T.~P.~Singh,
\PRL{101}, 181101 (2008).
\bibitem{vdH09}
R.~J.~van den Hoogen,
\JMP{50}, 082503 (2009).

\bibitem{BE}
T.~Buchert and J.~Ehlers,
\AaA{320}, 1 (1997).

\bibitem{EB}
J.~Ehlers and T.~Buchert,
\GRG{29}, 733 (1997).

\bibitem{buch1}
T.~Buchert,
\GRG{32}, 105 (2000).

\bibitem{buch2}
T.~Buchert,
\GRG{33}, 1381 (2001).

\bibitem{BBR}
J.~Behrend, I.~A.~Brown and G.~Robbers,
\JCAP{01}{2008}013.

\bibitem{L09}
J.~Larena,
\PR D{79}, 084006 (2009).

\bibitem{BBM}
I.~A.~Brown, J.~Behrend and K.~A.~Malik,
\JCAP{11}{2009}027.

\bibitem{CAL}
C.~Clarkson, K.~Ananda and J.~Larena,
\PR D{80}, 083525 (2009).

\bibitem{GMV}
M.~Gasperini, G.~Marozzi and G.~Veneziano,
\JCAP{02}{2010}009.

\bibitem{R10}
S.~R\"as\"anen,
\JCAP{03}{2010}018.

\bibitem{RSKB}
H.~Russ, M.~H.~Soffel, M.~Kasai and G.~B\"orner,
\PR D{56}, 2044 (1997). 

\bibitem{B80}
J.~M.~Bardeen,
\PR D{22}, 1882 (1980).

\bibitem{BKL}
J.~Bi\v{c}\'ak, J.~Katz and D.~Lynden-Bell,
\PR D{76}, 063501 (2007).

\bibitem{SY}
L.~Smarr and J.~W.~York,
\PR D{17}, 2529 (1978).

\bibitem{quasirev}
L.~B.~Szabados,
{\it Liv.\ Rev.\ Rel.}\ {\bf7}, 4 (2004).

\bibitem{K59}
A.~Komar
\PR{}{113}, 934 (1959).

\bibitem{H96}
M.~Heusler,
{\em``Black hole uniqueness theorems''},
(Cambridge University Press, 1996) pp.~14--17.

\bibitem{BY}
J.~D.~Brown and J.~W.~York,
\PR D{47}, 1407 (1993).

\bibitem{E00}
R.~J.~Epp,
\PR D{62}, 124018 (2000). 

\bibitem{CLN}
C.~M.~Chen, J.~L.~Liu and J.~M.~Nester,
\MPL A{22}, 2039 (2007).

\bibitem{NSV}
J.~M.~Nester, L.~L.~So and T.~Vargas,
\PR D{78}, 044035 (2008).

\bibitem{WCLN}
M.~F.~Wu, C.~M.~Chen, J.~L.~Liu and J.~M.~Nester,
\PL A{374}, 3599 (2010).

\bibitem{S08a}
R.~A.~Sussman,
\PR D{79}, 025009 (2009).

\bibitem{af09}
M.~M.~Afshar,
\CQG{26}, 225005 (2009).[arXiv:0903.3982]

\bibitem{BC2}
T.~Buchert and M.~Carfora,
\PRL{90}, 031101 (2003).

\bibitem{V12}
S.~Viaggiu,
2012 \CQG{29} 035016

\bibitem{VM}
S.~Viaggiu and M.~Montuori,
\IJMP D{22}, 1350065 (2013).

\bibitem{BC3}
T.~Buchert and M.~Carfora,
\CQG{25}, 195001 (2008).

\bibitem{WB}
A.~Wiegand and T.~Buchert,
\PR D{82}, 023523 (2010).

\bibitem{dnw}
J.~A.~G.~Duley, M.~A.~Nazer, and D.~L.~Wiltshire,
\CQG{30}, 175006 (2013).

\bibitem{McGaugh}
S.~S.~McGaugh,
\ApJ{683}, 137 (2008). 

\bibitem{LNW}
B.~M.~Leith, S.~C.~C.~Ng and D.~L.~Wiltshire,
\ApJ{672}, L91 (2008).

\bibitem{SW}
P.~R.~Smale and D.~L.~Wiltshire,
\MNRAS{413}, 367 (2011).

\bibitem{grb}
P.~R.~Smale,
\MNRAS{418}, 2779 (2011).

\bibitem{JRK}
S.~Jha, A.~G.~Riess and R.~P.~Kirshner,
\ApJ{659}, 122 (2007). 

\bibitem{guy05}
J.~Guy, P.~Astier, S.~Nobili, N.~Regnault and R.~Pain,
\AaA{443}, 781 (2005).

\bibitem{guy07}
J.~Guy, P.~Astier, S.~Baumont, \etal,
\AaA{466}, 11 (2007).

\bibitem{2mod}
D.~L.~Wiltshire,
in J.~A.~Auping (ed), {\it Proceedings of the International Conference on Two
Cosmological Models}, (Plaza y Vald\'es, Mexico City, 2012) pp.~361-384
[arXiv:1102.2045].

\bibitem{mg12}
D.~L.~Wiltshire,
in T.~Damour, R.~T.~Jantzen and R.~Ruffini (eds), {\it Proceedings of the 12th
Marcel Grossmann Meeting on General Relativity}, (World Scientific,
Singapore, 2012) pp.~434-452.

\bibitem{R07}
A.~G.~Riess, L.-G.~Strolger, S.~Casertano, \etal, 
\ApJ{659}, 98 (2007). 

\bibitem{Kessler09}
R.~Kessler, A.~Becker, D.~Cinabro, \etal,
\ApJ{} {\it Suppl.}\ {\bf185}, 32 (2009).

\bibitem{H09}
M.~Hicken, W.~M.~Wood-Vasey, S.~Blondin,
P.~Challis, S.~Jha, P.~L.~Kelly, A.~Rest and R.~P.~Kirshner,
\ApJ{700}, 1097 (2009).

\bibitem{GRB1}
B.~E.~Schaefer,
\ApJ{660}, 16 (2007).

\bibitem{GRB2}
N.~Liang, W.~K.~Xiao, Y.~Liu and S.~N.~Zhang,
\ApJ{685}, 354 (2008).

\bibitem{GRB3}
L.~Amati, C.~Guidorzi, F.~Frontera, M.~Della Valle, F.~Finelli, R.~Landi
and E.~Montanari,
\MNRAS{391}, 577 (2008).

\bibitem{GRB4}
L.Amati and M.~Della Valle,
\IJMP D{22}, 1330028 (2013).

\bibitem{pill}
R.~H.~Cyburt, B.~D.~Fields and K.~A.~Olive,
\JCAP{11}{2008}012.

\bibitem{bbn1}
D.~D.~Tytler, J.~M.~O'Meara, N.~Suzuki and D.~Lubin,
Phys.\ Scripta T {\bf85}, 12 (2000).

\bibitem{bbn2}
G.~Steigman,
\IJMP E{15}, 1 (2006).

\bibitem{rob13}
B.~F.~Roukema, J.~J.~Ostrowski and T.~Buchert,
\JCAP{10}{2013}043.

\bibitem{ZZ}
G.~B.~Zhao and X.~Zhang,
\PR D{81}, 043518 (2010). 

\bibitem{SCHMPS}
P.~Serra, A.~Cooray, D.~E.~Holz, A.~Melchiorri, S.~Pandolfi and D.~Sarkar,
\PR D{80}, 121302 (2009). 

\bibitem{GCC}
J.~A.~Gu, C.~W.~Chen and P.~Chen,
{\em New J.\ Phys.}\ {\bf11}, 073029 (2009).

\bibitem{ZC}
C.~Zunckel and C.~Clarkson,
\PRL{101}, 181301 (2008).

\bibitem{SSS}
V.~Sahni, A.~Shafieloo and A.~A.~Starobinsky,
\PR D{78}, 103502 (2008).

\bibitem{SSS2}
A.~Shafieloo, V.~Sahni and A.~A.~Starobinsky,
\PR D{80}, 101301 (2009).

\bibitem{AP}
C.~Alcock and B.~Paczy\'nski,
{\it Nature} {\bf281}, 358 (1979).

\bibitem{wz}
C.~Blake, K.~Glazebrook, T.~Davis, \etal,
\MNRAS{418}, 1725 (2011).

\bibitem{gal}
L.~Anderson, E.~Aubourg, S.~Bailey, \etal,
\MNRAS{427}, 343 (2013). 

\bibitem{boss}
A.~Slosar, V.~Irsic, D.~Kirkby, \etal,
\JCAP{04}{2013}026.

\bibitem{CBL}
C.~Clarkson, B.~Bassett and T.~C.~Lu,
\PRL{101}, 011301 (2008).

\bibitem{S62}
A.~Sandage,
\ApJ{136}, 319 (1962).
\bibitem{McV}
G.~C.~McVittie,
\ApJ{136}, 334 (1962).
\bibitem{L98}
A.~Loeb,
\ApJ{499}, L111 (1998).

\bibitem{ELT1}
P.~S.~Corasaniti, D.~Huterer and A.~Melchiorri,
\PR D{75}, 062001 (2007).
\bibitem{ELT2}
J.~Liske, A.~Grazian, E.~Vanzella, \etal,
\MNRAS{386}, 1192 (2008).

\bibitem{p93}
P.~J.~E.~Peebles,
{\em Principles of Physical Cosmology},
(Princeton University Press, 1993).

\bibitem{l05}
E.~V.~Linder,
\PR D{72}, 043529 (2005).

\bibitem{ltmc}
G.~Lavaux, R.~B.~Tully, R.~Mohayaee, and S.~Colombi,
\ApJ{709}, 483 (2010).

\bibitem{bcmj}
M.~Bilicki, M.~Chodorowski, G.~A.~Mamon, and T.~Jarrett,
\ApJ{741}, 31 (2011).

\bibitem{el06}
P.~Erdo\u{g}du, O.~Lahav, J.~P.~Huchra, \etal,
\MNRAS{373}, 45 (2006).

\bibitem{wfh09}
R.~Watkins, H.A.~Feldman, and M.J.~Hudson,
\MNRAS{392}, 743 (2009).

\bibitem{fwh10}
H.A.~Feldman, R.~Watkins, and M.J.~Hudson,
\MNRAS{407}, 2328 (2010).

\bibitem{kash08}
A.~Kashlinsky, F.~Atrio-Barandela, D.~Kocevski, and H.~Ebeling,
\ApJ{686}, L49 (2008).

\bibitem{kash10}
A.~Kashlinsky, F.~Atrio-Barandela, H.~Ebeling, A.~Edge, and D.~Kocevski,
\ApJ{712}, L81 (2010).

\bibitem{dn11}
A.~Nusser and M.~Davis,
\ApJ{736}, 93 (2011). 

\bibitem{ms13}
Y.Z.~Ma and D.~Scott,
\MNRAS{428}, 2017 (2013).

\bibitem{thf12}
S.J.~Turnbull, M.J.~Hudson, H.A.~Feldman,
M.~Hicken, R.P.~Kirshner, and R.~Watkins,
\MNRAS{420}, 447 (2012).

\bibitem{sfact}
U.~Feindt, M.~Kerschhaggl, M.~Kowalski, \etal,
\AaA{560}, A90 (2013).

\bibitem{wsmw}
D.~L.~Wiltshire, P.~R.~Smale, T.~Mattsson and R.~Watkins,
\PR D{88}, 083529 (2013). 

\bibitem{t08}
R.B.~Tully, E.J.~Shaya, I.D.~Karachentsev, H.~Courtois, D.D.~Kocevski,
L.~Rizzi and A.~Peel,
\ApJ{676}, 184 (2008). 

\bibitem{ls08}
N.~Li and D.J.~Schwarz,
\PR D{78}, 083531 (2008).

\bibitem{md07}
M.L.~McClure and C.C.~Dyer,
New Astron.\ {\bf12}, 533 (2007).

\bibitem{aa06}
H.~Alnes and M.~Amarzguioui,
\PR D{74}, 103520 (2006).

\bibitem{Sz}
P.~Szekeres,
Commun.\ Math.\ Phys.\ {\bf41}, 55 (1975).

\bibitem{bw09}
K. Bolejko and J.S.B.~Wyithe,
\JCAP{02}{2009}020.

\bibitem{bnw}
K.~Bolejko, M.A.~Nazer and D.L.~Wiltshire,
in preparation.

\bibitem{Piso}
P.~A.~R.~Ade, N.~Aghanim, C.~Armitage-Caplan, \etal, 
\AaA{571}, A23 (2014).

\bibitem{fgm06}
P.E.~Freeman, C.R.~Genovese, C.J.~Miller, R.C.~Nichol, and L.~Wasserman,
\ApJ{638}, 1 (2006).

\bibitem{Pboost}
N.~Aghanim, C.~Armitage-Caplan, M.~Arnaud, \etal, 
\AaA{571}, A27 (2014).

\bibitem{rs13}
M.~Rubart and D.J.~Schwarz,
\AaA{555}, A117 (2013).
\end{thebibliography}

\end{document}